\documentclass[10pt,a4paper,superscriptaddress,aps,prd,showkeys,showpacs,nofootinbib,reprint]{revtex4-1}
\usepackage[utf8]{inputenc}
\usepackage{verbatim}
\usepackage{amsmath,amsfonts,amssymb}
\usepackage[breaklinks,colorlinks]{hyperref}
\usepackage{natbib}
\usepackage{tikz}
\usepackage{float}
\usepackage{graphicx}
\usepackage{adjustbox}
\usepackage{tabularx}
\usepackage{amsbsy}
\hypersetup{%
,urlcolor=blue
,citecolor=blue
,linkcolor=blue
}

\def\aap{Astronomy and Astrophysics}
\def\apjl{The Astrophysical Journal Letters}
\def\physrep{Physics Reports}
\def\araa{Annual Review of Astronomy and Astrophysics}
\def\mnras{Monthly Notices of the Royal Astronomical Society}
\def\aj{Astronomical Journal}
\def\fcp{Fundamental Cosmic Physics}
\def\jcap{Journal of Cosmology and Astro-Particle Physics}
\def\corr{{\cal F}^{\rm eff}_\gamma}

\begin{document}

\title{Dark matter axion detection in the radio/mm-waveband}

\author{R.~A.~Battye}
\email[]{richard.battye@manchester.ac.uk}
\affiliation{%
Jodrell Bank Centre for Astrophysics, School of Natural Sciences, Department of Physics and Astronomy, University of Manchester, Manchester, M13 9PL, U.K.
}

\author{B.~Garbrecht}
\email[]{garbrecht@tum.de}
\affiliation{%
Technische Universität München, Physik-Department, James-Franck-Straße, 85748 Garching, Germany
}

\author{J.~I.~McDonald}
\email[]{jamie.mcdonald@tum.de}
\affiliation{%
Technische Universität München, Physik-Department, James-Franck-Straße, 85748 Garching, Germany
}

\author{F.~Pace}
\email[]{francesco.pace@manchester.ac.uk}
\affiliation{%
Jodrell Bank Centre for Astrophysics, School of Natural Sciences, Department of Physics and Astronomy, University of Manchester, Manchester, M13 9PL, U.K.
}
\author{S.~Srinivasan }
\email[]{sankarshana.srinivasan@postgrad.manchester.ac.uk}
\affiliation{%
Jodrell Bank Centre for Astrophysics, School of Natural Sciences, Department of Physics and Astronomy, University of Manchester, Manchester, M13 9PL, U.K.
}

\label{firstpage}

\date{\today}

\begin{abstract}
We discuss axion dark matter detection via two mechanisms: spontaneous decays and resonant conversion in neutron star magnetospheres. For decays, we show that the brightness temperature signal, rather than flux, is a less ambiguous measure for selecting candidate objects. This is owing principally to the finite beam width of telescopes which prevents one from being sensitive to the total flux from the object. With this in mind, we argue that the large surface-mass-density of the galactic centre or the Virgo cluster centre offer the best chance of improving current constraints on the axion-photon coupling via spontaneous decays. For the neutron star case, we first carry out a detailed study of mixing in magnetised plasmas. We derive transport equations for the axion-photon system via a controlled gradient expansion, allowing us to address inhomogeneous mass-shell constraints for arbitrary momenta. We then derive a non-perturbative Landau-Zener formula for the conversion probability valid across the range of relativistic and non-relativistic axions and show that the standard perturbative resonant conversion amplitude is a truncation of this result in the non-adiabatic limit. Our treatment reveals that that infalling dark matter axions typically convert non-adiabatically in magnetospheres. We describe the limitations of one-dimensional mixing equations and explain how three-dimensional effects activate new photon polarisations, including longitudinal modes and illustrate these arguments with numerical simulations in higher dimensions. We find that the bandwidth of the radio signal from neutron stars could be dominated by Doppler broadening from the oblique rotation of the neutron star if the axion is non-relativistic in the conversion region. Therefore, we conclude that the radio signal from the resonant decay is weaker than previously thought, which means one relies on local density peaks to probe weaker axion-photon couplings.
\end{abstract}



\maketitle

\section{Introduction}
\label{sec:intro}
Understanding the exact nature of dark matter remains one of the major challenges in particle physics and cosmology. One particularly simple solution to the dark matter problem is offered by the QCD axion which results from the breaking of Peccei-Quinn (PQ) symmetry~\cite{ref:PQ}, proposed as a resolution to the strong CP problem of Quantum Chromodynamics (QCD). There are a number of specific ways to incorporate the axion into the Standard Model of particle physics; the most common being the KSVZ~\cite{ref:K, ref:SVZ} and the DFSZ~\cite{ref:DFSZ, ref:Zhit} models. Soon after the realisation that the axion was a natural consequence of PQ symmetry, it was pointed out that it could be produced by the non-thermal misalignment mechanism~\cite{ref:misalign1, ref:misalign2, ref:misalign3} and that its relic abundance and low momentum would allow it to be a Cold Dark Matter (CDM) candidate. The axion has since been subject of extensive theoretical work and has been proposed as a candidate for a number of other cosmological phenomena (see \cite{ref:Marsh} for a recent review). In what follows, we will make the assumption that axions are responsible for all the CDM in the Universe and discuss their detection in the radio/mm-waveband.

A recent detailed calculation~\cite{ref:WS} of the misalignment production of axions yielded
\begin{equation}
 \Omega_{\rm a}h_{100}^2\approx 0.54g_{\star}^{-0.41}\theta_{\rm i}^2\left(\frac{f_{\rm a}}{10^{12}~\rm GeV}\right)^{1.19}\,,
\end{equation} 
where $g_{\star}\approx 10$ is the number of relativistic degrees of freedom during the realignment process, $\theta_{\rm i}$ is the initial angle of misalignment, $h_{100}$ is defined by the Hubble constant $H_0=100h_{100}\,{\rm km}\,{\rm sec}^{-1}\,{\rm Mpc}^{-1}$ and $f_{\rm a}$ is the axion decay constant which is related to the axion mass, $m_{\rm a}$, by $m_{\rm a}c^2 = 6\,{\rm \mu eV} \left(f_{\rm a}/10^{12}{\rm GeV}\right)^{-1}$ (see also \cite{ref:BaeMisalign,Kawasaki2013,ref:Marsh,Enander2017} for other recent treatments of this issue). Recent measurements of the Cosmic Microwave Background (CMB) by the \textit{Planck} satellite~\cite{ref:Planck, ref:Planck2018} yield an estimate for the CDM density, $\Omega_{\rm c}h_{100}^2 \approx 0.12$. Assuming that this is the case, taking into account the uncertainty in the value of $g_{\star}$ and the standard assumption $\langle\theta_{\rm a}^2\rangle=\pi^2/3$, we can predict a mass range of $19\,\mu{\rm eV}\leq m_{\rm a}c^2\leq 23\,\mu{\rm eV}$. 

This particular choice of $\theta_{\rm a}$ is based on a scenario where the value at each position in space is assigned randomly and eventually homogenised by expansion. We will use it in what follows as our baseline choice (as done by many authors) but we note that it is not really a firm prediction at all. In inflationary scenarios one would expect a random value anywhere in the range $0<\theta_{\rm a}\le\pi$. One might expect that, in order to avoid an anthropic solution to the strong CP problem, there is a lower limit for $\theta_{\rm a}$ and hence $10^{-2}<\theta_{\rm a}<\pi$. In this case, we come up with a wider prediction for the range of masses from misalignment, $6\times 10^{-3}\,\mu{\rm eV}<m_{\rm a}c^2< 6\times 10^2\,\mu{\rm eV}$. 

We note that there is a lower limit to the detection approaches we are advocating due to the emission from neutral hydrogen, which would prevent detection of the axion signal for $m_{\rm a}c^2<12\,\mu{\rm eV}$. This happens because there will be a degeneracy between the spectral line associated to the axion and the HI emission line with $\lambda$ $\approx 21$ cm. At higher redshifts, this value will shift to smaller frequencies (larger wavelengths) and it will make it more difficult to disentangle the signal due to the axion decay. We also note that the spectral lines from organic molecules, for example, $\rm {CO, CS, HCO, HCN, H_2O}$ and ${\rm NH_3}$ can also be a source of degeneracy at frequencies greater than 10 GHz, although the impact of these lines is less clear. 

PQ symmetry is a $U(1)$ symmetry and therefore one would expect cosmic strings to form via the Kibble Mechanism when the symmetry is broken. The expected relic abundance from this process is expected to dominate if the symmetry breaking transition takes place after inflation, and comprises two contributions from long strings and loops~\cite{ref:Battye1,ref:Battye} 
\begin{equation}
\label{eqn:strings}
 \Omega_{\rm a}h^2_{100}\approx \left [1+10J\left({\frac{\alpha}{\kappa}}\right)\right]\Delta\left(\frac{f_{\rm a}}{10^{12}\rm GeV}\right)^{1.18}\,,
\end{equation}
where $\alpha$ is the loop production size relative to the horizon, $\kappa$ quantifies the rate of decay of the string loops, $J(x)=x^{3/2}\left[1-(1+x)^{-3/2}\right]$ and $1/3<\Delta<3$ is the theoretical uncertainty associated with the QCD phase transition. This estimate was recently refined~\cite{ref:WS}, notably improving the estimate of $\Delta$ and making the assumption that $\alpha/\kappa=0.5\pm 0.2$ to deduce $100\,\mu{\rm eV}<m_{\rm a}c^2<400\,\mu{\rm eV}$ under the assumption that the axions are the cold dark matter. Note that this axion mass range cannot be probed by standard axion haloscope experiments. 

The axion couples to ordinary matter very weakly, most notably to photons and this is quantified by the axion-photon coupling constant $g_{\rm a\gamma\gamma}$ for the axion decaying spontaneously into two photons with a lifetime given by~\cite{ref:KT}
\begin{align}\label{eqn:DecayTime}
 \tau_{2\gamma} = &\, {\frac{64\pi\hbar}{g_{\rm a\gamma\gamma}^2m_{\rm a}^3c^6}}\,,\\
 \approx &\, 8\times 10^{35}\,{\rm sec}\left(\frac{g_{\rm a\gamma\gamma}}{10^{-10}~{\rm GeV^{-1}}}\right)^{-2}\left(\frac{m_{\rm a}c^2}{\rm 250~\mu eV}\right)^{-3}\,,\nonumber
\end{align}
with a rest-frame emission frequency of $f_{\rm emit}=m_{\rm a}c^2/(2h)$ which is $\approx 2.4\,{\rm GHz}$ for $m_{\rm a}c^2=20\,\mu{\rm eV}$ and $\approx 30\,{\rm GHz}$ for $m_{\rm a}c^2=250\,\mu{\rm eV}$ which correspond to the misalignment (with $\langle\theta_{\rm a}^2\rangle\approx 3$) and string prediction ranges, respectively. In what follows, we will use $m_{\rm a}c^2=250\,\mu{\rm eV}$ and $m_{\rm a}c^2=20\,\mu{\rm eV}$ as particular fiducial values in order to calculate specific numbers, but it is worth pointing out that we have argued that it is possible for there to be an axion signal anywhere in the frequency range $\sim 70\,{\rm MHz}$ to $\sim 100\,{\rm GHz}$.

For specific models there is a relation between $g_{\rm a\gamma\gamma}$ and $m_{\rm a}$, which depends on the choice of $E/N$, which is the ratio of electromagnetic and colour anomalies \citep{diCortona2016}
\begin{equation}\label{eq:QCDg}
 g_{\rm a\gamma\gamma} = 5.1 \times 10^{-14} \, \text{GeV}^{-1}\left(\frac{m_{\rm a}c^2}{250 \mu \text{eV}}\right)\left|\frac{E}{N} - 1.92 \right|\,.
\end{equation}
The KSVZ model has $E/N$ = 0, while DFSZ model has $E/N = 8/3$ making the latter more weakly coupled to photons. At present, the most sensitive experimental limits come from the ADMX haloscope collaboration which constrains $g_{\rm a \gamma\gamma} < 10^{-15}$\,GeV$^{-1}$ for $1.90\,\mu$eV $\leq m_{\rm a}c^2 \leq 3.69\,\mu$eV, under the assumption that the local galactic dark matter density $\rho_{\rm gal} c^2 \approx 0.45\,{\rm GeV}\,{\rm cm}^{-3}$~\cite{ref:ADMXl, ref:ADMXf}. This limit was further improved recently to rule out DFSZ axions in the narrow mass range $2.66\,\mu$eV $\leq m_{\rm a}c^2\leq 2.81\,\mu$eV~\cite{ref:ADMX2018} with a limit of $g_{\rm a\gamma\gamma}< 4\times 10^{-16}\,$ GeV$^{-1}$. A number of experiments have been proposed to speed up these searches so that much wider ranges of mass can be probed \citep{Majorovits2017, Brun2019, Droster2019, McAllister2017}. Typically these approaches find it more difficult, for practical reasons, to be sensitive to higher axion masses and therefore we believe that the strongest motivation for the ideas we present in this work is to search for axions in the multi GHz frequency range and hence we have centred the estimates presented in subsequent sections on $m_{\rm a}c^2=250\,\mu{\rm eV}$, although they apply more widely.

There is an upper limit $g_{\rm a\gamma\gamma}<0.66\times 10^{-10}\,{\rm GeV}^{-1}$ from the CAST solar axion experiment which applies for $m_{\rm a}c^2< 10^{-2}$\,eV~\cite{ref:CAST}. Given this limit, the predicted range of axion masses and the limits on the mass from terrestrial haloscopes, it seems sensible to search for astrophysical signals from dark matter axions in virialized halos (for example, galaxies and galaxy clusters) in the frequency range $f_{\rm obs}\approx 1-100\,{\rm GHz}$ which might be loosely described as the radio/mm-waveband and for decay times $\tau_{2\gamma}\sim 8\times 10^{35}\,{\rm sec}$ and higher with the aim of achieving a limit which is better than the limit from CAST~\footnote{There has been a previous attempt to obtain limits on dark matter axions using 6 days of integration on the dwarf galaxies Leo 1, LGS 2 and Pegasus using the Haystack 37\,m telescope \cite{ref:Blout}. A limit of $g_{\rm a\gamma\gamma} < 10^{-9}$\,GeV$^{-1}$ was published for axion masses $298\,\mu$eV $\leq m_{\rm a}c^2 \leq$ $363\,\mu$eV, but given the estimates we make for the strength of the signal in subsequent sections we believe that there must have been an error in the analysis. We will comment further on this at the end of section~\ref{subsec:SpontSensitivity}.}. There have been a number of recent studies~\cite{ref:KQ1, ref:KQ2, ref:Caputo, ref:Caputo1} of this subject in the context of future radio telescope, such as the Square Kilometre Array~\cite{ref:SKA} (see \cite{ref:redbook} for a recent summary of the current SKA science case in the context of cosmology), and one aim is to clarify and extend this work.

These studies have explored enhanced decay mechanisms such as the effects of astrophysical magnetic fields and stimulated emission due to the CMB. In \cite{ref:KQ1,ref:KQ2} it was suggested that magnetic fields of amplitude $\sim 10\mu{\rm G}$, already detected in galaxies and clusters, could lead to a strong and eminently detectable signal. However, \cite{ref:Sigl} pointed out that the decay lifetime into a single photon with $f_{\rm emit}=m_{\rm a}c^2/h$ expected for such a process is
\begin{equation}
 \tau_{\rm B} = \frac{m_{\rm a}}{2\pi^2 \hbar^2 cg_{\rm a\gamma\gamma}^2}\frac{\mu_0 V}{k^3|\hat{B}(k_{\rm a})|^2}\,,
\end{equation}
where $\hat{B}(k_{\rm a})$ is the Fourier transform of the magnetic field evaluated at a wavenumber corresponding to the inverse Compton wavelength of the axion $k_{\rm a}=m_{\rm a}c/\hbar$, $\mu_0$ is the vacuum permeability and $V$ is the volume over which the conversion takes place. The coherence length of the magnetic fields in typical halos is expected to be of the order of the size of the halo, which is $\sim 100\,{\rm kpc}$ for a galaxy. For dark matter axions, which we have already argued will have Compton wavelengths in the cm/mm range, and some decaying spectrum of magnetic turbulence (for example, a Kolmogorov spectrum $k^3|{\hat B}(k)|^2\propto k^{-2/3}$), one finds that $\tau_{\rm B}\gg\tau_{2\gamma}$. In fact, \cite{ref:Sigl} explained that there is a maximum possible flux density that one might expect via this mechanism, and it is far too weak to be detected. For this reason we will ignore this in what follows.

The decay of axions into two photons can be enhanced in the presence of a photon background and, by contrast to the enhancement due to magnetic fields, this may be very significant. References \cite{ref:Caputo, ref:Caputo1} have shown that the effective decay lifetime can be reduced to $\tau=\tau_{2\gamma}/(1+\corr)$ where $\corr$ is the photon occupation number associated to the relevant sources considered. Sources of photons include the CMB, the radio background and galactic emission with $\corr={\cal F}_{\rm CMB}+{\cal F}_{\rm radio}+{\cal F}_{\rm gal}+\ldots$ For the CMB, this is given by 
\begin{equation}
 {\cal F}_{\rm CMB}=2\left[\exp\left(\frac{m_{\rm a}c^2}{2k_{\rm B}T_{\rm CMB}}\right)-1\right]^{-1}\,,
\end{equation}
where $T_{\rm CMB}=2.725\,{\rm K}=235\,\mu{\rm eV}/k_{\rm B}$ which can be approximated by ${\cal F}_{\rm CMB}\approx 4k_{\rm B}T_{\rm CMB}/(m_{\rm a}c^2)$ for $m_{\rm a}c^2\ll 470~\mu{\rm eV}$ which can provide a potentially very significant enhancement of the signal. The CMB and the radio background are both isotropic sources, and so the factor ${\cal F}$ is easily worked out to be proportional to the brightness temperature measured by experiments \cite{ref:ARCADE2, ref:RadioBack}. 

The contribution from the radio background is very uncertain for a number of reasons. Firstly, making absolute measurement of the background temperature is inherently difficult. But perhaps more important is that this measurement is made from the point of view of telescopes on Earth and it may not be the same elsewhere in the Universe and also at higher redshifts. In principle, it would be necessary to model the sources contributing to the radio background and quantify the uncertainty in order set limits on $g_{\rm a\gamma\gamma}$.  

A dedicated study of specific sources, which might be easier to model than the overall background, could result in significant effective enhancement in values of $\cal F$ for the axion masses between 1 and 20 $\mu$eV/$c^2$. \cite{ref:Caputo1} suggested that ${\cal F_{\rm source}} \approx I_{\rm source}/E_{\nu}^3$ where $E_{\nu} = hf_{\rm obs}$ is the energy of the photons. We will adopt this relation for our later order-of-magnitude estimates of the signal from the galactic centre including the enhancement due to diffuse radio emission (eg. synchrotron emission) as well as the radio background. 

We note that there have been attempts to search for the axion signal in the infra-red waveband~\cite{ref:InfraredAxionSearch}. In particular axions with masses $m_{\rm a}c^2\approx 1-10\,{\rm eV}$ have been considered which could have been produced thermally - in the absence of strong non-thermal production mechanisms such as misalignment and string decay. Thermal production predicts 
\begin{equation}
 \Omega_{\rm a}h_{100}^2\approx \frac{m_{\rm a}c^2}{130~{\rm eV}}\left(\frac{10}{g_{\star}}\right)\,,
\end{equation}
and the published limit is $g_{\rm a\gamma\gamma}<10^{-12}\,{\rm GeV}^{-1}$ for axions in the mass range $4.5\,{\rm eV}<m_{\rm a}c^2<7.7\,{\rm eV}$\footnote{There is also a limit of $m_{\rm a}c^2< 0.529$ eV~\cite{ref:MelDi1} from Planck temperature and polarisation data. As these axions may be produced in the early universe also via thermal processes, they constitute a hot dark matter component with masses strongly degenerate with those of the active neutrinos, as their signature on observables is identical to neutrinos. Hence, when axions are relativistic, they contribute to the effective number of relativistic degrees of freedom $N_{\rm eff}$.}. In section~\ref{sect:decay} we will discuss applying exactly the same ideas in the radio/mm waveband. We note additionally that axions with large masses are also subject to constraints from astrophysics, specifically due to axion cooling competing with that from neutrinos in stars and supernovae; the most stringent limit being from the observations of the neutrino burst from SN 1987A, which appears to preclude axions in the mass range $10^{-3}\,{\rm eV}-2\,{\rm eV}$~\cite{ref:KT}. This is based on detailed modelling of the interaction of axions with stellar material and the detailed modelling of stars and hence could be considered to be less direct and more susceptible to uncertainties than other probes.

In the latter half of this paper, we discuss the resonant mixing of photons and axion dark matter in pulsar magnetospheres \cite{Pshirkov:2007st,ref:NS-Hook,ref:NS-Japan,ref:LaiHeyl}. The idea is a simple one: namely that in regions of the plasma where the photon plasma mass and axion mass become degenerate, there is enhanced conversion of dark matter axions to photons, just as in a regular haloscope whose density is tuned to a particular axion mass range. In addition, the ultra-strong magnetic fields of neutron stars also greatly enhance the overall magnitude of the effect. Our analysis falls into roughly two parts. The first focuses on theoretical fundamentals of axion electrodynamics in magnetised plasma, beginning with an examination of one-dimensional (1D) propagation in planar geometries (the standard approach to axion-photon mixing). We clarify two important aspects, firstly how to treat distinct and locally varying dispersion relations of the photon, which we do via a controlled gradient expansion, incorporating the mass-shell constraints systematically. Next we are able to unify two apparently disparate analytic results for the conversion amplitude. The first is the perturbative $\mathcal{O}(g_{\rm a \gamma \gamma}^2)$ formula for the conversion process of e.g., \cite{ref:NS-Hook}, while the second is non-perturbative and given by the well-known Landau-Zener formula~\cite{Brundobler,ref:LaiHeyl} derived by computing the S-matrix for conversion as dictated by the mixing equations. Our analysis unifies these two approaches and reveals the perturbative result to be a truncation of the full Landau-Zener formula in the non-adiabatic limit. For a given plasma background, this allows one to see precisely for what axion masses and momenta the evolution becomes non-adiabatic and therefore where a perturbative treatment is justified (see fig.~\ref{fig:GammaPlot}).

Next we question to what extent the 1D mixing equations (which dominate the literature on axion-photon conversion in stellar environments) are valid, and examine how three-dimensional (3D) effects excite a wider variety of plasma modes and polarisations.  This component of our work is important in illustrating the need for a more systematic analysis of 3D effects in axion electrodynamics in magnetised plasmas, as we show qualitatively that if one is not in a specialised 1D geometric setup, then new polarisation modes of the photon are excited. We discuss the difficulties in analytically solving such a system, and leave any further investigation of what this might imply for the overall signal for future work. 

We finish our study of conversion in neutron star magnetospheres with some observational considerations, reviewing telescope sensitivities and Doppler broadening of the signal from the motion of the star.  

The structure of the paper is as follows. In section \ref{sect:decay} we discuss axion observations in virialised structures and outline the targets with the best prospects for axion decay detection. We devote section \ref{sect:mixing} to the analysis of the evolution of the axion field in neutron star magnetospheres. After a formulation of the problem from first principles, we first investigate a one-dimensional set-up which paves the way for the study of the mixing in two and three dimensions. In this way, we can highlight differences and similarities arising from the geometrical set-up of the problem. We then proceed to estimate the single dish and interferometer sensitivities to the axion-photon parameter space in the context of the resonant conversion in section \ref{sect:resdecay}. We compare previous approaches to this work and explore the simplest way to optimise and to determine the best candidate neutron stars to target in an experiment. We conclude in section~\ref{sect:conclusions}. Some technical details are left in the appendices: in appendix~\ref{App:MassinBeam} we discuss how to evaluate the mass contained in a beam and in appendix~\ref{Density}, we give a detailed derivation of the Wentzel–Kramers–Brillouin (WKB) expansion of axion-photon mixing, with a careful discussion of dispersion relations and a derivation of the Landau-Zener formula. 

In sections~\ref{sect:decay} and \ref{sect:resdecay} we will include all factors due to fundamental physics and present quantities in SI units or other appropriately practical units, whereas in section~\ref{sect:mixing} we will present theoretical calculations using natural units $c=k_{\rm B}=\hbar=1$ with the Lorentz–Heaviside convention $\varepsilon_0 = \mu_0 =1$ for the vacuum permittivity and permeability.

\section{Detecting Dark Axions emitted by Virialised Halos}\label{sect:decay}

In this section we will derive estimates for the signal due to the spontaneous decay, present some estimates of what might be possible with current and planned facilities operating in the radio/mm-waveband, concluding that amounts of integration time required are too large to be feasible, and discuss how one might optimise the detection and improve current constraints on the axion-photon parameter space. In order to present estimates of the signal strength we will set up a strawman object which is a galaxy with a virial mass, $M_{\rm vir}=10^{12}\,M_{\odot}$, virial radius $R_{\rm vir}=100\,{\rm kpc}$ at a distance $d=5\,{\rm Mpc}$ and a velocity width of $200\,{\rm km}\,{\rm sec}^{-1}$ which corresponds to an object similar to the nearby galaxy Centaurus A \citep{ref:NearGalCat}. We have chosen these values to be broadly consistent with the model for the virial radius ($\propto M_{\rm vir}^{1/3}$) from a given mass that we will use later in the subsequent discussion. As part of that discussion, we focus on our suggestion that the basic signal strength will be relatively independent of the object mass. Such an object would be expected to have an average surface mass density $\Sigma_{\rm vir}\approx M_{\rm vir}/(\pi R_{\rm vir}^2)\approx 0.07\,{\rm kg}\,{\rm m}^{-2}$ over an angular diameter of $\theta_{\rm vir}=2R_{\rm vir}/d\approx 40\,{\rm arcmin}$. We will see that this value, which we will use in the subsequent signal estimates, is probably quite conservative and that values up to a thousand times larger than this might be accessible in some objects, albeit over smaller areas, typically in the centre of the object. The basic conclusion will be that it will be difficult to imagine a telescope with a single pixel receiver system achieving a limit on $g_{\rm a\gamma\gamma}$ better than that from CAST. In order to be competitive with the CAST limit, we find that it is easier to optimise future experiments if one quantifies the signal in terms of the brightness temperature, rather than the flux density. We show that the brightness temperature is proportional to the surface-mass-density $\Sigma_{\rm beam}$ associated with the telescope beam, which makes it clear that future experiments must target the centres of virialised objects where this quantity is the largest possible value. From our analysis, the main conclusion is that the larger surface-mass density at the galactic centre/Virgo cluster centre coupled with large amounts of radio emission at the relevant frequencies could enhance the signal enough to probe couplings below the CAST limit. 

\subsection{Estimates of the signal amplitude for axion decay from virialised halos}\label{subsec:SpontSignal}
Clearly the first and most important task in determining whether or not dark matter axions can be detected via spontaneous decays is to obtain a reliable estimate for the strength of the resulting signal. Let us consider a virialised halo of mass $M$ and at redshift $z$. We further assume that axions constitute its whole mass. The total bolometric flux from the object is 
\begin{equation}\label{eqn:BolometricFlux}
 \int \,S_{\rm tot} \,\mathrm{d}f_{\rm obs} = \frac{L_{\rm obs}}{4\pi [r(z)]^2} = \frac{N_{\rm a}E_{\rm obs}}{\tau_{\rm obs}}\frac{1}{4\pi r(z)^2}\,,
\end{equation}
where $r(z)$ is the comoving distance to redshift $z$, $S_{\rm tot}$ the total flux density, $E_{\rm obs}=2hf_{\rm emit}/(1+z)$ and $\tau_{\rm obs}=(1+z)\tau_{2\gamma}/(1+\corr)$ are the emitted photon energy and decay life-time in the observer's frame, respectively and $\mathcal{F}_{\gamma}^{\rm eff}$ is the photon distribution discussed in the previous section. The luminosity in the observer's frame is $L_{\rm obs}=N_{\rm a}E_{\rm obs}/\tau_{\rm obs}$ and $N_{\rm a}=M/m_{\rm a}$ is the number of axions in the halo. One can obtain an estimate of the observed flux density by assuming that all the flux is detected (the point source approximation) and that it is equally distributed across a bandwidth $\Delta f_{\rm obs}$, effectively assuming a top-hat line profile, in the observer's frame
\begin{equation}\label{Eqn:Flux}
 S_{\rm tot} =\frac{Mc^2}{4\pi [d_{\rm L}(z)]^2\tau_{2\gamma}\Delta f_{\rm obs}}(1+\corr)\,,
\end{equation}
where $d_{\rm L}(z)=(1+z)r(z)$ is the luminosity distance to redshift $z$. We note that this formula is equivalent to that for the emission of neutral Hydrogen due to the spin-flip transition under the exchange of $M$ with the neutral Hydrogen mass, $M_{\rm HI}$, and $\tau_{2\gamma}$ with the effective lifetime of the spin state.

Neither of the assumptions will be true in reality. The assumption of a top-hat frequency profile should only lead to a small correction if $\Delta f_{\rm obs}$ is set by the velocity width of the halo $\Delta v/c=\Delta f_{\rm obs}/f_{\rm obs}$. From first principles, this is set by the halo mass as $\Delta v\propto M^{1/3}$. In what follows, it will be convenient to specify the measured value of $\Delta v$ for a specific object rather than calculate it self-consistently from the halo mass. For typical values, and a halo at redshift $z$, we find
\begin{align}\label{Eqn:Bandwidth}
 \Delta f_{\rm obs} = &\,{\frac{f_{\rm emit}\Delta v}{c(1+z)}}\,,\\
 \approx &\, \frac{20\,{\rm MHz}}{{1+z}}\left(\frac{\Delta v}{200\,\rm km\,sec^{-1}}\right)\left(\frac{m_{\rm a}c^2}{250\,\rm \mu eV}\right)\,. 
 \nonumber
\end{align}

\begin{figure*}
 \centering
 \includegraphics[scale=1.2]{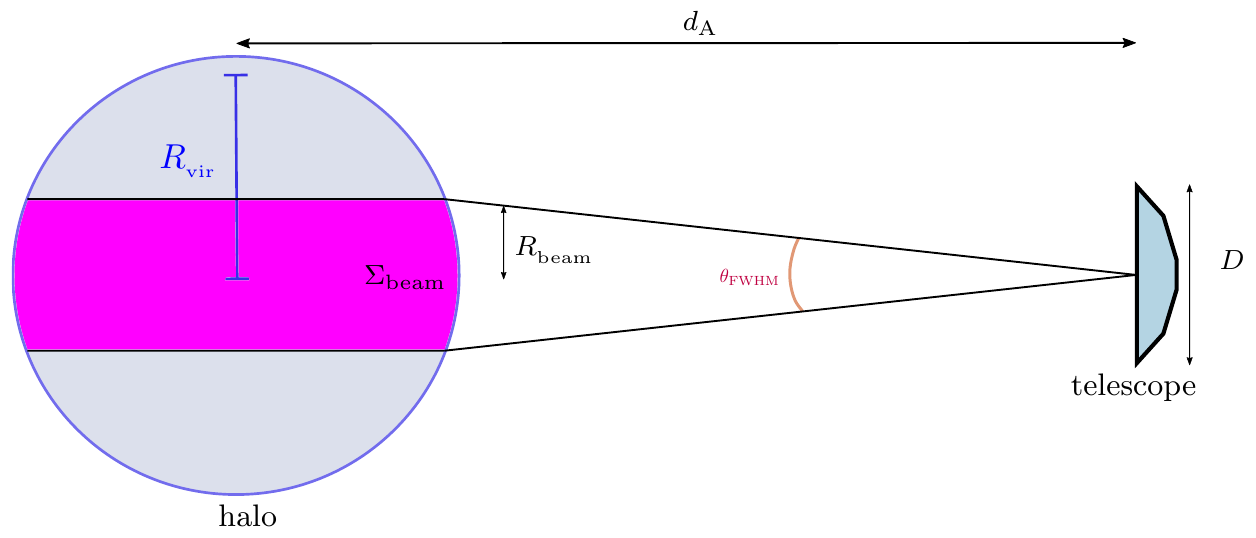}
 \caption{Schematic illustration of the telescope beam of width $R_{\rm beam}$ given in eq.~\eqref{eqn:rbeam} and virialised halo with surface density $\Sigma$ and virial radius $R_{\rm vir}$.}
 \label{fig:HaloBeam}
\end{figure*}

Typical receiver systems can produce spectra with the resolution in eq.~(\ref{Eqn:Bandwidth}) in all but the most extreme situations. The question of whether one is sensitive to flux from the entire halo is more complicated. Unless the telescope beam is larger than the projected angular size of the cluster, the total flux-density can be less than that of eq.~\eqref{Eqn:Flux} as illustrated in fig.~\ref{fig:HaloBeam}. Let us now estimate the importance of finite angular resolution.

We define $R_{\rm beam}$ as the {\em radius} corresponding to the Full-Width Half-Maximum (FWHM) angular {\em diameter} $\theta_{\rm FWHM}\approx \lambda_{\rm obs}/D$, where $\lambda_{\rm obs}$ is the observed wavelength and $D$ is the effective diameter of the observing telescope. In the case of a single dish telescope this is the actual size, whereas for an interferometer it will be given by the longest baseline. The beam radius can be estimated by $R_{\rm beam} = d_{\rm A}(z)\sin{(\theta_{\rm FWHM}/2)}$, where $d_{\rm A}(z)$ is the angular diameter distance which can be expanded for small $\theta_{\rm FWHM}$ to give
\begin{align}\label{eqn:rbeam}
R_{\rm beam} = &\,{\frac{hr(z)}{Dm_{\rm a}c}}\,,\\
\approx &\, 0.5\,{\rm kpc}\left(\frac{r(z)}{5\,{\rm Mpc}}\right)\left(\frac{D}{100\,{\rm m}}\right)^{-1}\left(\frac{m_{\rm a}c^2}{250\,\mu{\rm eV}}\right)^{-1}\,,\nonumber
\end{align}
where we have adopted a fiducial diameter of $100\,{\rm m}$ such as for the Green Bank Telescope (GBT). If $M_{\rm beam}\le M_{\rm vir}$ is the mass enclosed in the projected cylinder, then the observed flux density will be
\begin{align}\label{eqn:FluxValue}
 S_{\rm beam} \approx &\, 4\,\mu{\rm Jy}\left(1+\corr\right)\times 
 \left(\frac{\tau_{2\gamma}}{8\times 10^{35}~\rm s}\right)^{-1} \times\nonumber \\
 & \left(\frac{\Delta f_{\rm obs}}{20\,\rm MHz}\right)^{-1}
   \left(\frac{M_{\rm beam}}{10^{12}~M_{\rm \odot}}\right)\left(\frac{d_{\rm L}(z)}{5\,\rm Mpc}\right)^{-2} \,.
\end{align}
If we substitute (\ref{eqn:DecayTime}) and (\ref{Eqn:Bandwidth}) into (\ref{eqn:FluxValue}) we find that 
\begin{equation}\label{eqn:FluxValue_again}
 \begin{split}
  S_{\rm beam} \approx &\, 4\,\mu{\rm Jy}\left(1+\corr\right)\left(\frac{g_{\rm a\gamma\gamma}}{10^{-10}\,{\rm GeV}^{-1}}\right)^2\times\\
 & \left(\frac{m_{\rm a}c^2}{250\,\mu{\rm eV}}\right)^2
   \left(\frac{M_{\rm beam}}{10^{12}~M_{\rm \odot}}\right)\times\\
 & \left(\frac{\Delta v}{{200\,\rm km}{\rm sec}^{-1}}\right)^{-1}\left(\frac{d_{\rm L}(z)}{5\,\rm Mpc}\right)^{-2}\,.
 \end{split}
\end{equation}
From this we see that, if $\corr=0$, the expected flux density is $\propto m_{\rm a}^2$ for a fixed value of $M_{\rm beam}$. This reflects the fact that the size of the object which is inside the beam is dependent on $m_{\rm a}$ via the fact that $f_{\rm obs}\propto\theta_{\rm FWHM}$. This is an undesirable feature of using the flux density to assess the detectability of the axion signal, although it is possible to take into account the dependence of $M_{\rm beam}$ on $\theta_{\rm FWHM}$. Note that there will be additional dependence on $m_{\rm a}$ from $\corr$; for example, there is a component from the CMB which is $\propto m_{\rm a}^{-1}$.

It is possible to express the expected signal in terms of the intensity $I$, or equivalently the Rayleigh-Jeans brightness temperature 
\begin{equation}
I=\frac{2f_{\rm obs}^2k_{\rm B}T_{\rm RJ}}{c^2}\,,
\end{equation}
and we shall see that this is a much clearer way of quantifying the signal. For a source of axions at redshift $z$ with surface mass-density $\Sigma=\int\rho_{\rm a}\mathrm{d}l$, taking into account that the flux density is the integral of the intensity over the solid angle subtended by the source, the integrated line intensity is given by
\begin{equation}
 \int I_{\rm beam}\,{\rm d}f_{\rm obs}={\frac{c^2 \Sigma_{\rm beam}}{4\pi\tau_{2\gamma}(1+z)^4}}(1+\corr)\,,
\end{equation}
where the appropriate surface mass density is that integrated over the beam profile of the telescope, $\Sigma_{\rm beam}$. To obtain this expression, we used eq.~(\ref{Eqn:Flux}) and Etherington's reciprocity theorem $d_{\rm L}(z)=(1+z)^2d_{\rm A}$, as the solid angle of the object is defined as $\Delta\Omega=R^2/d_{\rm A}^2$. For the surface mass-density $\Sigma_{\rm beam}=\Sigma_{\rm vir}\approx 0.07\,{\rm kg}\,{\rm m}^{-2}$ of our strawman object, we can deduce an intensity 
\begin{equation}
 \begin{split}
  I_{\rm beam}\approx & {\frac{3\,{\rm mJy}\,{\rm sr}^{-1}}{(1+z)^4}}(1+\corr)\left(\frac{\tau_{2\gamma}}{8\times 10^{35}\,{\rm sec}}\right)^{-1}\times\\
  & \left(\frac{\Delta f_{\rm obs}}{20\,{\rm MHz}}\right)^{-1}\left(\frac{\Sigma_{\rm beam}}{0.07\,{\rm kg}\,{\rm m}^{-2}}\right)\,,
 \end{split}
\end{equation}
and a brightness temperature
\begin{align}\label{ref:brighttemp}
 T^{\rm beam}_{\rm RJ} \approx & {\frac{100\,{\rm pK}}{(1+z)^2}}(1+\corr)\left(\frac{\tau_{2\gamma}}{8\times 10^{35}\,{\rm sec}}\right)^{-1}\times\\
 & \left(\frac{\Delta f_{\rm obs}}{20\,{\rm MHz}}\right)^{-1}\left(\frac{m_{\rm a}c^2}{250\,\mu{\rm eV}}\right)^{-2}\left(\frac{\Sigma_{\rm beam}}{0.07\,{\rm kg}\,{\rm m}^{-2}}\right)\,.\nonumber
\end{align}
This can be simplified by substituting in eqs.~(\ref{eqn:DecayTime}) and (\ref{Eqn:Bandwidth}) to yield 
\begin{equation}\label{ref:brighttemp_again}
 \begin{split}
  T^{\rm beam}_{\rm RJ} \approx & {\frac{100\,{\rm pK}}{1+z}}(1+\corr)\left(\frac{g_{\rm a\gamma\gamma}}{10^{-10}\,{\rm GeV}^{-1}}\right)^2\times\\
  & \left(\frac{\Sigma_{\rm beam}}{0.07\,{\rm kg}\,{\rm m}^{-2}}\right)\left(\frac{\Delta v}{{200\,\rm km}{\rm sec}^{-1}}\right)^{-1}\,.
 \end{split}
\end{equation}

This expression does not have any explicit dependence on $m_{\rm a}$ and tells us that the key parameters dictating the signal strength are $g_{\rm a\gamma\gamma}$, $\Sigma_{\rm beam}/\Delta v$ and $\corr$. The only dependence on $m_{\rm a}$ is via the observation frequency and consequently the size of the area over which $\Sigma_{\rm beam}$ is computed. The size of the signal could be larger than this for our strawman object which is relevant to an average over the virial radius - see subsequent discussions.

\begin{figure*}[t]
 \centering
 \includegraphics[width = 0.65\textwidth]{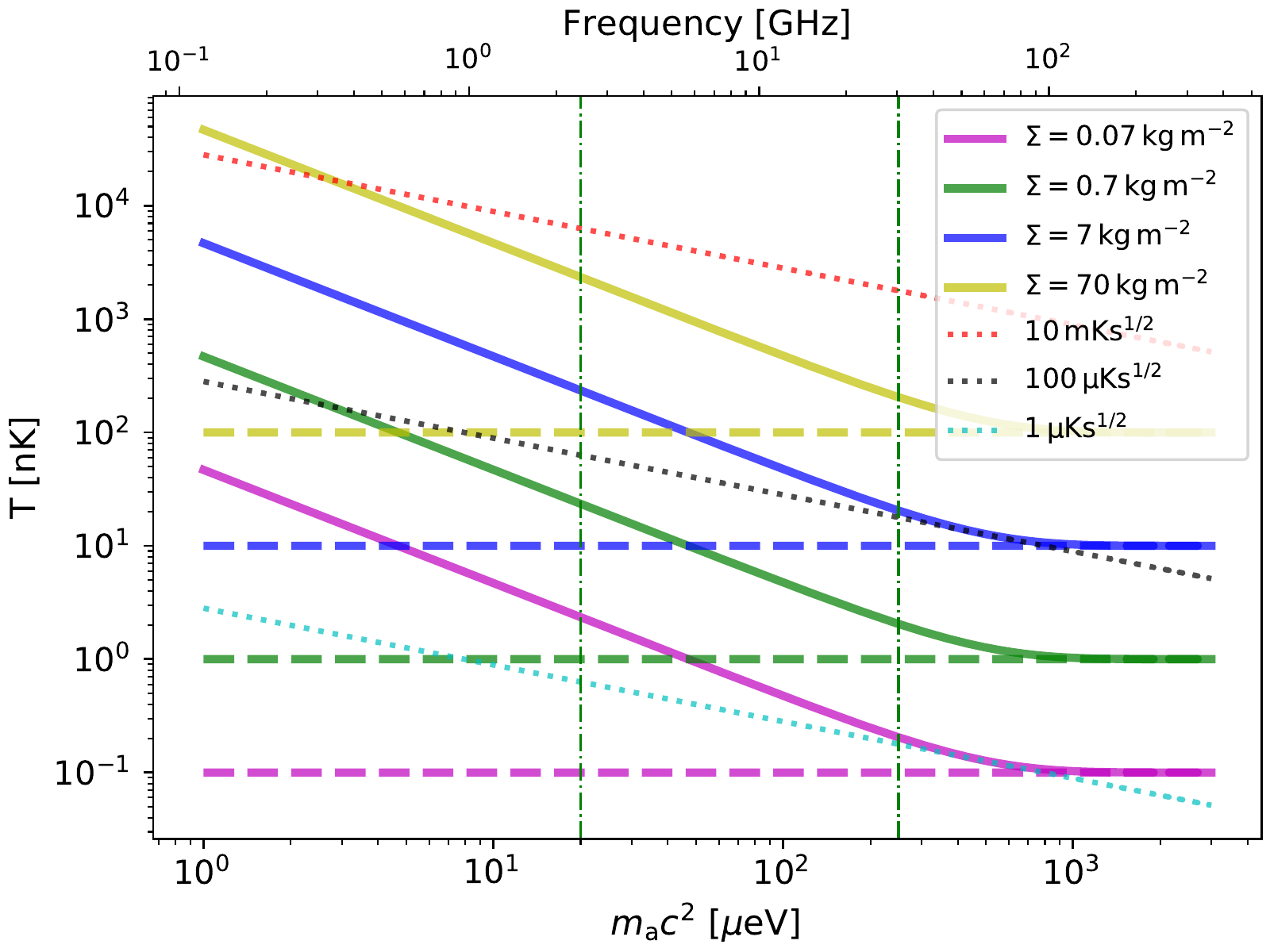}
 \caption{Estimates of the brightness temperature for a halo as a function of axion mass including spontaneous decay and the enhancement due to stimulated emission from the CMB (solid lines) and the pure spontaneous decay (dashed lines). We have fixed $g_{\rm a\gamma\gamma} = 10^{-10}\,{\rm GeV}^{-1}$ which is close to the CAST limit and is the goal signal level. We have also fixed $\Delta v=200\,{\rm km}\,{\rm s}^{-1}$ and used different values for $\Sigma_{\rm beam}=0.07$, $0.7, 7$ and $70\,{\rm kg}\,{\rm m}^{-2}$ which lead to brightness temperatures $\approx 100\,{\rm pK}$, $1$, $10$ and $100\,{\rm nK}$ respectively for $m_a\gg 470\,\mu{\rm eV}$ where spontaneous decay is dominant. For lower values of $m_{\rm a}$, we see the increase $\propto m_{\rm a}^{-1}$ due to stimulated emission from the CMB which could be added to other sources such as the radio background and galactic emission. We have also included some sample noise levels (dotted lines) due to 1 year of integration time with instantaneous sensitivities of $10\,{\rm mK}{\rm s}^{1/2}$, $100$ and $1\,\mu{\rm K}{\rm s}^{1/2}$ at $m_{\rm a}c^2=250\,\mu{\rm eV}$ with the scaling $m_a^{-1/2}$ necessary for a fixed velocity width. The two vertical lines represent $m_{\rm a}c^2 = 20\, \mu$eV and 250 $\mu$eV, respectively, which are illustrative values that we have used in the text.}
 \label{fig:CMB_Enhancement}
\end{figure*}

As a prelude to more detailed discussions of specific telescopes in the next subsection, we comment that a typical flux density of $S_{\rm beam}=4\,\mu {\rm Jy}$ might seem to be a quite accessible number for future large radio telescopes - many papers report detection of radio signals in the $\mu{\rm Jy}$ range using presently available facilities. Conversely a brightness temperature of $T_{\rm RJ}^{\rm beam}=100\,{\rm pK}$ is very low and much weaker than any value usually discussed. These numbers can be reconciled in realising that the flux density is averaged over a region $\Omega\approx \pi (R_{\rm vir}/d)^2\approx 1.2\times 10^{-3}\,{\rm sr}$ and it is also worth noting that most published radio detections are for bandwidths much larger than $20\,{\rm MHz}$. In the subsequent discussion we will argue that it is easier to understand whether the signal is detectable by considering the intensity or brightness temperature and that this gives a clearer picture of the potential for detection.

We can also calculate the background intensity due to all axions in the Universe with comoving density $\rho_{\rm a}$
\begin{equation}
 I_{\rm back}={\frac{c^2\rho_{\rm a}}{4\pi [r(z)]^2\tau_{2\gamma}f_{\rm emit}}}{\frac{\mathrm{d}V}{\mathrm{d}z\mathrm{d}\Omega}}\,,
\end{equation}
where $\tfrac{\mathrm{d}V}{\mathrm{d}z\mathrm{d}\Omega}=cr(z)^2/H(z)$ is the comoving volume element and $H(z)$ is the Hubble parameter at redshift $z$. Using this we can deduce a background brightness temperature
\begin{equation}
 T_{\rm RJ}^{\rm back} = \frac{3h^3c^5}{8\pi^2k_{\rm B}G}\frac{H_0\Omega_{\rm a}}{\tau_{2\gamma}}\left(\frac{1}{m_{\rm a}c^2}\right)^3\frac{(1+z)^2}{E(z)}\,.
\end{equation} 
Assuming that $\Omega_{\rm a}h_{100}^2 \approx 0.12$ and $h_{100} = 0.7$, we obtain 
\begin{align}\label{ref:backbrighttemp}
 T_{\rm RJ}^{\rm back} \approx &\, 0.3\,{\rm pK}{\frac{(1+z)^2}{E(z)}}\left(\frac{m_{\rm a}c^2}{250\,\rm \mu eV}\right)^{-3}\left(\frac{\tau_{2\gamma}}{8\times 10^{35}\,\rm s}\right)^{-1}\,, \nonumber\\
 \approx & 0.3\,{\rm pK}{\frac{(1+z)^2}{E(z)}}\left(\frac{g_{\rm a\gamma\gamma}}{10^{-10}\,{\rm GeV}^{-1}}\right)^2\,.
\end{align}
In making this background estimate we have ignored possible stimulated emission which would, of course, contribute at lower frequencies as was the case for the signal from virialised halos. The fact that this value is significantly lower than that for a halo means that there will be enough contrast to detect the signal from a halo against the background.

One can recover eq.~(\ref{ref:backbrighttemp}) by substituting the background value for $\Sigma/\Delta v$ into (\ref{ref:brighttemp_again}). This background value is given by
\begin{equation}
\frac{\mathrm{d}\Sigma}{\mathrm{d}v}=\rho_{\rm a}(z)\frac{\mathrm{d}l}{\mathrm{d}v}=\frac{(1+z)^3}{ E(z)} \frac{\rho_{\rm a}(0)}{H_0}\,,
\end{equation}
so that at $z=0$ this is $\rho_{\rm a}(0)/H_0\approx 1.2\times 10^{-9}\,{\rm kg}\,{\rm m}^{-3}\,{\rm s}$ using $\Omega_{\rm a}h_{100}\approx 0.17$. Note that one can make a rough estimate for the surface mass density of the background by multiplying the density of axions by the size of the Universe given by the Hubble radius, that is, $\Sigma_{\rm back}\approx \rho_{\rm a}c/H_0\approx 0.36\,{\rm kg}\,{\rm m}^{-2}$. This value is a factor of a few larger than the fiducial value we used for the halo surface mass density. To explain why this is the case, it is useful to notice that $\Sigma^{\rm halo}\approx \rho_{\rm a}\Delta_{\rm vir}R$, where $\Delta_{\rm vir}$ represents the virial overdensity of the halo. This quantity can be evaluated, given a cosmological model, using the virial theorem (see the Appendix in \cite{Pace2017a} for details on the implementation and \cite{Pace2019b} for a recent discussion on the topic), but here we will consider it to be of the order of 200 (higher values are also often used). The ratio between the two expressions, $\Sigma^{\rm back}/\Sigma^{\rm halo}\approx \tfrac{c/H_0}{\Delta_{\rm vir}R}\gg 1$ for our strawman object, but it is of the order of a few for $\Delta_{\rm vir}$ (a few hundred) and $R$ (a few Mpc).

In fig.~\ref{fig:CMB_Enhancement}, we present estimates of the brightness temperature expected from a halo with a fixed velocity width $\Delta v=200\,{\rm km}{\rm sec}^{-1}$ and a range of values for $\Sigma_{\rm beam}$ computed using (\ref{ref:brighttemp}). We have fixed $g_{\rm a\gamma\gamma}=10^{-10}\,{\rm GeV}^{-1}$ which is close to the upper limit from the CAST experiment (and hence the target goal) and have included the effects of stimulated emission by the CMB which leads to an increase $\propto m_{\rm a}^{-1}$ for $m_{\rm a}\ll 470\,\mu{\rm eV}$. We have chosen $\Sigma_{\rm beam}=0.07\,{\rm kg}{\rm m}^{-2}$ which is $\Sigma_{\rm vir}$ for our strawman object, along with ten, hundred and a thousand times this value. In subsequent sections, we will discuss that such values might be attainable by observing more concentrated regions of the halo close to their centres.

In addition we have also added noise curves for a total integration time of 1 year with instantaneous sensitivities of $10\,{\rm mK}{\rm s}^{1/2}$, 100 and $1\mu{\rm K}{\rm s}^{1/2}$ at $m_{\rm a}c^2=250\,\mu{\rm eV}$ with the scaling $\propto (m_{\rm a}/250\,\mu{\rm eV})^{-1/2}$ so that the noise level remains that for a fixed velocity width as $m_{\rm a}$ varies. We see that a sensitivity of $\sim 10\,{\rm mK}\,{\rm s}^{1/2}$ - which we will argue in section \ref{subsec:SpontSensitivity} is typical of a single pixel receiver at the relevant frequencies and bandwidths - is not sufficient to get anywhere near detecting the signal for $g_{\rm a\gamma\gamma}=10^{-10}\,{\rm GeV}^{-1}$, never mind that expected for the KSVZ and DFSZ models for typical values of $\Sigma_{\rm beam}$ as large as $7\,{\rm kg}\,{\rm m}^{-2}$. One might imagine that this can be reduced by having $N$ receivers/telescopes in which case the instantaneous sensitivity will be $\approx 10\,{\rm mK}\,{\rm s}^{1/2}/\sqrt{N}$. Looking at fig.~\ref{fig:CMB_Enhancement}, it appears that $N\sim 10^{2}$ would be necessary to probe signals created by $\Sigma_{\rm beam}\approx 70\,{\rm kg}\,{\rm m}^{-2}$, $\sim 10^{4}$ to probe $7\,{\rm kg}\,{\rm m}^{-2}$, $\sim 10^{6}$ to probe $0.7\,{\rm kg}\,{\rm m}^{-2}$ and $\sim 10^8$ for our strawman value of $0.07\,{\rm kg}\,{\rm m}^{-2}$. Therefore, it is clear that one would need to target sufficiently concentrated parts of haloes to probe this decay, which might be possible in haloes with supermassive black holes at their centres. While this enhancement would not allow one to probe the benchmark QCD models for the axion, one could at least probe the parameter space below the well-established CAST limit [see fig.(\ref{subsec:SpontSensitivity}) for sensitivity estimates].

\subsection{Sensitivity estimates for current and planned telescopes}\label{subsec:SpontSensitivity}

\begin{table*}[th]
 \centering
 \begin{tabular}{|c|c|c|c|c|c|c|}
  \hline
  Telescope  & $N$ & $A_{\rm eff}$ [$\rm m^2$] & $T_{\rm sys}$ [K] & Frequency [GHz] & $\theta_{\rm FWHM} $[arcmin] & $R_{\rm beam}$ [kpc]  \\
  \hline
  GBT &  1  & 5500 & 30 & 30  & 0.3 & 0.5\\
  FAST &  1  & 50000 & 20 &  2.4 & 1.4 & 2.1\\
  SKA1:Band 5 &  200 & 120 & 20  & 4.6-13.6 & 5.1-14.9 & 7.3-21.7 \\
  SKA2:Band 5 & 10000 & 120 & 20 &  4.6-13.6 & 5.1-14.9 & 7.3-21.7\\
  \hline
 \end{tabular} 
 \caption{Table of telescope parameters which we have used in section~\ref{subsec:SpontSensitivity} that are {\em indicative} of what might be possible using current and planned facilities. $N$ is the number of dishes, $A_{\rm eff}$ the effective collecting area, $T_{\rm sys}$ the overall system temperature (in Rayleigh-Jeans regime), $\theta_{\rm FWHM}$ the beam size and $R_{\rm beam}$ the radius corresponding to the beam size assuming a distance of 5 Mpc. GBT is the Green Bank Telescope and FAST is the Five hundred metre Aperture Spherical Telescope. They are currently operational and can cover a range of frequencies (up to $\approx 100\,{\rm GHz}$ in the case of GBT and up to $\approx 3\,{\rm GHz}$ for FAST). For the purposes of the discussion we have chosen to focus on one frequency for each and have chosen values of $T_{\rm sys}$ indicative of the noise levels that would be possible. We refer the reader to their webpages \url{https://greenbankobservatory.org} and \url{http://fast.bao.ac.cn/en/} for more detailed information about the capabilities. The Square Kilomtere Array (SKA) is currently being designed/built in two phases. Phase I is much more certain that phase II. Again we believe that our numbers are indicative of what might ultimately transpire.}
 \label{tab:Telescopes}
\end{table*}

In this section we assess the possibility of detecting the decay of dark matter axions emitted from virialised halos using current and planned telescopes operating in the radio/mm waveband. We have tabulated the numbers we have used in the sensitivity calculations below in table~\ref{tab:Telescopes}. Typically, previous analyses have focused on comparing the flux density to the expected telescope noise. As we have already alluded to and indeed we will explain below that it is best to frame the discussion of sensitivity in terms of the intensity, or more commonly the brightness temperature.

\subsubsection*{Flux Density Signal}

Having discussed the signal strength associated to axion decays in the previous sub-section, we turn now to another key parameter in determining the feasibility of detection - the integration time. The integration time required to detect a flux density $S_{\sigma}$ in a bandwidth $\Delta f_{\rm obs}$ can be deduced from the radiometer equation
\begin{equation}\label{Eqn:Radiometer}
t_{\rm int} = \left(\frac{2k_{\rm B}T_{\rm sys}}{A_{\rm eff}S_{\sigma}}\right)^2\frac{1}{\Delta f_{\rm obs}}\,,
\end{equation}
where $T_{\rm sys}$ is the system temperature, $S_{\sigma}$ is the flux density noise level and $A_{\rm eff}$ is the effective area. For a signal-to-noise ratio of unity, $S_{\rm beam} = S_{\sigma}$. For a single dish telescope with aperture efficiency $\eta$ (typically $\approx 0.5-0.7$), this is given by $A_{\rm eff}=\eta\pi D^2/4$. Using this, we can deduce that for a $1\sigma$ detection of the flux described by eq.~(\ref{eqn:FluxValue}) for a fiducial $M_{\rm beam}=10^{12}M_{\odot}$, the integration time is given by
\begin{align}\label{eqn:IntegrationTime}
 t_{\rm int} \approx & \frac{10\,{\rm days}}{(1+\corr)^2 (1+z)}\left(\frac{T_{\rm sys}}{30 \rm~K}\right)^2\left(\frac{A_{\rm eff}}{5500 ~\rm m^2}\right)^{-2}\times \nonumber\\
 & \left(\frac{\Delta v}{200\,{\rm km\,s^{-1}}}\right)\left(\frac{g_{\rm a\gamma\gamma}}{10^{-10}\rm \,GeV^{-1}}\right)^{-4}\times \\
 & \left(\frac{m_{\rm a}c^2}{250\,{\rm \mu eV}}\right)^{-5}
 \left(\frac{M_{\rm beam}}{10^{12}\,M_{\rm \odot}}\right)^{-2} \left(\frac{d_{\rm L}(z)}{5\,\rm Mpc}\right)^{4}\,,\nonumber
\end{align}
where the specific choice for $T_{\rm sys}$ and $A_{\rm eff}$ have been chosen to be indicative of what might be possible for observations at $30\,{\rm GHz}$ with a $100\,{\rm m}$ telescope such as the GBT which would have a resolution $\approx 20\,{\rm arcsec}$ operating in a band around $30\,{\rm GHz}$ and an axion mass $m_{\rm a}c^2\approx 250\,\mu{\rm eV}$. Despite this particular choice, the expression for $t_{\rm int}$ should be applicable to the whole range of frequencies observed by the GBT, and indeed any single dish radio telescope, provided $M_{\rm beam}$ is chosen appropriately. We chose the GBT to illustrate this since it is the largest telescope in the world operating at these frequencies and possibly as high as $\lesssim100\,{\rm GHz}$. Setting a 95\% exclusion limit - which is the standard thing to do in constraining dark matter - would require approximately 40 days. Detection at the $5\sigma$ level would take 25 times longer, that is 250 days of on-source integration time. Achieving an exclusion limit for the flux expected for the KSVZ model in this mass range would require ruling out $\tau_{2\gamma}\approx 6\times 10^{40}\,{\rm s}$ which would take $5\times 10^9$ times longer, and the level expected for DFSZ will be even lower, neither of which are practical. We note that ${\cal F}_{\rm CMB}\approx 0.5$ for $m_{\rm a}c^2=250\mu{\rm eV}$ and $\approx 12$ for $m_{\rm a}c^2=20\,\mu{\rm eV}$ which will reduce the required integration times, but probably not enough to make much difference to the conclusions.

Despite this, one might think that integration times of a few tens of days might allow one to impose stronger limits than the CAST bounds. However, the numerical value in \eqref{eqn:IntegrationTime} is quite misleading since such a telescope would have a resolution of $\approx 20\,{\rm arcsec}$ at these frequencies and therefore we would expect $M_{\rm beam}\ll M_{\rm vir}$. From eq.~(\ref{eqn:rbeam}) we have that $R_{\rm beam}\approx 0.5\,{\rm kpc}$ when the galaxy would be expected to have a total radius of $R_{\rm vir}\approx 100\,{\rm kpc}$, which is a factor of $200$ larger. 

We can obtain an estimate for the total halo mass contained within the beam by using the canonical halo dark matter distribution given by the Navarro-Frenk-White (NFW) profile \citep{ref:NFW} parameterized by the concentration parameter, $\hat{c}$, which is the ratio of the virial radius and the scale radius of the halo. It quantifies the amount of mass within the scale radius relative to that in the total halo, with large values of $\hat{c}$ having more mass {\em concentrated} in the centre than lower values. In appendix~\ref{App:MassinBeam} we have calculated for $\hat{c}R_{\rm beam}/R_{\rm vir} = R_{\rm beam}/r_{\rm s}\ll 1$, that is, a beam size much less than the characteristic scale of the NFW profile, the following estimate for the halo mass contained within the telescope beam:
\begin{figure}
 \centering
 \includegraphics[width =0.9 \linewidth]{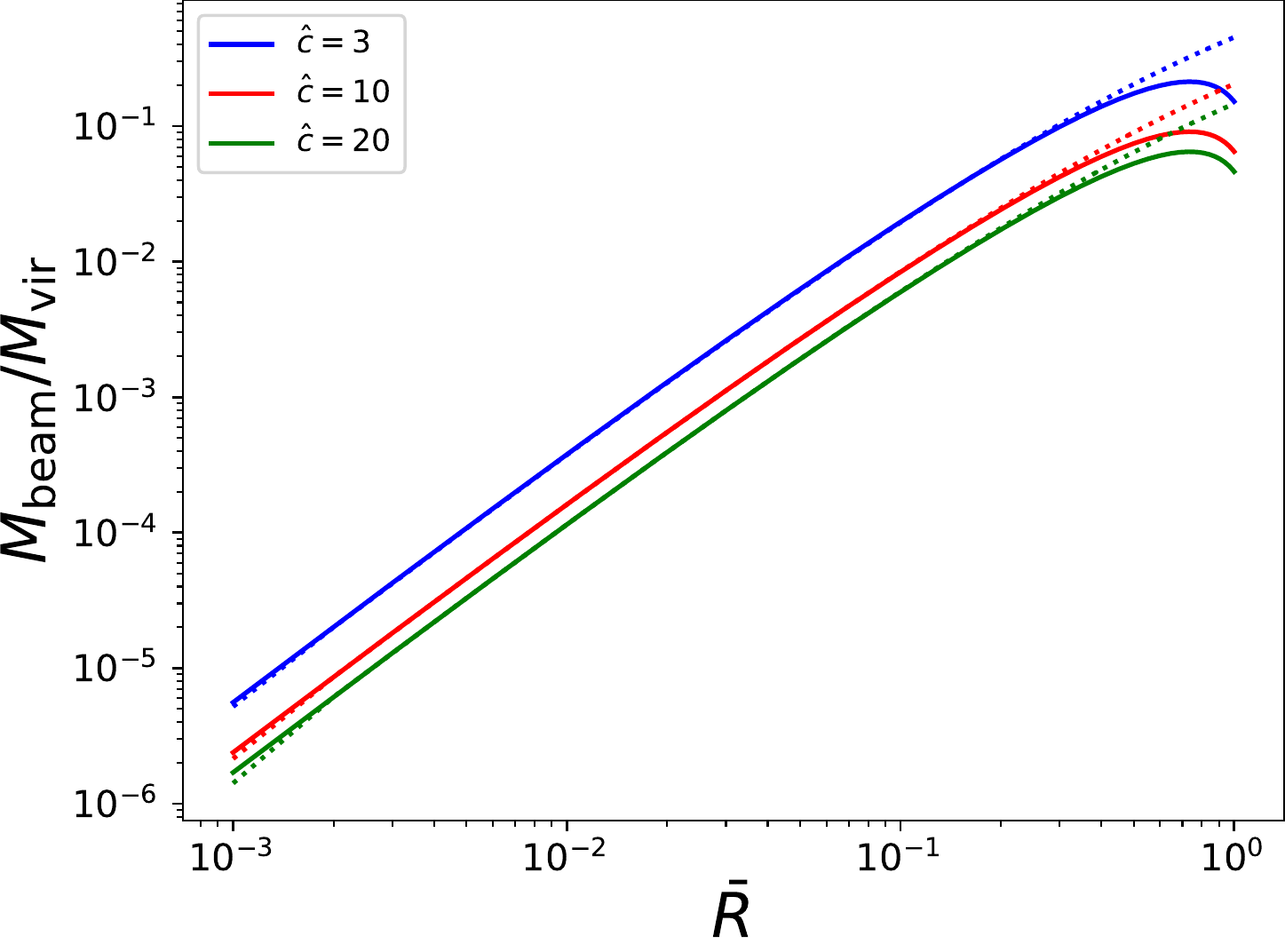}
 \caption{Projected mass within the beam as a function of $\bar{R} = cR_{\rm beam}/R_{\rm vir}$ assuming an NFW profile. From top to bottom, we consider three different concentration parameters, ranging from clusters to dwarf galaxies. The solid lines represent an analytic approximation for $\bar{R} \ll 1$, while the dotted lines are given by full numerical integration - see appendix \ref{App:MassinBeam} for details.}
 \label{fig:VirgoMass}
\end{figure}
\begin{equation}
 \frac{M_{\rm beam}}{M_{\rm vir}}= \frac{R_{\rm beam}^2}{R_{\rm vir}^2} \cdot \frac{\hat{c}^2}{2 f(\hat{c})}\log{\left(\frac{{2R_{\rm vir}}}{\hat{c}R_{\rm beam}}\right)} \,,
 \label{ref:mbeam}
\end{equation}
where $f(x)=\log(1+x)-\frac{x}{1+x}$. The behaviour of the beam mass is plotted in fig.~\ref{fig:VirgoMass}. Using this expression we deduce that $M_{\rm beam}\approx 0.8\times 10^{9}\,{\rm M_{\odot}}\,, 1.9\times 10^{9}\,{\rm M_{\odot}}$ and $6.2\times 10^{9}\,{\rm M_{\odot}}$ for ${\hat c}=3,5$ and 10, respectively. As one would expect, there is a trend for $M_{\rm beam}$ to increase as $\hat{c}$ increases, but even for relatively large values we find that in this case $M_{\rm beam}\ll M_{\rm vir}$. Clearly, this reduction in $M_{\rm beam}$ has a deleterious effect on the ability of a single dish telescope to even post an upper limit on the spontaneous decay of dark matter axions since $t_{\rm int}\propto M_{\rm beam }^{-2}$ with $t_{\rm int}\approx 3\times 10^4\,{\rm years}$ for $M_{\rm beam}=10^{9}M_{\odot}$. Therefore, one needs to be very careful in using (\ref{eqn:IntegrationTime}).

It is possible to think in terms of the flux density, but as we have explained above one has to be very careful to use the mass inside the beam radius and not the total mass of the object since they will typically be very different. Our view is that it is much easier to think in terms of the brightness temperature (or equivalently the intensity, although telescope sensitivities are more commonly expressed in terms of a brightness temperature). 

\subsubsection*{Brightness Temperature Signal}
The calculation of the noise temperature is simpler. The noise level in intensity is simply given by $I_{\sigma} = S_{\sigma}/\Omega_{\rm beam}$. Substituting for the intensity in terms of Rayleigh-Jean's law and setting $\Omega_{\rm beam} = \lambda^2/D_{\rm tel}^2$, we obtain the well-known Radiometer equation for brightness temperature
\begin{equation}
 T_{\sigma} = \frac{T_{\rm sys}}{\eta\sqrt{\Delta f_{\rm obs} t_{\rm int}}} \,,
\end{equation}
for a single telescope with system temperature $T_{\rm sys}$ and aperture efficiency $\eta$ observing in a bandwidth of $\Delta f_{\rm obs}$. The instantaneous sensitivity is just given by $T_{\rm sys}/(\eta\sqrt{\Delta f_{\rm obs}})\approx 10\,{\rm mK}\,{\rm s}^{1/2}\left(T_{\rm sys}/30\,{\rm K}\right)(\Delta f_{\rm obs}/20\,{\rm MHz})^{-1/2}$ for $\eta=0.7$ and hence the integration time required to detect a surface mass density of $\Sigma_{\rm beam}$, which is that averaged over the beam radius, at $1\sigma$ is 
\begin{equation}\label{eqn:tintegrate}
 \begin{split}
  t_{\rm int} \approx & 3\times 10^{8}\,{\rm years} \frac{(1+z)^3}{(1+\corr)^2}\left(\frac{T_{\rm sys}}{30\,{\rm K}}\right)^2\times\\
  & \left(\frac{g_{\rm a\gamma\gamma}}{10^{-10}\,{\rm GeV}^{-1}}\right)^{-4}\left(\frac{\Sigma_{\rm beam}}{0.07\,{\rm kg}\,{\rm m}^{-2}}\right)^{-2}\times\\
  &\left(\frac{\Delta v}{{200\,\rm km}{\rm sec}^{-1}}\right)\left(\frac{m_{\rm a}c^2}{250\mu{\rm eV}}\right)^{-1}\,.
 \end{split}
\end{equation}
Note that this is independent of the telescope collecting area, as one would expect for an unresolved detection, and also there is no explicit dependence on the distance, although there is a dependence on the redshift. Many of the other dependencies, for example, on $T_{\rm sys}$, $\Delta v$ and $g_{\rm a\gamma\gamma}$ are the same. Moreover, this expression makes it very obvious that the discussion above based on (\ref{eqn:IntegrationTime}) can be very misleading since the number at the front of the expression (remembering that the surface mass density of $0.07\,{\rm kg}\,{\rm m}^{-2}$ was chosen to correspond to the average across an object of mass $10^{12}\,M_{\odot}$ and radius $100\,{\rm kpc}$) is very much larger than in (\ref{eqn:IntegrationTime}).

The fact that $t_{\rm int}$ is dependent on $\Sigma_{\rm beam}$ has two advantages. The first is that it is clear that in order to increase the size of the signal and hence reduce $t_{\rm int}$ to a practical length of time one has to increase $\Sigma_{\rm beam}$. From our earlier discussion, we calculated, assuming an NFW profile, $M_{\rm beam}\sim 10^{9}M_{\odot}$ for our fiducial galaxy and telescope configuration for which $R_{\rm beam}\approx 0.5\,{\rm kpc}$, assuming a sensible range of concentration parameters. In this case the appropriate surface mass density would be\footnote{We note that (\ref{eqn:tintegrate}) and (\ref{eqn:IntegrationTime}) would be identical if $\Sigma_{\rm beam}$, $M_{\rm beam}$ and $R_{\rm beam}$ were chosen to be consistent with each other.} 
\begin{equation}\label{eqn:Sigma7}
\Sigma_{\rm beam}\approx 7\,{\rm kg}\,{\rm m}^{-2}\left(\frac{M_{\rm beam}}{2.3\times 10^{9}M_{\odot}}\right)\left(\frac{R_{\rm beam}}{0.5\,{\rm kpc}}\right)^{-2}\,.
\end{equation}
Of course this only gives one a factor of around $200$ improvement but it makes it clear in what direction one might have to go in optimising the signal strength. We will return to this issue in sect.~\ref{subsec:SpontTarget}.

The other advantage is that it makes clear what one would have to do to establish an upper bound on the signal: one would need an estimate of $\Sigma_{\rm beam}$ over the region which one was observing. Fortunately, the amplitude of any gravitational lensing signal that one might measure is directly related to the surface mass density. The measurement of the amplification and shear can be related to the surface mass density of the lenses. One of the largest surface mass densities measured from strong lensing on the scale of a few kiloparsecs (which corresponds to the typical beam sizes) is 50 $\rm kg\,{m^{-2}}$ \cite{ref:Winn2004SMD}. Such values are typically found towards the centre of virialised haloes. This motivates high resolution observations and detailed study of high-density sources with rich ambient radio emission for an accurate estimate of $\Sigma_{\rm beam}$ and ${\cal F}^{\rm eff}$.

The discussion so far has focused on the axion mass range $m_{\rm a}c^2\approx 250\,\mu{\rm eV}$, but we have also motivated searches at lower masses, for example, $m_{\rm a}c^2=20\,{\mu \rm eV}$ which corresponds to $f_{\rm obs}=2.4\,{\rm GHz}$. The Five hundred meter Aperture Spherical Telescope (FAST) might be a candidate large telescope for the detection of axions in this mass range. Despite its name, it can only illuminate beams with $D\approx 300\,{\rm m}$ corresponding to a resolution of $\approx 1.4\,{\rm arcmin}$ and $R_{\rm beam}\approx 2\,{\rm kpc}\ll R_{\rm vir}$. The bandwidth corresponding to $\Delta v=200\,{\rm km}\,{\rm sec}^{-1}$ at $z=0$ is $\Delta f_{\rm obs}=1.6\,{\rm MHz}$. The instantaneous sensitivity to such $T_{\rm sys}/(\eta\sqrt{\Delta f_{\rm obs}})\approx 20\,{\rm mK\,s}^{1/2}\left(T_{\rm sys}/20\,{\rm K}\right)(\Delta f_{\rm obs}/1.6\,{\rm MHz})^{-1/2}$ which is a little larger than for our estimate for the GBT at $30\,{\rm GHz}$ despite having a lower system temperature. The formula \eqref{eqn:tintegrate} should apply here as well with the values of $T_{\rm sys}$ and $\Sigma_{\rm beam}$ adjusted to take into account $R_{\rm beam}$ being a little larger. Ultimately, we come to the same conclusion.

If a focal plane array or phased array were fitted to the telescope, it might be possible to observe with $N$ beams and this would reduce the amount of integration time required by a factor of $1/N$. However, there are practical limitations on the size of array which one can deploy on telescope since the physical size of the region over which one can focus is limited; much more than $N\sim 100$ would be difficult to imagine. Moreover, the beams cannot point at the same region of the sky and just serve to increase the field-of-view. This does reduce the noise level, but over a wider area which would likely result in the decrease in the expected signal strength.

A number of recent works \citep{ref:KQ1,ref:KQ2,ref:Caputo,ref:Caputo1} have suggested that it might be possible to use the Square Kilometre Array (SKA) to search for axions. Naively the very large collecting area of the SKA in the formula (\ref{eqn:IntegrationTime}) would substantially reduce the necessary integration time. The proposed band 5 of the SKA, which has a frequency range of $4.6-13.7\,{\rm GHz}$, could potentially be of interest for the detection of axions in the mass range $40-110\,\mu{\rm eV}$. However, it is not valid to use the entire collecting area of the SKA in this way because the beam size, since it is an interferometer, is set by the longest baseline and this would be far too small. If one thinks in terms of brightness temperature, there is an extra factor, known as the filling factor, $\eta_{\rm FF}\ll 1$, which will increase the noise level $\propto \eta_{\rm FF}^{-1}$.

An interesting alternative approach would be to use each of the SKA telescopes as single telescopes in auto-correlation mode as it is envisaged for HI intensity mapping~\cite{ref:Bull}. The SKA dishes will have a diameter of $D=15\,{\rm m}$ and a sensitivity defined by $A/T_{\rm sys}\approx 6\,{\rm m}^2\,{\rm K}^{-1}$. Operating in band 5, this will have a resolution of $\theta_{\rm FWHM}\approx 15\,{\rm arcsec}$ at the lower end of the band and $\approx 6\,{\rm arcsec}$ at the higher end. In the first instance the SKA - SKA phase 1, sometimes called SKA1 - will have $\approx 200$ such dishes but may eventually - SKA2 - have $\approx 10000$. As before, the integration time for the telescopes decreases by a factor of $N$, the number of telescopes, but unlike a phased array on a single telescope they can co-point at the same region of sky which is advantageous. With 200 telescopes, we estimate an integration time of about 1.5$\times 10^6$ years, while for $10^4$ telescopes, we obtain $t_{\rm int} \approx 3\times 10^4$ years. This estimate will be smaller for lower masses (around 2 orders of magnitude at $m_{\rm a}c^2 = 20\,{\rm \mu eV}$) due to the enhancement from the stimulated decay. However, this will be mitigated to some extent by the factor $m_{\rm a}c^2$ in the denominator of \eqref{eqn:tintegrate}. The values used are for a strawman object, while if we use the surface mass density of (\ref{eqn:Sigma7}), we would estimate integration times $\approx 10^4$ times smaller, which might bring this in the realms of possibility.  We note that our integration time estimate for dwarf galaxies is consistent with that of reference \cite{ref:Caputo1} up to a factor of a few, although it is difficult to make a precise comparison. We believe that any minor discrepancies might be due to the fact that observational measurements of the size of the individual dwarf galaxies might lead to slight overestimation of the signal from them. This point is borne out in fig.~(\ref{fig:tintPlots}), where we obtain slightly lower integration times for higher mass objects when we determine object size from the virial overdensity parameter, via the relationship between the virial mass and radius.  

We have already mentioned that \cite{ref:Blout} published an upper limit for $g_{\rm a\gamma\gamma}$ based on 6 days of observations using the Haystack radio telescope for axions in the mass range around $m_{\rm a}c^2\approx 300\,\mu{\rm eV}$. In \cite{ref:Blout} they state that $T_{\rm sys}\approx 100\,{\rm K}$ and we estimate $A_{\rm eff}\approx 750\,{\rm m}^2$ (assuming $\eta\approx 0.6$) and hence flux density and brightness temperature sensitivities of $100\,{\rm mJy}\,{\rm s}^{1/2}$ and $40\,{\rm mK}\,{\rm s}^{1/2}$, respectively, in an observing bandwidth of $\Delta f_{\rm obs}\approx 4\,{\rm MHz}$. They assume a mass of $\approx 10^{7}M_{\odot}$ and a diameter of $\approx 10\,{\rm kpc}$ for the dwarf galaxies which they probe at distances in the range $d\approx 200\,{\rm kpc}$ with velocity width of $\Delta v\approx 30\,{\rm km}\,{\rm s}^{-1}$ equivalent to $\Delta f_{\rm obs}\approx 3.6$\,MHz. For $\tau_{2\gamma}=5\times 10^{33}\,{\rm s}$, which corresponds to their upper limit of $g_{\rm a\gamma\gamma}<10^{-9}\,{\rm GeV}^{-1}$, we predict a flux density of $S\approx 4\,{\rm mJy}$ which would take $3\times 10^{3}\,{\rm s}$ to obtain a 95\% exclusion limit. However, the typical angular diameter of these objects is $\approx 3\,{\rm deg}$, which is very much larger - by around more than a factor of 100 - than the beam size which would mean that $M_{\rm beam}\ll M_{\rm vir}$. For the reasons explained earlier, it is clear that they must have made some error in their calculations and this limit should be discounted.

\subsection{Optimising Target Objects}\label{subsec:SpontTarget}

\begin{figure*}[!t]
 \centering
 \includegraphics[width=0.45\textwidth]{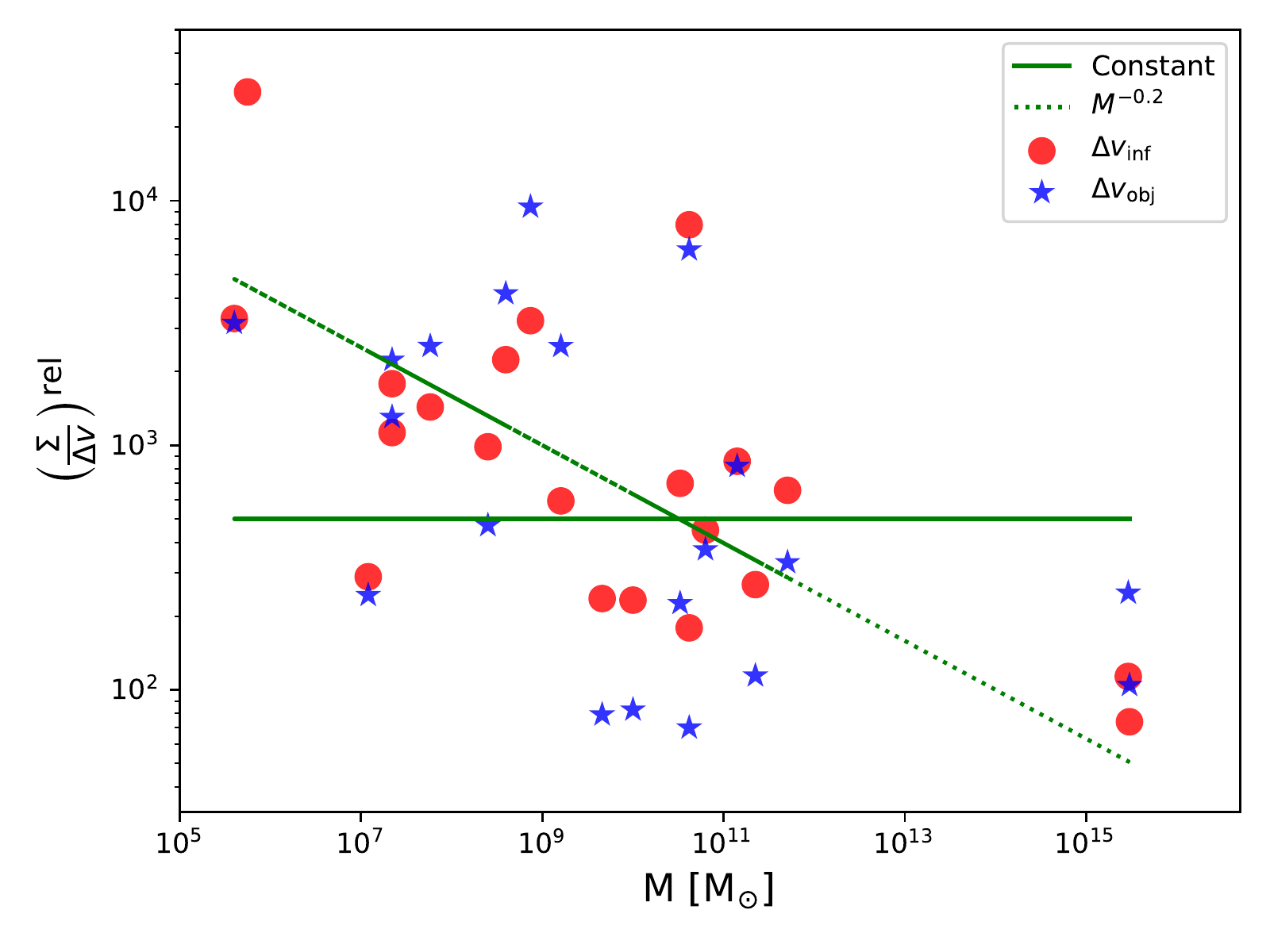}
 \includegraphics[width=0.45\textwidth]{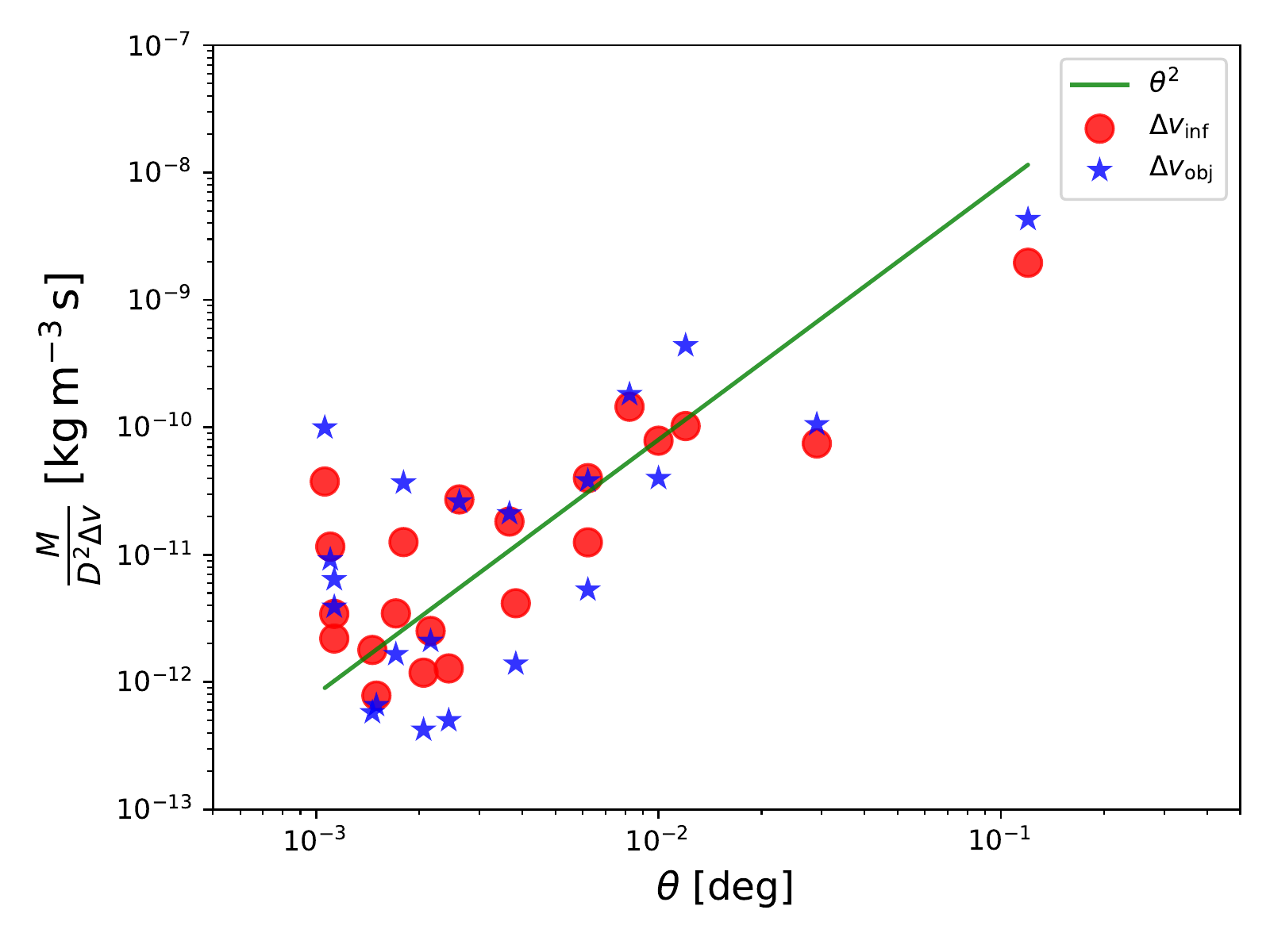} \\
 \caption{In the {\em left} panel. Signal strength as given by $\Sigma/\Delta v \propto T_{\rm RJ}$. We assume an \textit{identical} object and beam size  $\Sigma = M_{\rm obj}/(\theta_{\rm obj} D_{\rm obj})^2$ taking values from table~\ref{ClusterCalculations}. Note we normalised $\Sigma$ by the background value $1.2\times 10^{-9}\,{\rm kg}\,{\rm m}^{-3}\,{\rm s}$. The trend appears relatively flat for the data in the table - the solid green line - and is compatible with the simple argument presented in the text, albeit with a somewhat higher value ($\approx 500$) relative to the background value. Possibly there is a trend with mass which we denoted with a line $\propto M^{-0.2}$ which could be due to the concentration parameter varying as a function of mass and the fact that the angular sizes are probably the scale radius for some fitted profile function rather than the virial radius. We note that much of this trend is driven by the outliers at low mass, ultra-faint dwarf spheroidal, and high mass, the galaxy clusters, Virgo and Coma. In the {\em right} panel, we present the quantity in \eqref{eqn:CaputoQuantity} for the data in table~\ref{ClusterCalculations} which clearly increases like $\theta_{\rm obj}^2$ as denoted by the line in the plot. Note that the starred data points, which use observational measurements of the velocity width $\Delta v_{\rm obj}$, and the circular points, which correspond to the inferred width $\Delta v_{\rm inf} \approx \left(GM_{\rm obj}/R_{\rm obj}\right)^{1/2}$, show the same trend.}
 \label{fig:ScatterPlots}
\end{figure*}

In the previous two sections we have explained that, if one targets a halo with surface mass density $\Sigma_{\rm beam}\approx 0.07\,{\rm kg}\,{\rm m}^{-2}$ and velocity width $\Delta v\approx 200\,{\rm km}\,{\rm s}^{-1}$, the signal from spontaneous decay combined with stimulated emission from the CMB for $g_{\rm a\gamma\gamma}=10^{-10}\,{\rm GeV}^{-1}$ is too weak to be detected even for an array of receivers with $N\lesssim 10^{6}$. We came to this conclusion by estimating the integration time required to detect the signal focusing on the expression for the signal expressed in terms of the brightness temperature (\ref{ref:brighttemp_again}).

\subsubsection{Maximising brightness temperature}
Examination of this equation makes it clear that the largest possible signal is obtained by maximising $\Sigma_{\rm beam}/\Delta v$. If the object is such that $\theta_{\rm FWHM}\approx \theta_{\rm vir}$, we estimate the quantity to be $\approx 3.5\times 10^{-7}\,{\rm kg}\,{\rm m}^{-3}\,{\rm s}^{-1}$ for the strawman object used in the previous section which is around 300 times larger than the background value. This value is based on what we think, at a level of better than a factor two, are realistic values, but precise knowledge of it is absolutely critical to any attempt to improve the CAST limits of $g_{\rm a\gamma\gamma}$ using this approach. In this section, we will discuss, using theoretical arguments and comparing to observations, the range of values for $\Sigma_{\rm beam}/\Delta v$ that might be available for us to be observed in the Universe.

Consider now the possibility that the effective beam size is sufficiently large to capture the full object flux so that 
$S_{\rm beam}=S_{\rm tot}$. From the beam geometry, one expects that $S_{\rm tot} \propto M_{\rm vir}$ - the scenario considered by \cite{ref:Caputo}. Indeed this setup can be realised by considering the resolution of the SKA dishes at 2.4 GHz ($m_{\rm a}c^2 = 20\,{\rm \mu eV}$) for which most of our candidate objects (table \ref{ClusterCalculations}) are within the beam of the telescope. Put simply, this means that we are in the regime where the surface mass density within the beam is that of the whole object, i.e., $\Sigma_{\rm beam}=\Sigma_{\rm vir}$. Similarly, $M_{\rm beam} = M_{\rm vir}$. Throughout the subsequent discussion we therefore identify $\Sigma_{\rm beam} = \Sigma_{\rm vir}$ and phrase our analysis purely in terms of $\Sigma_{\rm vir}$.

One might wonder how $\Sigma_{\rm vir}/\Delta v$ depends on the size of the object. If we consider a halo with virial overdensity $\Delta_{\rm vir}\sim 100$, then $M_{\rm vir}=\tfrac{4\pi}{3}\Delta_{\rm vir}\rho_{\rm a} R_{\rm vir}^3$, where $\rho_{\rm a}=\Omega_{\rm a}\rho_{\rm crit}$ is the background density of axions and $\rho_{\rm crit}$ is the critical density. An estimate for the velocity width, up to order one factors, is $\Delta v=(GM_{\rm vir}/R_{\rm vir})^{1/2}$ and hence we find that 
\begin{equation}\label{eqn:SigmaQuantity}
\frac{\Sigma_{\rm vir}}{\Delta v}\approx 0.7\left(\frac{\Delta_{\rm vir}\rho_{\rm a}}{G}\right)^{1/2}\approx 3.5\times 10^{-7}\,{\rm kg}\,{\rm m}^{-3}\,{\rm s}\,,
\end{equation}
which is independent of the size of the object - that is, there is no dependence on $M_{\rm vir}$ or $R_{\rm vir}$. If $\Delta_{\rm vir}$ is universal and independent of the size of the object, as it is supposed to be almost by definition, then the expected brightness temperature averaged over a virialised halo will be independent of the size and hence the optimal detection for a specific halo size and telescope configuration would be obtained by matching the size of the object approximately to the telescope beam width. This is the standard practice to optimise detection efficiency in all branches of astronomy.

This suggestion, that there is no optimal size of object, appears to be contrary to the conclusions of \cite{ref:Caputo}, who claimed that the optimal detection would be for dwarf spheroidal galaxies, that is, the very lowest mass halos. They came to this conclusion considering the quantity 
\begin{equation}\label{eqn:CaputoQuantity}
 \frac{1}{\Delta v}\int \mathrm{d}\Omega \mathrm{d}l\rho_{\rm a}\propto \frac{M_{\rm beam}}{d^2\Delta v}\propto S_{\rm beam}\,,
\end{equation}
where $d$ is the distance to the object and the angular integration is over the angular size of the object - or, as they state it, for a telescope beam which has the same size as the object. This quantity is $\propto S_{\rm beam}$ defined in \eqref{eqn:FluxValue} which is equivalent to (\ref{ref:brighttemp}) if one is careful with the choice of $\Sigma_{\rm beam}$. But we have already explained that one can come to the wrong conclusion if one uses the wrong value of $M_{\rm beam}$ for a specific halo and that it is actually better to think in terms of the surface mass density $\Sigma_{\rm beam}$.

In fig.~\ref{fig:ScatterPlots}, we have plotted the quantities in \eqref{eqn:CaputoQuantity} and \eqref{eqn:SigmaQuantity} using the data in table~\ref{ClusterCalculations} which is similar to, but not exactly the same as, that used in \cite{ref:Caputo}. In particular, we have added some galaxies and galaxy clusters to the dwarf galaxies which they focus on that enables us to probe a wider lever arm in mass. The table contains values for the distance to and the mass of the object $D_{\rm obj}$ and $M_{\rm obj}$, respectively, the angular size $\theta_{\rm obj}$ and the velocity width $\Delta v_{\rm obj}$. These are inferred in a heterogeneous way, but should at least be indicative of some truth. We would not necessarily expect these values to be those for a virialised halo and therefore we denoted them with the suffix ``obj" to distinguish them as being observationally determined. From the observed information, we can infer the radius, $R_{\rm obj}=\theta_{\rm obj}/(2D_{\rm obj})$ and also check consistency with our analytic estimates above by inferring $\Delta v_{\inf}=(GM_{\rm obj}/R_{\rm obj})^{1/2}$, as well as calculating the surface mass density appropriate to an average over the object radius, $\Sigma_{\rm obj}=M_{\rm obj}/(\pi R_{\rm obj}^2)$.

Firstly, we find in the right panel of fig.~\ref{fig:ScatterPlots} that \eqref{eqn:CaputoQuantity} which was plotted in \cite{ref:Caputo} is indeed $\propto\theta_{\rm obj}^2$ as claimed. But on the basis of the theoretical argument above, this is exactly what one would expect for the total flux density $S_{\rm tot}\propto \Sigma_{\rm ave}\theta^2/\Delta v$, where $\Sigma_{\rm ave}$ is some average surface mass density for the objects, and hence, while it provides some confidence that the modelling is correct, it does not yield any obvious information about which objects would be optimal.

In the left panel of fig.~\ref{fig:ScatterPlots} we have plotted $\Sigma_{\rm obj}/\Delta v$ for the data presented in table~\ref{ClusterCalculations}, using both $\Delta v_{\rm obj}$ and $\Delta v_{\inf}$ with consistent results. We find that the data are compatible with $\Sigma_{\rm beam}/\Delta v$ being a constant over eight orders of magnitude and for it to be $\approx 500$ times the background value - slightly higher than for our strawman object - within the kind of uncertainties that we might expect coming from a heterogeneous sample such as the one which we have used. Visually, there could be some evidence for a trend $\sim M^{-0.2}$ which we have also included to guide the eye, but the evidence for this is largely due to a few outliers at the low- and high-mass ends where perhaps the observational estimates are most uncertain. So it could be that there is some preference for lower mass halos over high mass halos, but the effect is not very dramatic. Note that on the $y-$axis, we plot $\left(\Sigma/\Delta v\right)^{\rm rel} \equiv \frac{\Sigma/\Delta v}{1.2\times 10^{-9}\,{\rm kg\,m^{-3}\,s}}$, where the denominator is the value associated to the background.

It could be that the possible trend seen in the left panel of fig.~\ref{fig:ScatterPlots} is related to the concentration parameter of the halo. It is likely that the observationally determined angular size, $\theta_{\rm obj}$, is not the virial radius but some scale radius from a fitting function used in conjunction with images. If this is the case, then we might expect a weak trend with mass.

The concentration parameter has been computed in numerical simulations and is usually assumed to be universal for halos of a given mass, $M$. A recently proposed expression is~\cite{ref:MultiDark}
\begin{equation}
 {\hat c}(M,z)={\hat c}_0(z)\left(\frac{M}{ M_0}\right)^{-\gamma(z)}\left[1+\left(\frac{M}{M_1(z)}\right)^{0.4}\right]\,,
\end{equation}
where $M_0=10^{12}h_{100}^{-1}M_{\odot}$ and $\hat{c}_0(z)$, $\gamma(z)$ and $M_1(z)$ are fitted parameters which are redshift dependent. We will focus on low redshifts where $\hat{c}_0(z)\approx 7.4$, $\gamma(0)\approx 0.12$ and $M_1(0)=5.5\times 10^{17}h_{100}^{-1}M_{\odot}$. 
From this we see that at $z=0$, $\hat{c}\propto M^{-0.12}$, that is, lower mass halos typically are more concentrated than higher mass halos, and therefore there will be more mass inside the scale radius, and for observations focusing on the region inside this scale radius $\Sigma_{\rm beam}$ might be larger.

\begin{table*}
 \centering
 \begin{tabular}{|l|c|c|c|c|c|}
 \hline
 Object  & $D_{\rm obj}$ & $M_{\rm obj}$ ($\rm M_{\odot}$) & $\theta_{\rm obj}$ & $\Delta v_{\rm obj}$ [$\rm km\,s^{-1}$] & Reference(s)  \\
 \hline
 Leo 1 &  $250$  kpc & $2.2\times 10^{7}$ & $12.6$ arcmin & 8.8 & \cite{ref:Mateo} \\
 NGC 6822 &  $490$ kpc & $1.6\times 10^9$ & $40$ arcmin & 8 & \cite{ref:Mateo} \\
 Draco &  $82$ kpc & $2.2\times10^{7}$ & $28.3$ arcmin & 9.5 & \cite{ref:Mateo} \\
 Wilman 1 & 45 kpc & $4\times 10^5$ & 9 arcmin & 4 & \cite{ref:DwarfGalMass}  \\
 Reticulum 2 &  30 kpc & $5.6 \times 10^5$ & $3.6$ arcmin & 3.3 &\cite{ref:Beasts, ref:Reticulum2} \\
 Sextans B & 1345 kpc& $3.9\times 10^8$ & 3.9 arcmin & 18 & \cite{ref:Mateo}\\ 
 Pegasus & 955 kpc& $5.8 \times 10^{7}$ & 3.9 arcmin & 8.6 & \cite{ref:Mateo} \\
 Antlia & 1235 kpc& $1.2\times 10^7$ & 5.2 arcmin & 6.3 & \cite{ref:Mateo} \\
 NGC 205 & 815 kpc& $7.4\times 10^{8}$ & 6.2 arcmin& 16 & \cite{ref:Mateo}\\
 \hline
 NGC 5128 & 3.8 Mpc & $5.1\times 10^{11}$ & 34.7 arcmin & 477 & \cite{ref:NearGalCat} \\
 NGC 5194 & 15.8 Mpc & $4.2\times 10^{10}$ & 8.4 arcmin & 175 & \cite{ref:NearGalCat} \\
 Maffei2 & 2.8 Mpc & $4.2\times 10^{10}$ & 3.8 arcmin & 306 & \cite{ref:NearGalCat} \\
 IC2574 & 4.0 Mpc &  $4.6\times 10^9$ & 13.2 arcmin & 107 & \cite{ref:NearGalCat} \\
 SexA & 1.3 Mpc & $2.5\times 10^8$ & 5.9 arcmin & 46 & \cite{ref:NearGalCat} \\
 NGC 3556 & 9.9 Mpc & $3.3\times 10^{10}$ & 5.0 arcmin & 308 & \cite{ref:NearGalCat} \\
 IC 0342 & 3.3 Mpc & $1.4\times 10^{11}$ & 21.4 arcmin & 181 & \cite{ref:NearGalCat} \\
 NGC 6744 & 8.3 Mpc & $2.2\times 10^{11}$ & 21.4 arcmin & 323 &\cite{ref:NearGalCat}\\
 ESO 300-014 & 9.8 Mpc & $10^{10}$ & 7.1 arcmin & 130 & \cite{ref:NearGalCat}\\
 NGC 3184 & 11.1 Mpc & $6.3\times 10^{10}$ & 7.4 arcmin & 128 & \cite{ref:NearGalCat}\\
 \hline
 Virgo & $18$ Mpc & 2.9$\times 10^{15}$ &  7 degrees & 1100 & \cite{ref:VirgoMass1,ref:VirgoMass2}  \\
 Coma & 100 Mpc & 3$\times 10^{15}$ & 100 arcmin & 1100 & \cite{ref:ComaMass, ref:ComaDistance}\\
 \hline
 \end{tabular}
 \caption{Table of masses ($M_{\rm obj}$), distances ($D_{\rm obj}$), angular sizes ($\theta_{\rm obj})$ and velocity widths ($\Delta v_{\rm obj})$ extracted from the literature and used in fig.~\ref{fig:ScatterPlots}. In each case we have specified the reference of the paper from which the numbers are extracted/calculated. From paper to paper the methods employed are different and hence the overall sample is relatively heterogeneous. For each object we can infer a radius $R_{\rm obj}=\theta_{\rm obj}D_{\rm obj}/2$ and a velocity width $\Delta v_{\rm inf}=(GM_{\rm obj}/R_{\rm obj})^{1/2}$. We find that $\Delta v_{\rm obj}$ is strongly correlated with $\Delta v_{\rm inf}$ as we would expect and indeed that $M_{\rm obj}$ is also correlated with $R_{\rm obj}$.}
\label{ClusterCalculations}
\end{table*}

\begin{figure}[t]
 \centering
 \includegraphics[width = 0.45\textwidth,height=0.45\textwidth]{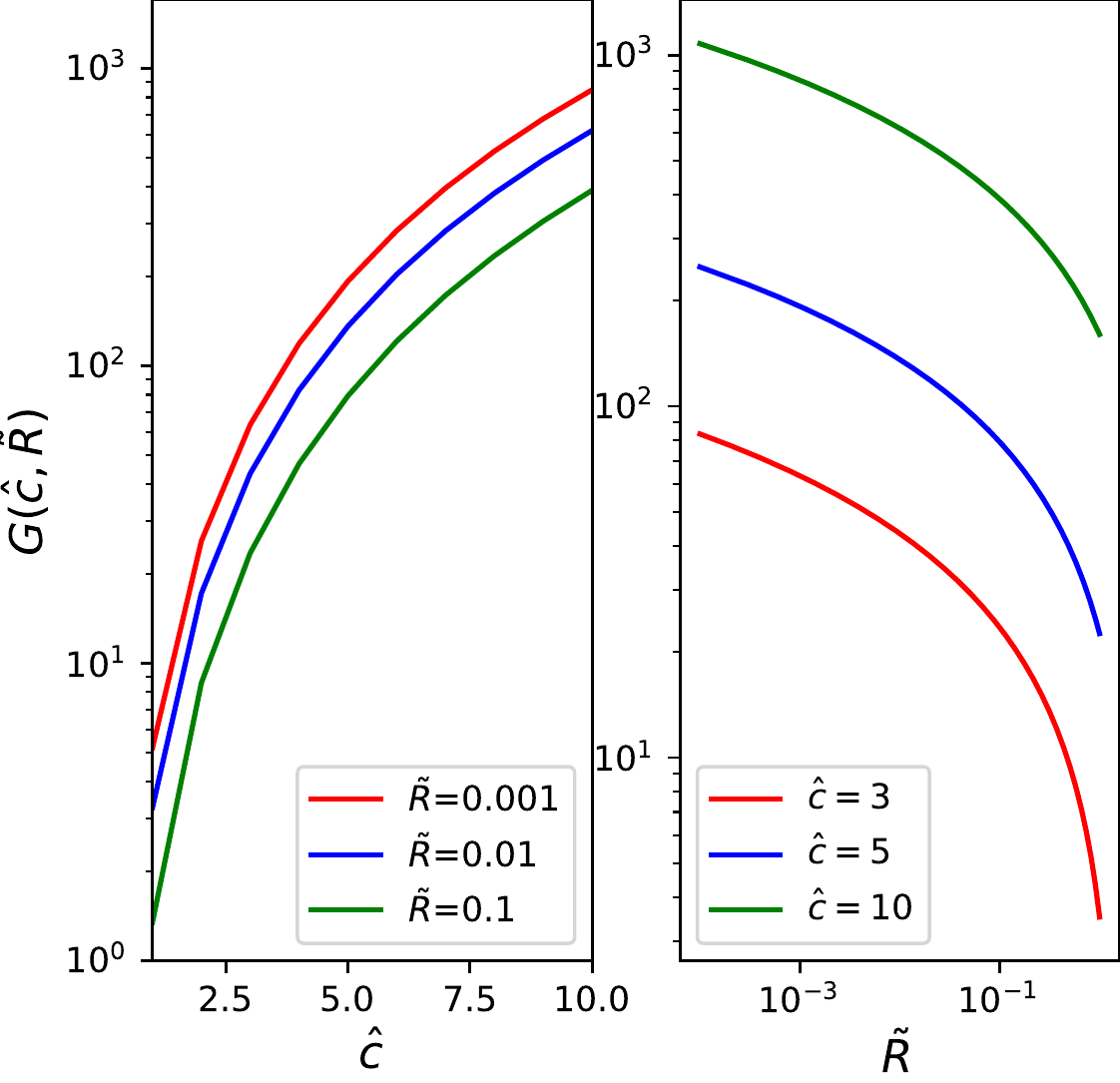}
 \caption{The function $G(\hat{c}, \tilde{R})$ as a function of its arguments. In the left panel, we plot $G$ as a function of $\hat{c}$ for different values constant $\tilde{R}$, and vice versa in the right panel.}
 \label{fig:GFunction}
\end{figure}

This leads us on to an important caveat in this discussion: one does not have to choose to focus on trying to detect the entire signal from a halo and indeed it will be optimal, as well as practical, to not do this. Using (\ref{ref:mbeam}), we can eliminate $M_{\rm beam}$ and $M_{\rm vir}$ in terms of $\Sigma_{\rm beam}$ and $\Sigma_{\rm vir}$. To do this we first recall the definition of the beam surface-mass density (see appendix \ref{App:MassinBeam})
\begin{align}
\Sigma(R_{\rm beam}) &= \int_{R_{\rm beam}}^{\hat c} \frac{r\rho(r)}{\sqrt{r^2 - R_{\rm beam}^2}} dr\,, \\
M_{\rm beam} &= \, 2\pi\int_0^{R_{\rm beam}}R\Sigma(R)\mathrm{d}R\,,
\end{align}
where $r$ is the radial coordinate of the object in question and $R_{\rm beam}$ is the projected distance which we identify to be given by the beam size. 
Explicitly for an NFW profile $\rho(r)=\rho_{\rm s}F(r/r_{\rm s})$ with $F(y)=y^{-1}(1+y)^{-2}$, where $r_{\rm s}$ is the scale radius, $R_{\rm vir}$ the virial radius and the ratio of the two ${\hat c}=R_{\rm vir}/r_{\rm s}$. Next we can expand these integrals in small beam radius limit $\bar{R}_{\rm beam}\ll1$ to find the relation
\begin{equation}\label{eq:SigmaBeamNFW}
 \Sigma_{\rm beam}\simeq G\left(\hat{c},{\frac{R_{\rm beam}}{R_{\rm vir}}}\right)\Sigma_{\rm vir}\,,\qquad \bar{R}_{\rm beam} \ll 1\,,
\end{equation}
for an NFW profile $G(x,y)=x^2\log(2y/x)/f(x)$ for $y/x\ll 1$. We anticipate that one could derive a similar expression for any halo profile. 

We plot the function $G(\hat{c}, \frac{R_{\rm beam}}{R_{\rm vir}})$ as a function of $\hat{c}$ and $\tilde{R} = R_{\rm beam}/R_{\rm vir}$, in fig.~\ref{fig:GFunction} which indicates that enhancements of up to 1000 might easily be possible and that these are likely to be larger in lower mass objects than those of higher mass. Therefore, at a first glance it would appear that, for a fixed experimental set up ($R_{\rm vir}/R_{\rm beam}$ fixed), one should search for an object with the largest concentration, a general result which we already anticipated in section~\ref{subsec:SpontSignal}. However, one should also note that for small $\tilde{R}$, which is fixed by the resolution of the telescope, the enhancement across the different concentration parameters is comparable. Furthermore, for a fixed resolution $\theta$, $R_{\rm beam}/R_{\rm vir}$ is significantly smaller for larger mass halos, since $R_{\rm vir}$ is much larger. As a result, $\Sigma_{\rm beam}$ is larger for larger mass halos. 

In conclusion, we have argued that maximising $\Sigma_{\rm beam}/\Delta v$ will give the largest possible brightness temperature signal. Theoretical arguments suggest that if the beam encloses the virial radius of a particular object, this will be independent of mass and a very rudimentary search of the literature for specific values suggests that this could be true. However, for fixed observational setup, and hence fixed resolution, one might find a significant enhancement of the signal due to the fact that the surface mass density will increase as one probes the more central regions of a halo. These are likely to be larger for larger mass objects since the telescope beam probes denser regions of larger mass halos. This is the reason we have presented our sensitivity estimates as a function of $\Sigma_{\rm beam}$ and results for range of values $\Sigma_{\rm beam}=0.07-70\,{\rm kg}\,{\rm m}^{-2}$ in fig.~\ref{fig:CMB_Enhancement}.

\subsubsection{Minimising Integration Time}

\begin{figure*}[!t]
 \centering
 \includegraphics[width = 0.74\textwidth]{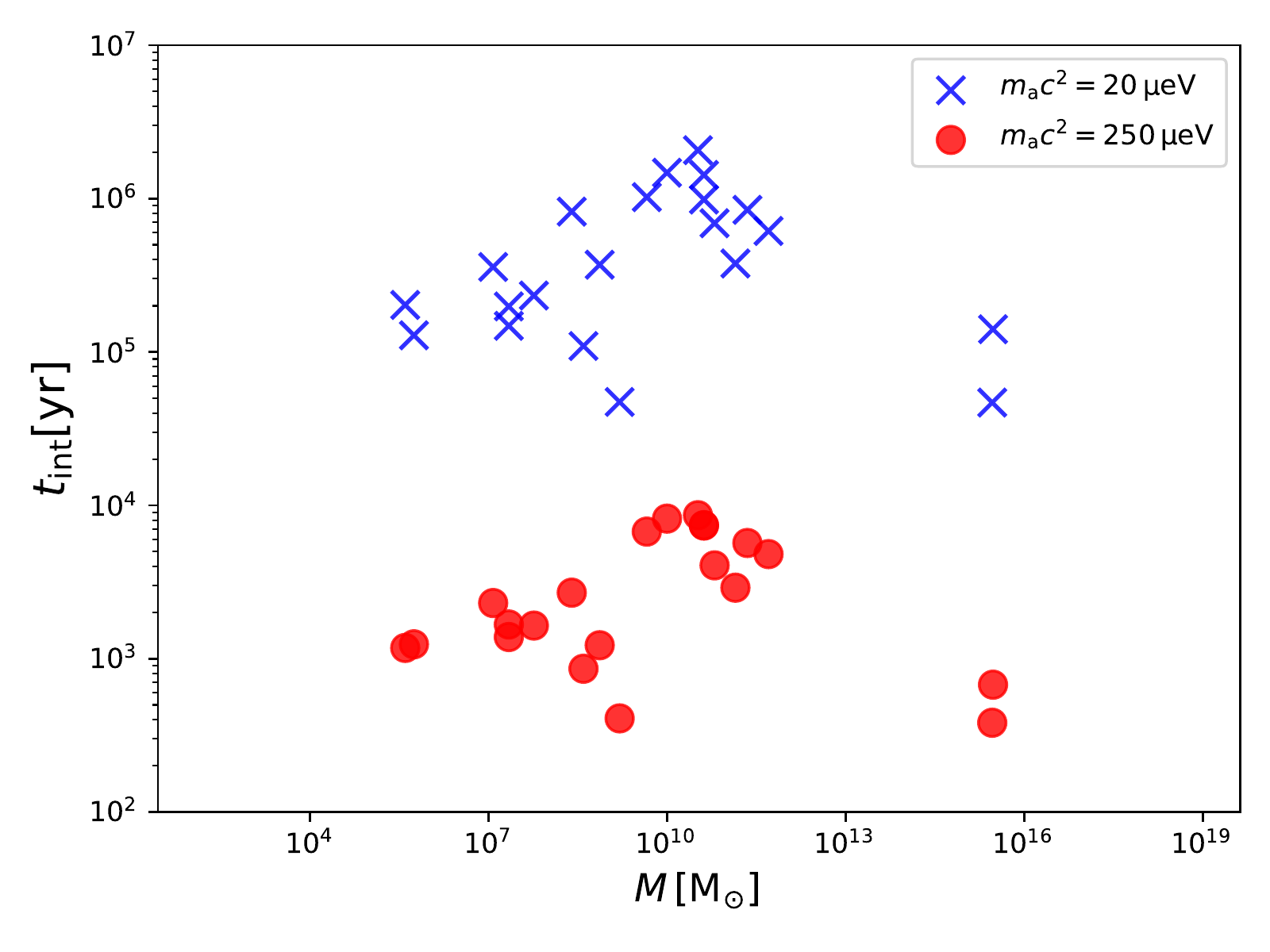}
 \caption{The integration time for the 1$\sigma$ detection of the brightness temperature signal for the objects in table \ref{ClusterCalculations}, assuming a single-pixel detector in a GBT-like telescope and the stimulated enhancement from both the CMB and the radio background. In this case, we have used \eqref{eq:SigmaBeamNFW} to evaluate $\Sigma_{\rm beam}$ assuming the resolution of the GBT, that is, the virial mass and the virial radii are related by the virial overdensity parameter, $M_{\rm vir} \propto R_{\rm vir}^3 \Delta_{\rm vir}$. Note that we assume the fiducial signal strength corresponding to $g_{\rm a\gamma\gamma} = 10^{-10}\,{\rm GeV^{-1}}$.}
 \label{fig:tintPlots}
\end{figure*}

From \eqref{eqn:IntegrationTime} and \eqref{eqn:tintegrate} we see that the integration time can be expressed either in terms of $M_{\rm beam}$ or $\Sigma_{\rm beam}$. Here we shall use the latter measure. We have just seen how brightness temperature is proportional to $\Sigma_{\rm beam}/\Delta v$ and therefore largest when this ratio is maximal. However, whilst brightness temperature is a key observable, the ultimate arbiter of feasibility of detection is of course the integration time. From \eqref{eqn:IntegrationTime} we see the integration time has a slightly different dependence on the halo parameters $\Sigma_{\rm beam}$ and $\Delta v$ to that of the brightness temperature, scaling instead as $t_{\rm int} \propto \frac{1}{\Delta v} (\Sigma_{\rm beam}/\Delta v)^{-2}$, with the additional factor of $1/\Delta v$ arising from the bandwidth of the signal. In light of the different parametric dependence of the integration time and brightness temperature on the halo parameters $\Sigma_{\rm beam}$ and $\Delta v$, and from table \ref{ClusterCalculations} since $\Delta v$ varies significantly between objects, formally maximising $\Sigma_{\rm beam}/\Delta v$ (brightness temperature) is slightly different to minimising $\Delta v/\Sigma_{\rm beam}^2$ (integration time). Thus, it is natural to re-run the analysis of the previous discussion and check whether there is also no preferred object group for $t_{\rm int}$.

We can then estimate the beam surface mass density $\Sigma_{\rm beam}$ using the NFW profile as found in   \eqref{eq:SigmaBeamNFW} and take values of $\Delta v$ from table \ref{ClusterCalculations} as before. Thus, we must know $R_{\rm vir}$, $\hat{c}$ and $\Delta v$. We can infer the virial radius from the mass of the object $M_{\rm vir}= M_{\rm obj} =\frac{4\pi}{3}\Delta_{\rm vir}\rho_{\rm a} R_{\rm vir}^3$, using the values in the table. The results for the integration time for different objects are plotted in fig.~\ref{fig:tintPlots}. We have assumed the resolution of the GBT, i.e., $\theta_{\rm FWHM} \approx 10^{-4}$ at 30 GHz.  

At $m_{\rm a}c^2 = 250\, {\rm \mu eV}$, the stimulated enhancement factor is quite small. However, the decay time $\tau_{2\gamma}$ is significantly smaller than at $m_{\rm a}c^2 = 20\,{\rm \mu eV}$. The values of $\cal F^{\rm eff}_{\gamma}$ at lower mass aren't large enough to compensate for the increase of the decay time. Note that $\Sigma_{\rm beam}$ is roughly a factor of 2-3 smaller for lower mass, since the resolution is a factor $\approx 12$ larger. Therefore, the integration time is lower at larger masses. As mentioned before, we see that the larger mass halos give a slightly lower integration time, since we are probing smaller values of $\tilde{R}$, i.e., denser regions of the halo. The Virgo cluster at $m_{\rm a}c^2 = 250 \,{\rm \mu eV}$ has an integration time of around 350 years. Ideally, one would want to find objects where $1 + {\cal F}_{\gamma} \gg 1$ at $m_{\rm a}c^2 \geq 100\,{\rm \mu eV}$. Therefore, this motivates a more detailed study of the radio emission from the centre of the Virgo cluster. 

In \cite{ref:Caputo1} it was suggested that the Galactic Centre could be a target since it would benefit from a large signal enhancement from the CMB, the measured radio background, but perhaps most importantly from the diffuse radio emission associated with the high density region and supermassive black hole located there. The size of the enhancement in this direction, ${\cal F}_{\gamma}^{\rm GC}$, due to the photon occupation number density, will depend on the resolution of the telescope used in the measurement since ${\cal F} \approx I_{\nu}/E^3$. Hence, we need to estimate the intensity of radio emission from the Galactic Centre.

A measurement of the flux density of Sagittarius A$^{\ast}$ at 30 GHz is presented in the Planck Point Source  Catalogue~\cite{ref:PSC} and we will assume an intensity power law spectral index $\alpha = -2.8$ indicative of synchrotron emission and compatible with the spectrum of the Galactic Centre~\citep{ref:PlanckIV_2018}. For any observation for which this source is effectively point-like, the intensity can be estimated as $I=S/\Omega_{\rm beam}\times (f/30{\rm GHz})^{-2.8}$ where $S\approx 200 {\rm Jy}$ is the  flux density from the catalogue, $f$ is the frequency of observation and $\Omega_{\rm beam}$ is the area of the beam, which scales with frequency like $f^{-2}$.

For a GBT-like instrument, this gives us an intensity estimate $\approx 5\times 10^{5} \,{\rm Jy}\,{\rm sr}^{-1}$ and hence the enhancement is
\begin{equation}
 {\cal F_{\gamma}^{\rm GC}} \approx 50 \left(\frac{250\,{\rm \mu eV}}{m_{\rm a}c^2}\right)^{0.8} \,.
\end{equation}
Clearly, this suggests that the galactic centre might be a good candidate to target for future studies. Of course, we are assuming in this calculation that the synchrotron index is the dominant contributor to the frequency dependence of the signal, which might be an oversimplification. However, this estimate clearly demonstrates that one can achieve similar sensitivity to the galactic centre with just a 100 m single-dish telescope rather than an array of many dishes used in auto-correlation mode, as done in reference \cite{ref:Caputo1} (which indicates that our order-of-magnitude estimate approximately agrees with their analysis). To make an accurate estimate of the stimulated enhancement factor, a dedicated study of the synchrotron, free-free as well as anomalous microwave emission(s) needs to be carried out, ideally on a pixel-by-pixel basis, from high-resolution observations of the galactic centre. 

\subsection{Observational conclusions}
In the previous sections we have argued that the brightness temperature is a more robust quantity to measure, since one does not have to optimise to a specific solid angle for a given resolution. As a result, we have concluded that the appropriate quantity to optimise is $\Sigma_{\rm beam}/\Delta v$. Higher resolution measurements of objects can benefit from an enhancement in the measured $\Sigma_{\rm beam}$. For a flux density measurement, such an arrangement would result in $M_{\rm beam}\ll M_{\rm vir}$, which, of course, implies a weaker signal. Therefore, for single dish observations, the clear way forward is to target smaller regions of the Universe where one may obtain an enhancement for the surface mass density. Clearly, for such observations, one will require higher resolution which is easy for instruments like the GBT. 

\begin{figure*}
 \centering
 \includegraphics[width = 0.45\textwidth]{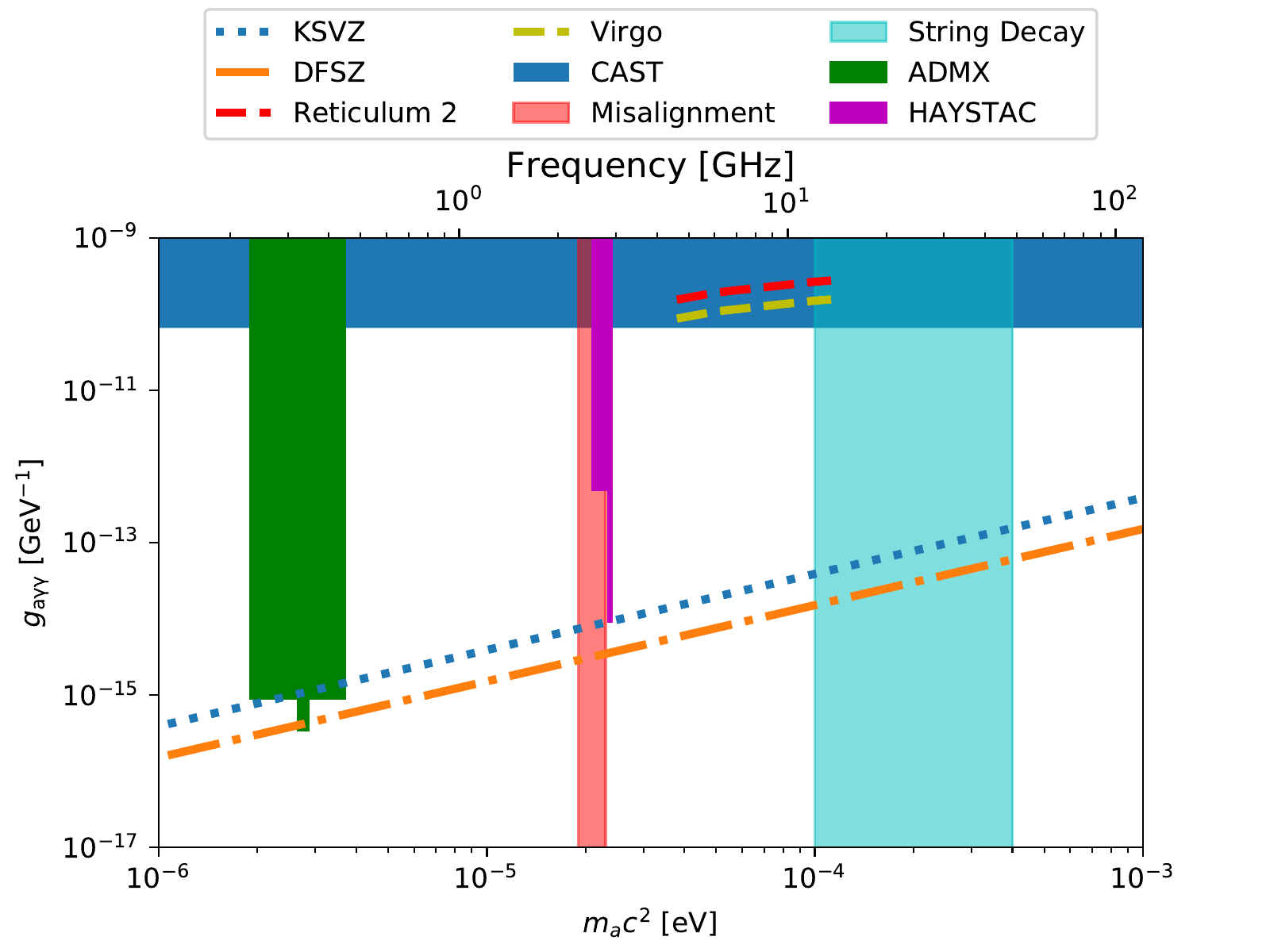}
 \includegraphics[width = 0.45\textwidth]{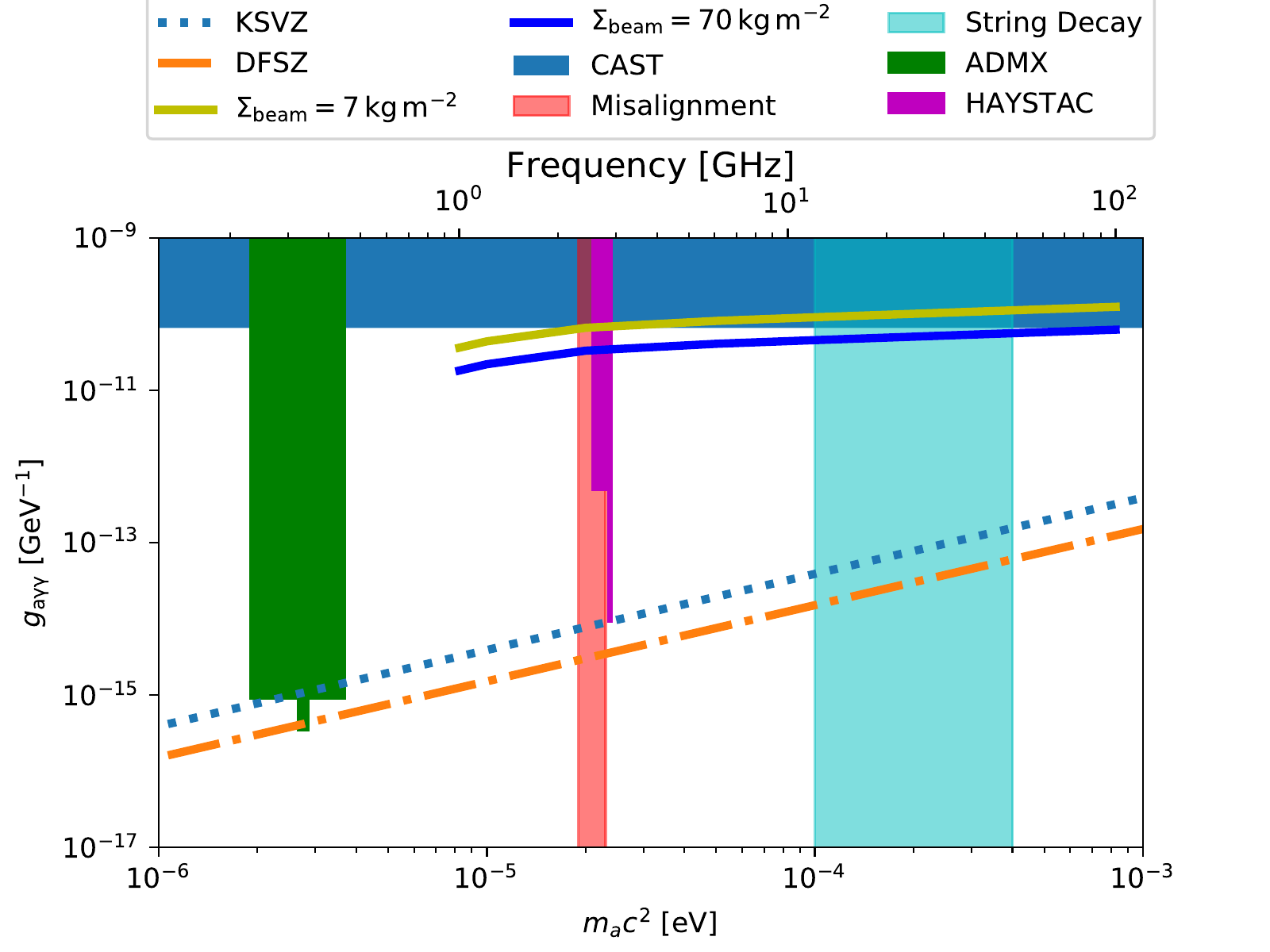}
 \caption{The sensitivity to axion-photon coupling as a function of axion mass observing a source with surface mass density $\Sigma_{\rm beam}$ and a velocity dispersion of $200\,{\rm km\,s^{-1}}$. In the left panel, we assume $N = 10^4$ telescopes (SKA2:Band 5), used in single-dish mode for an integration time of 4 days and a system temperature of $30\,{\rm K}$. The frequency coverage is as given on table \ref{tab:Telescopes}. We include e the enhancement due to the CMB and the radio background in this case, but note that the enhancement from the radio background is very uncertain. In the right panel, we show the sensitivity from observations of the galactic centre between 1 and 100 GHz, assuming a 100 m single-dish telescope with a system temperature of $30\,{\rm K}$, such as the GBT, and an integration time of 4 days. We included estaimtes of the stimulated emission enhancement from the CMB, the radio background and the synchrotron emission from the supermassive black hole, Sagittarius A$^{\ast}$ discussed in the text. We note that in reality, the system temperature for most radio telescope receivers varies with frequency, which would need to be modelled in an experiment. The sky-blue shaded region is the parameter region excluded by the CERN Axion Solar Telescope (CAST) \cite{ref:CAST} \footnote{We thank Igor Irarstorza for sharing the CAST data.}. The green and magenta exclusions are from the ADMX \cite{ref:ADMX2018} and HAYSTAC \cite{ref:HAYSTAC} haloscope experiments. We also highlight the axion mass ranges predicted by the misalignment mechanism (red) and the string decay (cyan).}
 \label{fig:SensitivityVir}
\end{figure*}

We have also discussed the stimulated decay enhancement of the signal and noted that this enhancement is substantial at lower mass. A future experiment would greatly benefit from a dedicated study of specific sources for which high intensity radio emission has been measured. In our previous section, we motivated the Virgo cluster and the galactic centre. Note that for our sensitivity estimates for the galactic centre, we have assumed a constant $\Sigma_{\rm beam}$ for all axion masses, since the presence of the black hole results in a density spike at the galactic centre out to a few parsecs from the position of Sagittarius $\rm A^{\ast}$.
For the radio background, we use the power law derived in \cite{ref:ARCADE2}, given by
\begin{equation}
 T_{\rm ARCADE-2}\approx 1.2\,{\rm K}\left(\frac{1\,{\rm GHz}}{f_{\rm obs}}\right)^{2.62}\,.
\end{equation}
Substituting this expression back in, one obtains 
\begin{equation}
 {\cal F_{\gamma}^{\rm RB}} \approx 1.6\times 10^3 \left(\frac{1\,{\rm GHz}}{f_{\rm obs}}\right)^{3.62} \,.
\end{equation}
We note that this is probably an over-estimate of ${\cal F}^{\rm RB}$ since the ARCADE measurement would require an additional population of radio sources at the relevant frequencies. In principle, there is also a free-free component as well as anomalous microwave emission from the galactic plane, some of which will contribute to the photon occupation number associated to the galactic centre. We remark that while a complete study of the sensitivity to the galactic centre is outside the purpose of this work, our order of magnitude estimate motivates a more detailed future study. 

In the near future, the SKA will go into operation. With $1\,{\rm km}^2$ of collecting area, the SKA brings the possibility of very high radio sensitivity. However, we note an sparse interferometer is, by construction, most suited to measuring flux densities with high resolution. One can use Rayleigh-Jeans law to convert the noise level on the flux density, which is set by the collecting area into a brightness temperature temperature sensitivity 
\begin{equation}
T_{\sigma} = \frac{T_{\rm sys}}{\eta_{\rm FF}\sqrt{\Delta f_{\rm obs}t_{\rm obs}}}\,.
\end{equation}
The factor $\eta_{\rm FF} \equiv (N A_{\rm eff})/D_{\rm baseline}^2\ll 1$ is known as the filling factor and this increases the expected noise level for the brightness temperature. Here, $N$ is the number of telescopes in the interferometric setup and $A_{\rm eff}$ is the effective collecting area of each telescope. However, if the telescopes are all used in single dish mode, then the integration time for a measurement decreases by a factor $N$ since all the telescopes can point at the same region of the sky. 

The high resolution associated with interferometers also means that their large collecting area is offset by the small beam size, again decreasing $M_{\rm beam}$ by several orders of magnitude. As mentioned before, the flux density sensitivity can be increased by using the telescope in single-dish mode, which results in a factor of $N$ decrease in integration time.  

We conclude from our analysis that the brightness temperature is the appropriate quantity to optimise radio telescope searches for the spontaneous decay. In fig.~\ref{fig:SensitivityVir} we show our estimates of the radio sensitivity to the spontaneous decay. In both the panels, we have set the integration time, $t_{\rm int}$ to be 4 days. The left panel shows the SKA2:Band 5 sensitivity operating in the single dish mode for the Virgo cluster and the Reticulum 2 dwarf galaxy using the numbers explained in the caption. Note that in principle, the sensitivity to the Virgo cluster could be significantly better, as we assume there is no radio emission from the centre of Virgo at frequencies larger than 10 GHz. In the right panel, we show the sensitivity to the galactic centre, assuming $\Sigma_{\rm beam} \approx 7$ and $70\,{\rm kg\,m^{-2}}$ and a single pixel detector in a GBT-like telescope. It is clear the galactic centre is a promising target for future experiments, which motivates a more detailed study of the different sources of stimulated enhancement.

\section{Resonant mixing in neutron stars}\label{sect:mixing}
There has recently been renewed interest in the possibility of detecting radio signals from the resonant conversion of dark matter axions in neutron star magnetospheres \cite{ref:NS-Hook,ref:NS-Japan}, originally proposed in \cite{Pshirkov:2007st} together with a number of follow-up studies \cite{Camargo:2019wou,Safdi:2018oeu,Edwards:2019tzf}. The conversion happens in some small critical region within the magnetosphere where the plasma mass $\omega_{\rm pl}$ is approximately equal to the axion mass $m_{\rm a}$. This part of the magnetosphere -- whose width $\propto 1/\left|\nabla \omega_{\rm pl}\right|$ is determined by the gradients of the background plasma -- acts essentially as a stellar haloscope. The characteristic frequencies for non-relativistic axions are given by the axion mass. The emitted radiation then results in a radio line peaked at frequencies $\omega \simeq m_{\rm a}$. 

The effect is similar to the Mikheyev–Smirnov–Wolfenstein (MSW) mechanism for neutrino inter-conversion \cite{Kuo:1989qe} where a finite density of background charge carriers can endow neutrinos with an effective mass so that when the mass-splitting becomes small, flavour mixing is enhanced. Relativistic axion-photon mixing in neutron stars has also been studied in \cite{ref:LaiHeyl} where, by contrast with the dark matter axion case, it was assumed that all particles are in the \textit{weak dispersion} regime $\omega \simeq \left| \textbf{k} \right|$, as in earlier references \cite{Raffelt:1987im}. 

The principal aim of this section is to re-examine the canonical assumptions made in the study of axion-photon mixing in a medium and determine to what extent they can be justified in a neutron star setup. Our analysis focuses on the following points:

\begin{enumerate}
 \item Unlike for simple haloscopes with constant magnetic fields and uniform plasma densities, magnetospheres are inhomogeneous with a non-trivial 3D structure. We, therefore, examine to what extent the axion-Maxwell equations can be reduced to a two-flavour mixing system in a 1D planar geometry, whose evolution depends on a single integration parameter along the line of sight.
 \item We go beyond refs.~\cite{ref:NS-Hook,ref:LaiHeyl} and perform a controlled gradient expansion (appendix \ref{Density}) of the mixing equations similar to ref.~\cite{Prokopec:2003pj}. This allows us to obtain in a systematic way the leading order WKB behaviour of the mixing system and has the particular advantage of providing a careful treatment of dispersion relations which are in general distinct for the axion and photon away from the resonance region. Our treatment is also valid away from purely relativistic/non-relativistic regimes, with our final form of the first order mixing equations valid for arbitrary values of the momenta.
 \item We establish in which regions of the axion phase-space $(k,m_{\rm a})$ the evolution can be considered non-adiabatic. This determines when the $a \rightarrow \gamma$ conversion can be treated perturbatively in the coupling $g_{\rm a \gamma \gamma}$ and where a non-perturbative Landau-Zener formula \cite{Brundobler,ref:LaiHeyl} for two-level mixing must be applied.
 \item We examine the role of higher dimensional structure in producing a longitudinal mode $\nabla \cdot \textbf{E}\neq 0$ for the photon and to what extent geometry affects the decoupling of polarisations $\textbf{E}_\parallel$ and $\textbf{E}_{\perp}$, parallel and normal to the background magnetic field.
\end{enumerate}

\subsection{Axion Electrodynamics}
Our starting point is the standard Lagrangian for the axion and photon, with medium effects described by a current $j^{\mu}$:
\begin{equation}\label{eqn:ClassicalEulerHeisenberg}
 \begin{split}
  \mathcal{L} = & -\frac{1}{4}F_{\mu\nu}F^{\mu\nu} - A_{\mu}j^{\mu} \\
  & + \frac{1}{2}\left(\partial_{\mu}a\partial^{\mu}a - m_{\rm a}^2a^2 \right) + \frac{1}{4}g_{\rm a\gamma\gamma}aF_{\mu\nu}\tilde{F}^{\mu\nu} \,,
 \end{split}
\end{equation}
where $F^{\mu\nu}$ and $\tilde{F}^{\mu\nu}$ are the electromagnetic field tensor and its dual, respectively. The equations of motion for the electromagnetic (EM) fields are given by
\begin{align}
\nabla \cdot \textbf{E}& = \rho - g_{\rm a \gamma \gamma} \textbf{B} \cdot \nabla  a\,, \label{Gauss}\\
\nabla \times \textbf{B} - \dot{\textbf{E}} &= \textbf{J} +  g_{\rm a \gamma \gamma}\dot{a} \textbf{B} - g_{\rm a \gamma \gamma} \textbf{E}\times \nabla a\,,\label{curlB}\\
\nabla \cdot \textbf{B} &=0\,, \\
\dot{\textbf{B} } + \nabla \times \textbf{E}& =0 \label{Bianchi2}\,.
\end{align}		
Next, we linearise the equations of motion about the background solutions satisfying the $g_{\rm a \gamma \gamma}=0$ equations of motion by setting $\textbf{E} \rightarrow \textbf{E}_0 + \textbf{E}$ and $\textbf{B}\rightarrow \textbf{B}_0 + \textbf{B}$, with a corresponding ansatz for $\rho$ and $\textbf{J}$. We also neglect the background electric field, setting $\textbf{E}_0=0$, since for neutron stars the magnetic component typically dominates in the magnetosphere, see, e.g., \cite{Melrose:2016kaf}. The electromagnetic fluctuations must be self-consistently accompanied by perturbations of charge carriers in the plasma via Lorentz forces. This can be modelled via an Ohm's law relation between the current and electric fluctuations $\textbf{E}$ and $\textbf{J}$,
\begin{equation}\label{Ohm}
 \textbf{J} = \sigma \cdot \textbf{E}\,,
\end{equation}
where the three-by-three matrix $\sigma$ is the conductivity tensor. Note that together with current conservation $\dot{\rho} + \nabla \cdot \textbf{J}=0$, this closes the system of equations. To obtain a simple system of mixing equations, we specialise to a stationary background throughout the remainder of this section assuming $\textbf{B}_0$ and $\sigma$ to be time-independent, as would be the case for an aligned rotator neutron star model. One then obtains the following system of mixing equations for $\textbf{E}$ and $a$,
\begin{align}
\square \, a + m_{\rm a}^2 a & = g_{\rm a \gamma \gamma}\textbf{E} \cdot \textbf{B}_0\,,\label{axionEOM}\\
\square \, \textbf{E} + \nabla (\nabla \cdot \textbf{E}) + \sigma \cdot \dot{\textbf{E}} &= -g_{\rm a \gamma \gamma} \ddot{a}\textbf{B}_0\,, 
\label{EOMPerts}
\end{align}
where \eqref{EOMPerts} was obtained by taking the curl of \eqref{Bianchi2} and combing with \eqref{curlB} and \eqref{Ohm}. We have thus completely parametrised the axion-photon fluctuations in terms of two physical fields, $\textbf{E}$ and $a$. Note that the magnetic component is determined immediately from integration of \eqref{Bianchi2}. We see from \eqref{EOMPerts} that, in general, different polarisations of $\textbf{E}$ will mix owing to the presence of a longitudinal mode $\nabla \cdot \textbf{E} \neq 0$, which can be sourced via the axion [see eq.~\eqref{Gauss}] or when $\sigma$ has off-diagonal components. Note, furthermore, that in a stationary background, the fields have simple harmonic time-dependence $\sim e^{-\imath\omega t}$. The conductivity in a magnetised plasma takes the form \cite{Gurevich2006}
\begin{equation}\label{sigma}
 \sigma(\omega) = \frac{\imath\,e^2 n_e}{m_e} R_B(\theta)
  \left(
        \begin{array}{ccc}
         \frac{\omega}{\omega^2  -\omega_{\rm B}^2 } & \,  \frac{i\omega_{\rm B}}{\omega^2  -\omega_{\rm B}^2 } & \, 0 \\
         -\frac{i\omega_{\rm B}}{\omega^2  -\omega_{\rm B}^2 } &  \frac{\omega}{\omega^2  -\omega_{\rm B}^2 } &\, 0 \\
         0 & \, 0 & \, \frac{1}{\omega}
        \end{array}
  \right)R_B(\theta)^{-1}\,,
 \end{equation}
 where $\theta=\theta(\textbf{x})$, $\omega_{\rm B}=eB_0/m_e$ is the gyrofrequency, $R_B(\theta)$ is the local rotation matrix which rotates $\textbf{B}_0$ into the $z$-direction and $B_0=|\mathbf{B}_0|$. We assume furthermore that $\omega \ll \omega_{\rm B}$, which is easily satisfied for neutron stars with $B \simeq 10^{9}$-$10^{14}\text{G}$ and frequencies $\omega \simeq m_{\rm a} \sim \mu\text{eV}$ associated to non-relativistic axions. In this case, one has $\sigma(\omega)\cdot \textbf{E} = \imath(\omega_{\rm pl}^2/\omega) \textbf{E}_{\parallel}$, where $\textbf{E}_{\parallel}$ is the component of $\textbf{E}$ along $\textbf{B}_0$.

\subsection{Resonant mixing in 1D}\label{sec:1D}
Here we spell out what are the precise physical assumptions needed to reduce the plasma \eqref{axionEOM}-\eqref{EOMPerts} to a simple 1D problem. 

Consider first a planar geometry in which all background fields depend on a single parameter $z$, i.e., $\textbf{B}_0=\textbf{B}_0(z)$. Then, since $\textbf{B}_0$ is transverse ($\nabla \cdot \textbf{B}_0 = 0$), it follows immediately that $\textbf{B}_0$ has no polarisation in the $z$-direction. Consider also that the wavefronts propagate in the same direction, such that $a = a(z)$ and $\textbf{E} = \textbf{E}(z)$. Crucially, these geometric assumptions ensure
\begin{equation}
 \textbf{B}_0(z) \cdot \nabla (\nabla \cdot \textbf{E}(z)) = 0\,,
\end{equation}
since by construction there are no gradients in the direction of $\textbf{B}_0$. Thus, by geometric considerations and assumptions, we are able to exclude the effects of a longitudinal component $\nabla \cdot \textbf{E}$ from the mixing equations. One can then project \eqref{EOMPerts} onto $\textbf{B}_0$ to arrive at the following set of mixing equations,
\begin{equation}
\left(
\begin{array}{cc}
 \partial_z^2 - m_{\rm a}^2 + \omega^2 & \quad \omega g_{\rm a \gamma \gamma} B_0(z)\\ 
 \omega g_{\rm a \gamma \gamma} B_0(z) & \partial_z^2 - \omega_{\rm pl}^2(z) +\omega^2
\end{array}
\right)
\left(
\begin{array}{cc}
a\\
\mathcal{E}
\end{array}
\right) = 0\,, \label{eq:2by2}
\end{equation}
where $\mathcal{E} = E_{\parallel}/\omega$, $E_{\parallel}=\textbf{E}\cdot\textbf{B}_0/|\textbf{B}_0|$ is the component of $\textbf{E}$ parallel to $\textbf{B}_0$ and $\omega_{\rm pl}^2 = e^2 n_{e}/ m_{\rm e}$ is the plasma frequency. The remaining component $E_{\perp}$ normal to $\textbf{B}_0$, from Gauss' law can be seen to satisfy $\partial_z E_{\perp} = 0$ and thus by boundary conditions must vanish. Thus, in such a geometry, the mixing simplifies to only two degrees of freedom. To fully solve these equations, one should ensure that solutions have the appropriate ingoing and outgoing waves at infinity,
\begin{align}
z \rightarrow -\infty: \quad 
&\left(
\begin{array}{c}
a \\
\mathcal{E}
\end{array}
\right) = \left(
\begin{array}{c}
a_I \, e^{\imath k_{\rm a} z} \\
0
\end{array}
\right)
+
 \left(
 \begin{array}{c}
 a_R \,e^{-\imath k_{\rm a} z} \\
 \gamma_R \ e^{-\imath k_\gamma z}
 \end{array}
 \right)\,,\\
 z \rightarrow \infty : \quad 
 &\left(
 \begin{array}{c}
 a \\
\mathcal{E}
 \end{array}
 \right) = 
 \left(
 \begin{array}{c}
 a_T \,e^{\imath k_{\rm a} z} \\
 \gamma_T \ e^{\imath k_{\gamma} z}
 \end{array}
 \right)\,, \label{1DMixing}
\end{align}
where $a_I$ is the amplitude of the incident wave and $\gamma_R$ and $a_R$, $\gamma_T$ and $a_T$ are the amplitudes of the reflected and transmitted waves, respectively.

There are two principal analytic formulae which describe the resonant conversion, one of which, as we now show, is the truncation of the other. The first result \cite{ref:NS-Hook,Camargo:2019wou,Safdi:2018oeu,Edwards:2019tzf} is perturbative, whilst the second explicitly solves the mixing equations with appropriate boundary conditions, providing a non-perturbative conversion amplitude in $g_{\rm a \gamma \gamma}$ - this is the Landau-Zener formalism \cite{ref:LaiHeyl,Brundobler}.

The first step in deriving analytic results is to reduce the system to a first order equation. This involves two stages, firstly a gradient expansion with respect to background fields and secondly imparting information about local dispersion relations into the resulting equations. A somewhat heuristic derivation of a first order equation is given in the classic reference \cite{Raffelt:1987im} for relativistic particles $k \gg \omega_{\rm pl}, \, m_{\rm a}$ with trivial dispersion $\omega \simeq k$. This is the so-called ``weak dispersion" regime also examined in \cite{ref:LaiHeyl}. However, here we deal with non-relativistic dark matter axions which have $\omega \simeq m_{\rm a}$, and since we are interested also in a photon whose dispersion varies locally according to $\omega^2 = k^2 + \omega_{\rm pl}^2$, a more subtle analysis is required. We therefore derive explicitly in appendix~\ref{Density} the following first-order analogue of \eqref{eq:2by2},
\begin{equation}
\frac{\mathrm{d}}{\mathrm{d}z}
\left( 
\begin{array}{c}
\psi_{\rm a}\\
\psi_\gamma
\end{array}
\right) = \frac{\imath}{2 \bar{k}(z)} \left(
\begin{array}{cc}
m_{\rm a}^2 & \omega g_{\rm a\gamma \gamma} B_0(z)\\
\omega g_{\rm a\gamma \gamma}B_0(z) & \omega_{\rm pl}^2(z)
\end{array}
\right)
\left( 
\begin{array}{c}
\psi_{\rm a}\\
\psi_\gamma
\end{array}
\right)\,, \label{eq:Schrodinger}
\end{equation}
with $\bar{k} \equiv \sqrt{\omega^2 - \bar{M}^2}$ and where the key difference from refs.~\cite{ref:NS-Hook} or \cite{ref:LaiHeyl} is the realisation that the distinct axion and photon mass-shell conditions express themselves in a local \textit{average momentum} associated to the average $\bar{M}^2 = (M_1^2 + M_2^2)/2 = (\omega_{\rm pl}^2 + m_{\rm a}^2)/2$ of the two eigenmasses,
\begin{equation}
 M_{1,2}^2 = \frac{1}{2}\left\{m_{\rm a}^2 + \omega_{\rm pl}^2 \pm \left[ (m_{\rm a}^2 - \omega_{\rm pl}^2)^2 + 4 B_0^2g_{\rm a\gamma\gamma}^2 \omega^2 \right]^{1/2} \right\}\,.
\end{equation}
In particular, it also varies throughout space. Note that in the relativistic limit $\bar{k} \rightarrow \omega$ reproduces the weak dispersion equations of \cite{ref:LaiHeyl} and at the critical point, one can set $\bar{k} \rightarrow k$ to the axion momentum $\omega^2 = k^2 + m_{\rm a}^2$, giving the localised version of ref.~\cite{ref:NS-Hook} about $z=z_{\rm c}$, where $z_{\rm c}$ is the location of the resonance at which $m_{\rm a} = \omega_{\rm pl}$. Here $\psi_{\rm a}$ and $\psi_\gamma$ appearing in eq.~\eqref{eq:Schrodinger} can be viewed as axion and photon states which have been put on-shell. For compactness of notation we also define
\begin{equation}
\Delta_{\rm a} = m_{\rm a}^2/2\bar{k}\,, \quad 
\Delta_\gamma = \omega_{\rm pl}^2/2\bar{k}\,, \quad 
\Delta_B = \omega g_{\rm a\gamma \gamma} B_0/2\bar{k}\,.
\end{equation}

\subsubsection{Perturbative calculation}

As was done in \cite{ref:NS-Hook} following the approach of \cite{Raffelt:1987im}, these equations can be solved perturbatively. Following the latter of these references, by going to the interaction picture, one can derive the following conversion probability
\begin{equation}\label{eq:Pintegral}
P_{\rm a \rightarrow \gamma} = \left| \int_{-\infty }^\infty dz^{\prime} \Delta_B (z^{\prime}) e^{\imath \int_0^{z^{\prime}} dz^{\prime\prime} \left[\Delta_\gamma(z^{\prime\prime}) - \Delta_{\rm a}(z^{\prime\prime})\right]} \right|^2\,.
\end{equation}
The exponent is stationary at the resonance, allowing one to perform the integral using the stationary phase approximation to get
\begin{equation}
P_{\rm a \rightarrow \gamma} = \frac{2 \pi \Delta_B^2(z_{\rm c})}{ |\Delta^{\prime}_{\gamma}(z_{\rm c})|} \equiv 2 \pi \gamma\,. \label{eq:gammaDef}
\end{equation}
where $z_{\rm c}$ is defined by $\omega_{\rm pl}(z_{\rm c}) = m_{\rm a}$ and the prime represents the derivative with respect to $z$. In order to make contact with the Landau-Zener formula for the conversion probability of ref.~\cite{ref:LaiHeyl}, we note that by using the definition of the mixing angle
\begin{equation}
\tan{2 \theta} = \frac{\omega B_0(z)g_{\rm a \gamma \gamma}}{m_{\rm a}^2 - \omega_{\rm pl}^2}\,, \label{eq:Mangle}
\end{equation}
we can write
\begin{equation}
\gamma = 2\pi \frac{\Delta M^2(z_{\rm c})/2\bar{k}_{\rm c}}{4 |\theta^{\prime}(z_{\rm c})|} + \mathcal{O}\left(\bar{k}^{\prime}(z_{\rm c}), B_0^{\prime}(z_{\rm c}) \right)\,,
\end{equation}
where $\Delta M^2 = M_1^2 - M_2^2$ is the mass-splitting in the mass-diagonal basis. Thus, up to gradients in the dispersion relation and the magnetic field, the result is precisely that of \cite{ref:LaiHeyl}. Note that by looking at the exponent in the stationary phase approximation, the width of the corresponding Gaussian gives the characteristic width $\Delta z_{\rm c}$ of the resonant region
\begin{equation}
 \left(\Delta z_{\rm c}\right)^2 = \frac{\pi}{|\Delta^{\prime}_{\gamma}(z_{\rm c})|}\,.
\end{equation}
We mimic the $\sim 1/r^3$ behaviour of the near-field dipole of the neutron star by taking 
\begin{equation}\label{eq:B1D}
 B_0(z) = \frac{B_{\ast} R^3}{z^3}\,,
\end{equation}
and use the Goldreich-Julian density \cite{ref:GJ} for the plasma frequency, with $n_e = \Omega B_0(z)$ and $\Omega$ the rotation frequency of the neutron star, from which it follows that
\begin{equation}\label{eq:deltazc}
  \Delta z_{\rm c} \simeq \sqrt{\frac{2\pi z_{\rm c} \bar{k}}{3 m_{\rm a}^2}}\,, \qquad 
 z_{\rm c} = R \left[\frac{B_{\ast} \Omega e^2}{m_e m_{\rm a}^2} \right]^{1/3}\,.
\end{equation}
This allows one to write the conversion probability explicitly as
\begin{equation}\label{eqn:Hook}
P_{\rm a\rightarrow\gamma} = \frac{1}{2}\,\frac{\omega^2}{\bar{k}^2(z_{\rm c})} g_{\rm a \gamma \gamma}^2\, B(z_{\rm c})^2 \Delta z_{\rm c}^2\,.
\end{equation}
There is a pleasing interpretation of this result in terms of a resonant forced oscillator solution - as can be seen from the form of \eqref{EOMPerts}. The photon field $\mathcal{E} = E_{\parallel}/\omega$ can be viewed as a harmonic oscillator with local ``frequency" $k_{\gamma} = \sqrt{\omega^2 - \omega_{\rm pl}^2}$ which becomes equal to that of the axion forcing $k_{\rm a} = \sqrt{\omega^2 - m_{\rm a}^2}$ when $\omega_{\rm pl} = m_{\rm a}$. Since the particular solution to the forced resonant oscillator grows linearly with $z$ behaving as $\sim z e^{\imath k_{\gamma} z}$ and since the overall magnitude of the forcing is set by $\omega g_{\rm a \gamma \gamma} B_0$, the total resonant growth in the photon amplitude is then given by multiplying the size of the region (linear $z$ behaviour) by the magnitude of the forcing - which gives precisely the amplitude-squared of \eqref{eqn:Hook}.

\subsubsection{Landau-Zener}
It is also interesting to quote the well-known Landau-Zener expression for the conversion probability in a two-state system \cite{Brundobler} which is obtained by linearising $\Delta_{\gamma}$ in \eqref{eq:Schrodinger} about $z=z_{\rm c}$ and neglecting gradients in the mixing $\Delta_B$, leading to (see appendix \ref{Density}) 
\begin{equation}\label{eq:LZ}
 P_{\rm a\rightarrow \gamma} = 1 - e^{-2 \pi \gamma }\,, \qquad 
 \gamma = \frac{\Delta_B^2(z_{\rm c})}{\left|\Delta^{\prime}_\gamma(z_{\rm c})\right|}\,.
\end{equation}
The physical interpretation of this result is that $\gamma$ controls the \textit{adiabaticity} of the evolution - i.e., how rapidly the background is varying. Formally this corresponds to the size of background plasma gradients. We see immediately that the perturbative result \eqref{eq:gammaDef} (refs.~\cite{ref:NS-Hook,Raffelt:1987im}) is precisely the truncation of the Landau-Zener probability \eqref{eq:LZ} (\cite{ref:LaiHeyl,Kuo:1989qe}) in the non-adiabatic limit for small $\gamma$. 

It is intriguing to note the link between these results. Of course mathematically speaking, the stationary phase approximation used to compute \eqref{eq:Pintegral} amounts to a linearisation of the plasma mass about the critical point and our use of the Landau-Zener result is formally valid in the limit for which the mass-splitting $m_{\rm a}^2 - \omega_{\rm pl}^2$ varies linearly with $z$ implying the same implicit assumption. However, given that the derivation of each of these results seems a priori to be quite different - it is striking to see that their agreement is exact in the $\gamma \ll 1$ limit.

\begin{figure}
 \centering
 \includegraphics{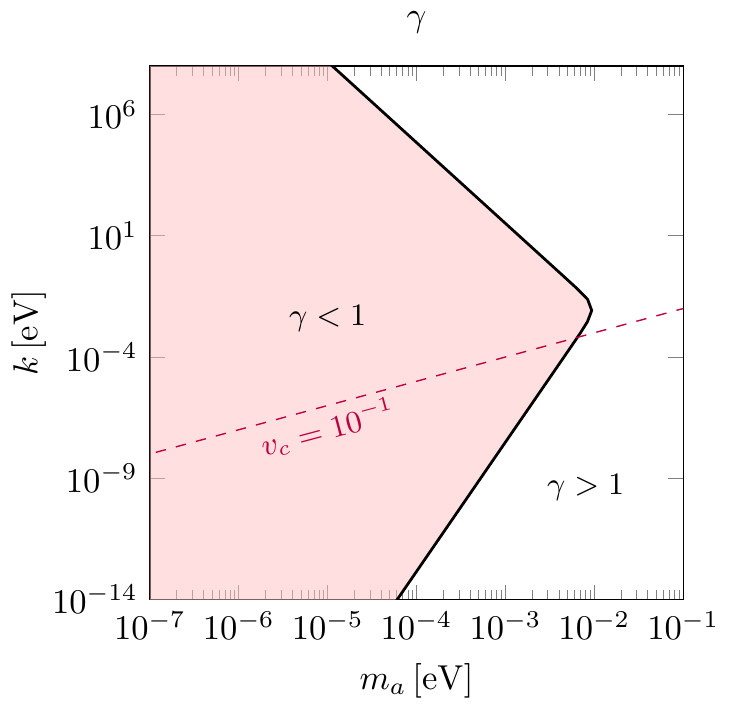}
 \caption{The adiabaticity parameter $\gamma$ of \eqref{eq:LZ} for the QCD axion with $g_{a \gamma \gamma}$ given by \eqref{eq:QCDg} with $E/N=8/3$. We considered a magnetic field \eqref{eq:B1D} with $B_{\ast}=10^{14}\text{G}$, a rotation period $P=0.1 {\rm s}$ with $R=10\,{\rm km}$. We also show the velocity at the critical point $v_{\rm c} =k_{\rm c}/m_{\rm a}$ for the value $10^{-1}$ which can be reached via gravitational acceleration.}
 \label{fig:GammaPlot}
\end{figure}

The size of $\gamma$ -- and therefore the regime in which a perturbative treatment is appropriate -- is given in fig.~\ref{fig:GammaPlot} for the QCD axion with canonical neutron star parameters. Note that our systematic treatment of mass-shell constraints allows us to study $\gamma$ across the full range of relativistic and non-relativistic axion parameter space.

\begin{figure}[t]
\centering
\includegraphics[width = 0.45\textwidth]{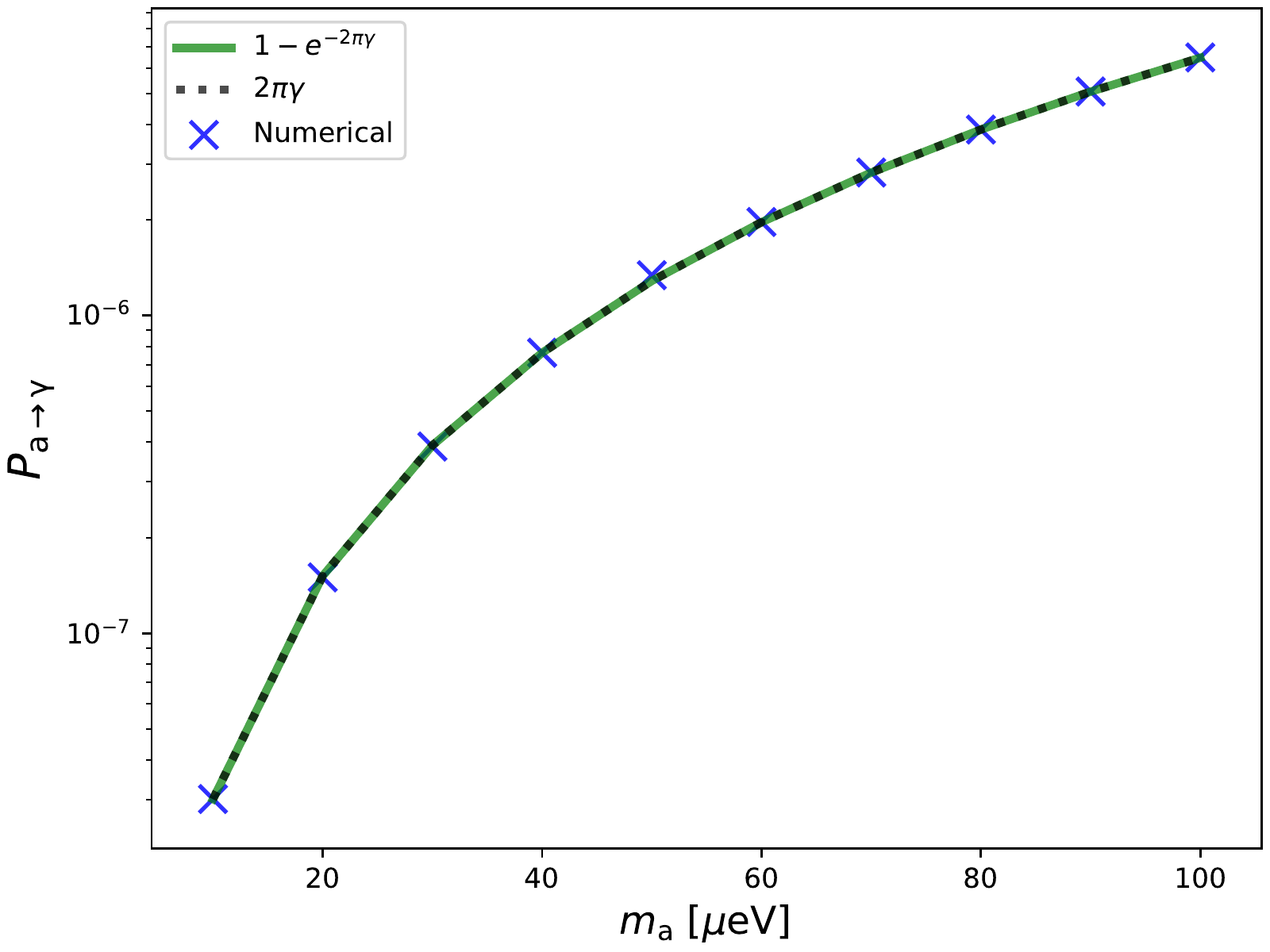}
\caption{The analytic Landau-Zener probability, adiabaticity parameter $\gamma$ and the numerical solution of the full second order equations of motion as a function of the axion mass. Here, we assume $g_{\rm a\gamma\gamma} = 10^{-12}$ GeV$^{-1}$, $B_0 = 10^{14}$ G and $k_{\rm a} = 0.1 m_{\rm a}$. For these parameters, we find from energy conservation that the average velocity of the axion at the resonant conversion region $v_{\rm c} \approx \frac{2GM}{z_{\rm c}}$ is roughly 10\% of the speed of light.}
\label{fig:massplot}
\end{figure}

Fig.~\ref{fig:massplot} summarises our results for conversion in 1D and compares the full numerical results of the second order equation \eqref{eq:2by2} against analytic approximations. The numerical conversion probability was computed by assuming an incident axion from $z \rightarrow - \infty$ with the magnetic field background $\eqref{eq:B1D}$ and solving the equations for the photon up to a finite depth inside the region of plasma overdensity defined by $\omega_{\rm pl}>\omega$ in which the photon amplitude becomes exponentially suppressed. This was implemented numerically as a Dirichlet and Neumann boundary condition by setting the electric field and its first derivative to zero at some finite depth inside the $\omega_{\rm pl} > \omega$ region. 

Figs.~\ref{fig:GammaPlot} and \ref{fig:massplot} show that the conversion of dark matter axions in neutron star magnetospheres typically involves non-adiabtic evolution for which a perturbative treatment in $g_{\rm a \gamma \gamma}$ is valid. The fact one does not stray into the adiabatic regime arises from two considerations. Firstly, for asymptotic values of the axion velocity $v_{\rm a} \equiv k_{\rm a}/m_{\rm a}$ given by $10^{-3}$, gravitational acceleration can bring these up to around $10^{-1}$ shown by the purple line in fig.~\ref{fig:GammaPlot}. Secondly there is an upper limit on the axion mass beyond which the resonance region would be pushed inside the neutron star. These two facts together restrict one to the non-adiabatic region of dark matter axions.

Of course there are some caveats to the above assumptions. Firstly axions with very high or very low momenta can in principle be pushed into the adiabatic regime. However, the gravitational acceleration of the neutron star puts a lower bound $v_{\rm c} \geq GM/z_{\rm c}$, which is saturated by axions which are asymptotically at rest. Meanwhile for large $v$, the distribution is exponentially suppressed by the velocity dispersion $v_0$. 

\subsection{Mixing in higher dimensions}
Firstly we review some of the canonical assumptions made in reducing the system of equations \eqref{axionEOM}-\eqref{EOMPerts} to a simple 1D form and then explain why these assumptions may break down in more complicated geometries.

\subsubsection{Polarisation, geometry and the longitudinal mode}\label{sec:1DLimitations}
It is common to assume a transverse photon \cite{Raffelt:1987im,Kartavtsev:2016doq}, such that $\nabla \cdot \textbf{E} = 0$. In this instance, the purely transverse field $\textbf{E}$ can be projected onto the magnetic field in such a way that the polarisation normal to $\textbf{B}_0$ decouples
\begin{equation}
\nabla \cdot \textbf{E} = 0 \quad \Rightarrow \quad
\left.
\begin{array}{c}
\! \! \! \square \, a + m_{\rm a}^2 a = g_{\rm a \gamma \gamma} E_{\parallel} B_t,\\
\square \, E_{\parallel} + \omega_{\rm pl}^2 E_\parallel = -g_{\rm a \gamma \gamma} \ddot{a} B_t
\end{array}
\right.\,,
\end{equation}
where $B_{\rm t}$ is the projection of $\textbf{B}_0$ onto the (assumed to be transversely polarised) $\textbf{E}$ field. This form is valid either for isotropic conductivities or $\omega, \omega_{\rm pl} \ll \omega_{\rm B}$ such that $\sigma\cdot\textbf{E} = \imath \omega_{\rm pl}^2/\omega E_{\parallel}$. Under such an assumption the system will reduce to the mixing of two scalar degrees of freedom, $E_{\parallel}$ and $a$. 

However, in the absence of special geometric considerations described in sec.~\ref{sec:1D} the presence of plasma and the axion itself will source a longitudinal component $\nabla \cdot \textbf{E} \neq 0$, as can be seen explicitly from Gauss' equation \eqref{Gauss}, which, by using current conservation in a stationary background together with Ohm's law reads
\begin{equation}
\nabla \cdot \textbf{E} = \nabla \cdot \left(\frac{\sigma}{\omega} \cdot \textbf{E}\right)- g_{\rm a \gamma \gamma} \textbf{B}_0 \cdot \nabla a\,.
\end{equation}
If one chooses a geometry such that the axion field has no gradients in the direction of $\textbf{B}_0$, then the longitudinal mode will not be excited by the axion. If, for instance, the axion gradients are negligible over the scale of the experiment in question, it will have no effect, and in a homogeneous background one would then have $(1-\omega_{\rm pl}^2/\omega^2)\nabla \cdot \textbf{E} = 0$, allowing to neglect the longitudinal mode. However, for the neutron star case, these simplifications need not apply so straightforwardly as we now demonstrate explicitly.

\subsubsection{2D example}\label{eq:2DExample}
Several authors have studied axion-electrodynamics in non-planar geometries \cite{Ouellet:2018nfr,Knirck:2019eug}. We also note in passing a more detailed examination of axion-plasma effects \cite{Mendonca:2019eke,Tercas:2018gxv}. 
We solve \eqref{EOMPerts} in a stationary background, working perturbatively in $g_{\rm a \gamma \gamma}$ such that the back-reaction onto the axion beam can be neglected. We, therefore, consider the axion as a fixed source, to solve for $\textbf{E}$
\begin{align}\label{eq:EStationary}
-\nabla^2 \, \textbf{E} + \nabla (\nabla \cdot \textbf{E}) + \omega_{\rm p}^2 \textbf{E} - \omega^2 \textbf{E} & = \omega^2 g_{\rm a \gamma \gamma} a \textbf{B}_0\,,
\end{align}
with $a = a_0 e^{\imath \textbf{k}_{\rm a} \cdot \textbf{x}}$. 
This form implicitly assumes that enough time has elapsed since the axions last scattering that an initially localised axion packet will have dispersed to scales much larger than the neutron star via quantum diffusion by the time it approaches the resonant region, justifying the infinite transverse extent of the axion wave-fronts in \eqref{eq:EStationary}. Such an approximation is easily justified by the low density of particles in the inter-stellar medium, and the weakness with which they couple to the axion \cite{ref:Marsh}. The wave-optics picture used above can be viewed as summing over all possible rays parallel to $\textbf{k}_{\rm a}$ which pass-through the neutron star, since by virtue of the uncertainty principle, only the axion's momentum $\textbf{k}_{\rm a}$, not its location, is known.

To solve \eqref{eq:EStationary}, we implement a finite element method solver by constructing a mesh over a given integration region. In order to resolve the wave front structures, the characteristic length of the mesh elements must be less than the wavelength $\lambda_{\rm a} = 1/k$ of the axion. Furthermore, the size of the resonance region can be written as $r_{\rm c} = v_{\rm a} \lambda_{\rm a} [r_{\rm c}/(3 \lambda_{\rm a})]^{1/2}$ where $v_{\rm a}$ is defined by $v_{\rm a} = k/m_{\rm a}$. Since the critical radius $r_{\rm c}\gg \lambda_{\rm a}$ is set by neutron star scales, we see that $\Delta r_{\rm c} \gg \lambda_{\rm a}$. In other words, the effective haloscope size is many orders of magnitude larger than the axion wavelength. This should be contrasted with the results of \cite{Knirck:2019eug}, where axion wavelengths are comparable to the  size of the experiment. This hierarchy presents a numerical challenge in that one must integrate over many wavelengths along the conversion region, with a sufficiently high resolution over each wavelength, resulting in a large number of mesh cells. The situation is clearly exacerbated in higher dimensions, where even more cells will be required. 

Therefore, we consider a 2D setup, which allows one to study the mixing process in non-planar geometries, whilst keeping the computational cost low. We consider the following ``2D dipole" magnetic field:
\begin{align}
 & \textbf{B}_0 = \nabla \times \textbf{A}\,, \quad 
   \textbf{A} = f(r,\theta)\hat{\textbf{z}}\,, \label{eq:Bfield2D}\\
 & f(r,\theta) = e^{\imath m\theta}B_{\ast}\left(\frac{R}{r}\right)^m\,, \qquad 
   R \leq r < \infty\,,\nonumber
\end{align}
which satisfies $\nabla \cdot \textbf{B}_0 = 0$ automatically and $\nabla \times \textbf{B}_0 = 0$ by virtue of $f(\theta,\varphi)$ being a solution to the cylindrical Laplace equation. We take $m=1$ to mimic a dipole-like configuration explicitly, then
\begin{equation}\label{eq:Bcar}
\textbf{B}_0(x,y) = B_{\ast} R \left \{\frac{-2xy}{(x^2+y^2)^{2}}, \frac{x^2-y^2}{(x^2+y^2)^{2}} \right\}\,.
\end{equation}
The conductivity is taken in the high-magnetisation limit $\omega_{\rm B} \gg \omega$, and mimics the Goldreich-Julian density of an aligned rotator by projecting on the direction of the magnetic dipole:
\begin{equation}\label{eq:plasma2D}
\sigma = \frac{\imath \omega_{\rm pl}^2}{\omega} \left(\hat{\textbf{B}}_0 \otimes \hat{\textbf{B}}_0\right)\,, \qquad 
n_{\rm GJ}^{\rm 2D} = \Omega(\hat{\textbf{y}}\cdot \textbf{B}_0)\,.
\end{equation}

\begin{figure*}
\centering
\includegraphics[width = 0.75\textwidth]{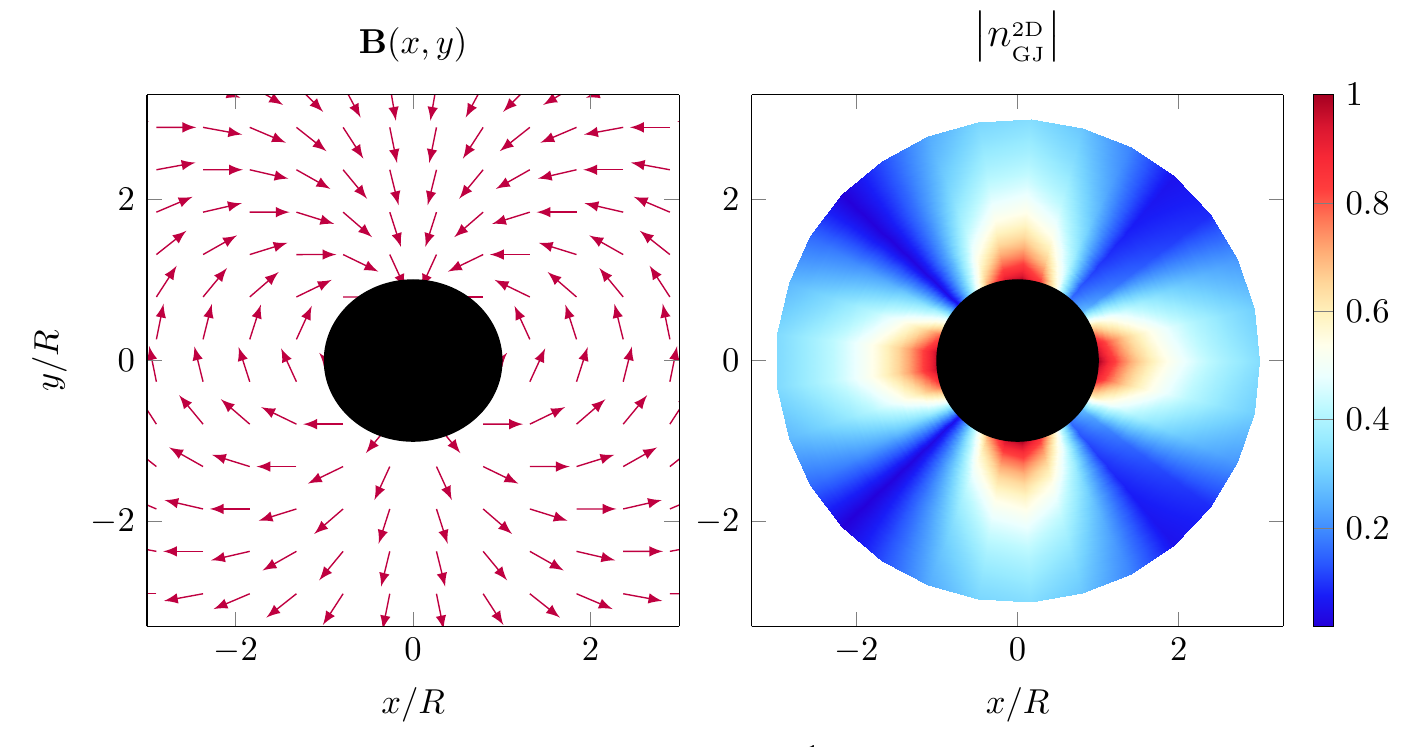}
\caption{Magnetic field and 2D Goldreich-Julian density of eqs.~\eqref{eq:Bfield2D}-\eqref{eq:plasma2D} normalised to surface values.}
\label{fig:BandGJ}
\end{figure*}
\begin{figure*}
 \centering
 \includegraphics[width = 0.75\textwidth]{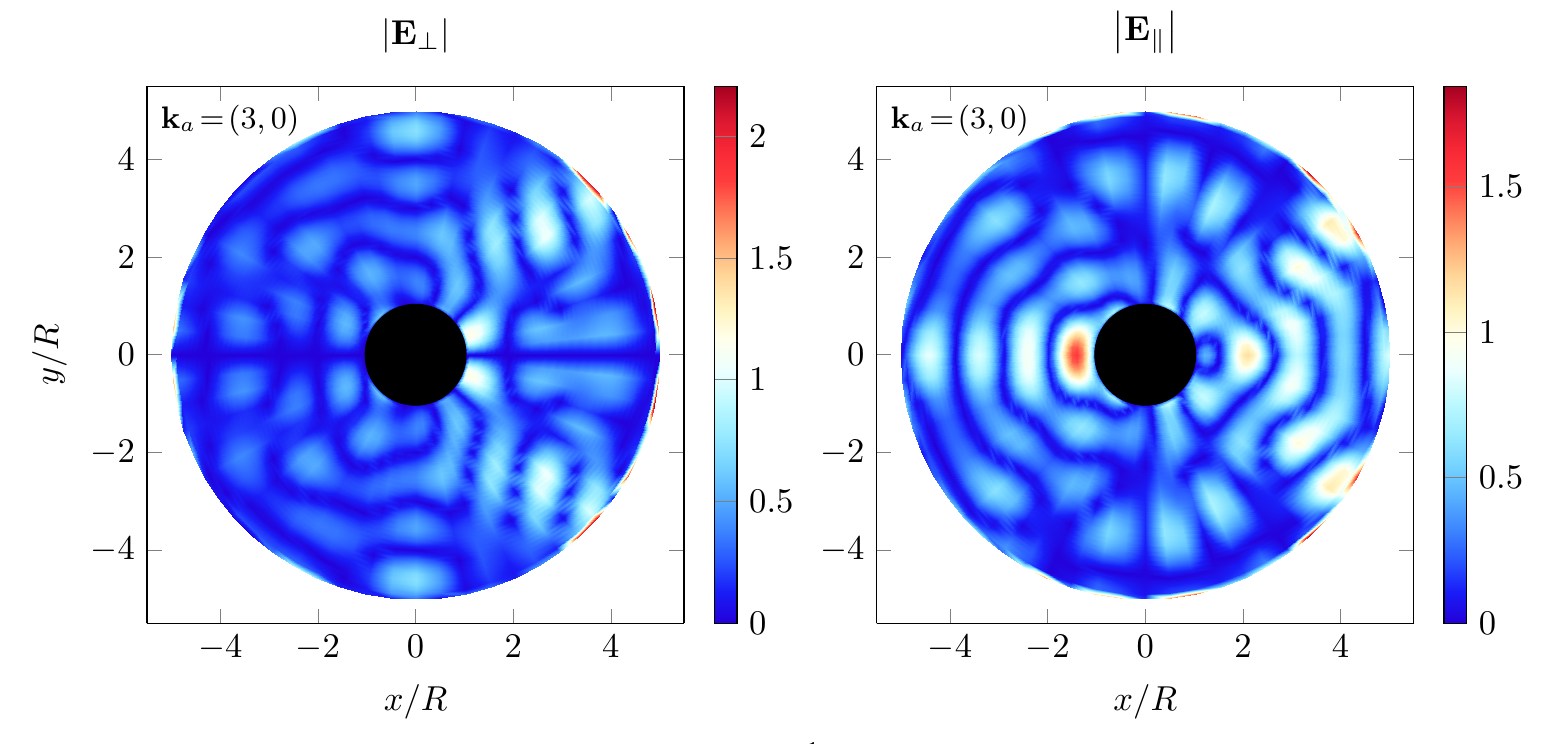}
 \caption{The electric fields components perpendicular and normal to $\textbf{B}_0$ from solving \eqref{eq:EStationary} (normalised to $B_{\ast} a_0g_{\rm a \gamma \gamma}$) with the profiles in \eqref{eq:plasma2D} and the boundary conditions 
 \eqref{eq:BCRadiation}-\eqref{eq:BCPerfectConductor}. We took the values $\textbf{k} = (0,3)$ and $m_{\rm a} = 1$ in units of $R^{-1}$.}
 \label{fig:Efields}
\end{figure*}

\begin{figure}[t]
 \centering
 \includegraphics[scale=1]{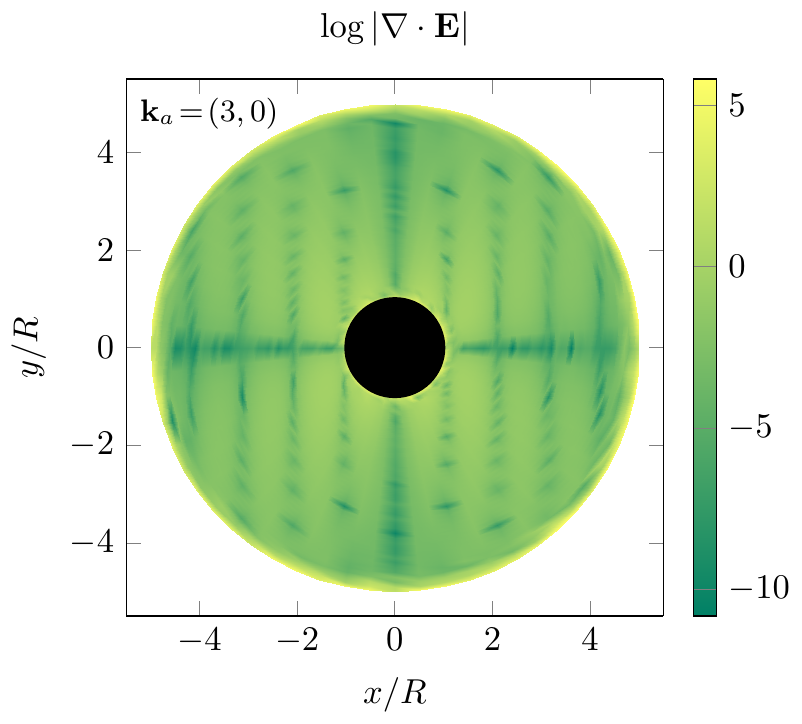}
 \caption{The divergence of the electric field (arbitrary units) with values as in fig.~\ref{fig:Efields}.}
 \label{fig:DivE}
\end{figure}

The resulting background configurations are shown in fig.~\ref{fig:BandGJ}. We implement the following boundary conditions which are correspondingly of Robin and Dirichlet type
\begin{align}
 & \textbf{n} \times \nabla \times \textbf{E} - \imath \omega \textbf{n} \times (\textbf{n} \times \textbf{E}) = 0\,, &
 r \rightarrow \infty\,, \label{eq:BCRadiation} \\
 & \textbf{E} = 0\,, & r = R\,, \label{eq:BCPerfectConductor}
\end{align}
where $\textbf{n}$ is the unit normal to the outer boundary of the integration region. The first of these implements purely outgoing waves so that the outer boundary is absorptive. It can be derived by considering asymptotic solutions of the vector Helmholtz equation \cite{Peterson,Knirck:2019eug}. The second condition assumes a perfect conductor at $r < R$ so that $\textbf{E}$ vanishes for $r \leq R$, with no surface charges at $r=R$ such that the electric field continuity conditions $\textbf{n} \cdot \left[\textbf{E}(r \rightarrow R^+) - \textbf{E} (r \rightarrow R^-) \right]= 0$ and $\textbf{n} \times \left[\textbf{E}(r \rightarrow R^+) - \textbf{E} (r \rightarrow R^-) \right] =0$ on the inner boundary at $r=R$. The results are shown in figs.~\ref{fig:Efields} and \ref{fig:DivE}.

It is clear to see that in those regions where $\nabla a \cdot \textbf{B}_0 \sim \textbf{k}_{\rm a} \cdot \textbf{B}_0 \neq 0$ one has $\nabla \cdot \textbf{E} \neq 0$ whose profile tracks those axion wave-fronts parallel to $\textbf{B}_0$. Since the decoupling procedure of the different polarisations breaks down in a non-planar geometry, we also see in fig.~\ref{fig:Efields} that $\textbf{E}_{\perp}$ enters in the mixing equations and becomes dynamical. 

In general then, we see that a simple decoupling of polarisations need not hold in non-planar geometries, suggesting that the 2-component mixing equations applied to a neutron star context in \cite{ref:NS-Hook} are at best an order of magnitude approximation. That said, since the sourcing of $\nabla \cdot \textbf{E}$ arises via axion gradients, it may be that for sufficiently non-relativistic axions these terms could be neglected in a controlled way, however such an analysis is beyond the scope of the present work. In addition, even if one can decouple polarisations, one still has to contend with multi-directional gradients, such that the 1D Landau-Zener formulae would need to be adapted to a 3D setting. 

Ultimately, it may be that accurate results can only be obtained by full 3D simulations of the mixing equations as in \cite{Knirck:2019eug}. However, as discussed in previous paragraphs, resolving the wave-front structure across the resonance region requires a large number of mesh cells. One remedy could be a coarse-graining procedure in which one tracks the field amplitudes, but integrates out structures below wavelength scales. This is akin to a gradient expansion used to derive \eqref{eq:Schrodinger} in appendix~\ref{Density} and would entail performing the same expansion in 3D on \eqref{eq:EStationary} to derive a three dimensional set of transport equations similar to those of \cite{Prokopec:2003pj,Stirner:2018ojk} encountered in flavour mixing in leptogenesis or neutrino oscillations in supernovae.

\section{Estimating the Signal and Radio Sensitivity for the Resonant Decay}\label{sect:resdecay}
 The radio sensitivity to resonant conversion in neutron star magnetospheres has been previously discussed in \cite{ref:NS-Hook, Safdi:2018oeu}. In particular, \cite{Safdi:2018oeu} discussed the radio sensitivity to neutron star populations. The conclusion of their study was that the radio lines from the individual ``brightest" neutron stars (where bright here means where the resonant conversion is the strongest) offer better sensitivity to the axion-photon coupling than a population. An important factor contributing to this is that the frequency width of the signal in the case of observing a population of stars is proportional to the inverse of the velocity dispersion, compared to the inverse square of the velocity dispersion in the case of single neutron stars, which increases integration time considerably. Therefore, one needs to increase the field of view of the observation considerably to observe a large enough population in order to get a larger signal compared to the isolated bright neutron star case. Unfortunately, one is then limited by the fact that telescopes with large collecting areas have higher resolution, which lead to smaller fields of view. We remark that it might be possible to design a bespoke instrument optimised to try and maximise this signal, but such an undertaking is beyond the scope of this current work. 
 
In this section, we work out the flux density associated with the resonant mixing and explore the possibility of a detection with current and future telescopes. We discuss the impact of the velocity dispersion near a neutron star and Doppler broadening due to its overall motion in section \ref{subsec:ResSensVelocity}. We then discuss the single-dish sensitivity to the axion-photon decay in neutron stars and describe potential neutron star targets in section \ref{subsec:ResSensHook}, where we also compare and contrast our sensitivity calculations to that of \cite{ref:NS-Hook}. Our sensitivity estimates are for the resonant production in a single neutron star, where we assume that the signal is a spectral line broadened by the velocity dispersion of the axions. We also forecast sensitivities of single-dish telescopes (for which we assume the GBT or the Arecibo telescope to be typical examples) and interferometers, like the the SKA.

\subsection{Velocity Dispersion and Doppler Broadening}\label{subsec:ResSensVelocity}
Since the pulsar magnetosphere in general is not a stationary configuration, the energy of test particles moving in this background is not conserved.
While a somewhat rich structure is indicated by simulations~\cite{Philippov:2014mqa,Kalapotharakos:2017bpx}, we consider here the minimal model of an oblique rotating magnetic dipole field that also determines the electron density according to Goldreich and Julian~\cite{ref:GJ} and hence the critical surface. In order to arrive at an analytically transparent picture, we make some additional simplifying assumptions.

For an oblique rotator, the intersection of a plane perpendicular to the rotation axis with the critical surface takes the shape of an ellipse. 
When the lengths of the semi-major and semi-minor axes are $a$ and $b$, respectively, the numerical eccentricity is $\varepsilon=\sqrt{1-b^2/a^2}$. This ellipse rotates at an angular velocity $\Omega$ about its middle point.

Consider a corotating point on the critical surface, where an axion may be converted into a photon.
We can further distinguish the cases of reflection and transmission. For reflection, an infalling axion reaches the critical surface from the outside and the photon is subsequently reflected when further climbing the potential barrier made up by the plasma. For transmission, the axion is coming from the inside region and the photon then continues to travel outbound.  The instantaneous velocity of
the tangential plane of the critical surface in general is not parallel to the plane itself (unless the point considered is aligned with one of the axes of the ellipse or in the degenerate case of a circle). Physically, a particle that interacts with the critical surface transfers momentum to the magnetosphere, corresponding to the Doppler effect from the reflection by a moving mirror\footnote{We thank Georg Raffelt for bringing this issue to our attention.}.

\begin{figure}[t]
\begin{center}
\includegraphics[scale=0.35]{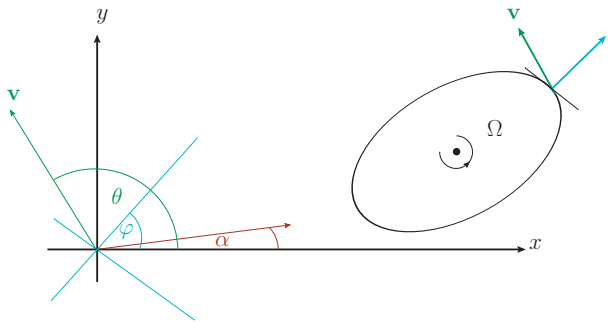}
\end{center}
\caption{
\label{fig:doppler}
Parametrization of the Doppler shift on a moving, misaligned mirror.}
\end{figure}

We therefore calculate the reflection or transmission of a ray of a particle of mass $m_{\rm a}$ in the $xy$-plane that approaches the origin at an angle $\alpha$ (all angles refer here to the $x$-axis), where it falls on a plane whose normal vector points in the direction $\varphi$.
Upon reflection or transmission, the particle is converted into a massless state.
The plane moves at a constant velocity $v$ in the direction of the angle $\vartheta$, see fig.~\ref{fig:doppler}.
The calculation can be carried out by first boosting the four-momentum of the massive initial state from the rest frame of the observer to the rest frame
of the critical surface. In that frame, the zero-component of the four-momentum is conserved as well as
the spatial components of the momentum parallel to the surface. The component perpendicular to the
surface is then found by imposing the energy-momentum relation of a massless particle. The final answer
is obtained when boosting back to the frame of the observer.

To clarify this approach, we first quote the result for the situation where the surface moves toward
the incoming massive particle, $\alpha=\vartheta=\varphi=0$, such that we obtain
\begin{align}
k^{0\prime}=\frac{c \sqrt{m_{\rm a}^2+k^2}+k v}{c\mp v}\,,
\end{align}
where $k$ and $k^{\prime}$ are the moduli of the wave vectors of incoming and reflected wave, respectively.
Throughout this section, an upper sign refers to the case of reflection and a lower one to transmission.
Clearly, when setting $m_{a}=0$, we obtain the classic result for Doppler shift
for reflection as well as zero change in the frequency for transmission. We may therefore anticipate
that for non-relativistic axions, the Doppler shift for axions leaving the magnetosphere is not suppressed
compared to infalling axions.

To arrive at a conservative estimate of the Doppler broadening in the magnetosphere, we now assume that the shape is only mildly elliptical such that the misalignment angle $\varphi-\vartheta+\pi/2\ll 1$ between the tangential plane and its velocity corresponds to a small parameter that we can expand in.
The Doppler shift then takes the simple form
\begin{align}
\frac{k^{\prime 0}}{k^0} = &\, 1\mp \frac{v}{c}\frac{\sqrt{k^2+m_{\rm a}^2-k^2\cos^2(\alpha-\vartheta)}\pm k\sin(\alpha-\vartheta)}{\sqrt{k^2+m_{\rm a}^2}\sqrt{1-\frac{v^2}{c^2}}}\times \nonumber\\
&\, \left[2(\varphi-\vartheta+\pi/2) + {\cal O}\left[(\varphi-\vartheta+\pi/2)^2\right]\right]\,,
\end{align}

In the limit of a relativistic incident particle, $m_{\rm a}/k\to 0$, this reduces to
\begin{align}
\frac{k^{\prime 0}}{k^0} = &\, 1-\frac{v}{c}\frac{\sin(\alpha-\vartheta)}{\sqrt{1-\frac{v^2}{c^2}}}
2(\varphi-\vartheta+\pi/2) + \\
&\, {\cal O}\left[(\varphi-\vartheta+\pi/2)^2\right]\,,\nonumber
\end{align}
for reflections and $k^{\prime 0}/k^0\approx 1$ for transmissions.
In the opposite limit, $m_{\rm a}\gg k$, we find
\begin{align}
 \frac{k^{\prime 0}}{k^0} = &\, 1\mp\frac{v}{c}\frac{1}{\sqrt{1-\frac{v^2}{c^2}}}
2(\varphi-\vartheta+\pi/2) + \\
&\, {\cal O}\left[(\varphi-\vartheta+\pi/2)^2\right]\,,\nonumber
\end{align}
which is the expression useful for the present context.

In order to estimate the average effect for the conversion in the magnetosphere, we note that, for a given eccentricity, the angle $\varphi-\vartheta+\pi/2$ can assume values between $\pm\varepsilon^2/4$ within one rotation . Furthermore, depending on the impact parameter, the angle $\alpha-\theta$ approximately takes values between $-\pi/2$ (for trajectories that come very close to the core of the pulsar) and $\pi$ (for trajectories that just about touch the critical surface on the far side of the pulsar). A full quantitative analysis involving the axion and photon trajectories should be straightforward, but it is probably not of obvious benefit since the oblique rotator model of the magnetosphere is likely to be oversimplified, and hence we just make an estimate of the size of the effect. Assuming further $v/c\ll 1$, we estimate that
\begin{equation}\label{doppler:estimate}
 \left\langle\left|\frac{k^{\prime0}}{k^0}-1\right|\right\rangle\sim\frac{\Omega r_{\rm c} \varepsilon^2}{c}\approx 6\times 10^{-4} \left(\frac{\Omega}{1\,{\rm Hz}}\right) \left(\frac{r_{\rm c}}{200\,\rm km}\right)\varepsilon^2\,.
\end{equation}
This is to be compared with the width from the velocity dispersion of the axion dark matter
\begin{equation}
\frac{1}{2} v_0^2/c^2\approx 8\times 10^{-7} \left(\frac{v_0}{100\,{\rm km}{\rm s}^{-1}}\right)^2\,.
 \end{equation}
We see that the impact of Doppler broadening depends very strongly on the axion velocity in the resonant conversion region.
When the axion is non-relativistic, the Doppler broadening dominates in the width of the spectral line over the velocity dispersion.
For axions that are  relativistic at the point of conversion, there is the interesting possibility that the Doppler broadening
for transmissions is strongly suppressed, which may be of importance for line searches. For our subsequent estimates, we use the
the non-relativistic expression~(\ref{doppler:estimate}) for the Doppler broadening.
 
As stated above, the oblique rotator model with the electron density proposed
by Goldreich and Julian is chosen here because it is analytically tractable.
Eventually, it should be replaced with a more realistic model of the magnetosphere.
Even for the Goldreich--Julian model, we have made simplifying assumptions that we now comment on. 

First, the conversion from
the axion to the photon takes place during some finite time during which the
location $z_{\rm c}$ of the critical surface, where the conversion takes place, changes its position due to acceleration.
The width (\ref{eq:deltazc}) of the surface in which
the conversion occurs can be estimated as $\Delta z_{\rm c}\sim (z_{\rm c} m_{\rm a})^{1/2} v_{\rm c}/c$. Assuming that the converting axion passes through this region at a speed $v_{\rm c}/c\sim0.1$ (created in the gravitational potential of the neutron star), it is clear that this takes a time much smaller than the rotation time, $2 \pi/\Omega$, for axions in the GHz mass range. 

Second, more important are corrections that should arise from the fact that
the outgoing photon can only be considered relativistic when the Lorentz factor
$\gamma\approx m_{\rm a}/\omega_{\rm pl}$ is large, which occurs for large $(z/z_{\rm c})^{3/2}$. Integration of time along the photon trajectory implies that the point
$z$ is reached after the time $z/c+{\cal O}(z_{\rm c}/c)$. If during that time the
background plasma changes significantly because of the rotation of the pulsar, one
should anticipate order-one corrections to the Doppler effect. 

Finally, due to the curvature of the contours of equal plasma mass and the finite distance to be traversed before it becomes relativistic, there should be corrections
due to the continuous refraction of the escaping photon. For axions traversing the 
critical surface at a small angle, these can also be of order one.

In case there is additional structure in the magnetosphere beyond the Goldreich-Julian model, then the estimate~(\ref{doppler:estimate}) should be considered as conservative
when applied within its range of validity, which is $\varepsilon\ll 1$.
This is because for structures in the magnetosphere that are indicated by simulations,
the critical surfaces appear to move at large velocities $\sim \Omega r_{\rm c}$.
It would, therefore, be desirable to numerically compute the broadening for realistic magnetosphere models on a full statistical average of axion trajectories and, if possible, to devise of methods of correcting for the Doppler effect.
We stress that since the estimate in \eqref{doppler:estimate} is significantly larger than the background velocity dispersion, the amplitude of the radio signal will be weaker, as we will show in the subsequent sections of the paper. 

\subsection{Single-Dish Sensitivity to Resonant Conversion}\label{subsec:ResSensHook}
From \eqref{eqn:BolometricFlux}, the flux density reads
\begin{equation}\label{eqn:FluxNS}
 S = \frac{c^2}{4\pi\Delta f_{\rm obs}r(z)^2}\int \frac{\rho_{\rm a}}{\tau_{\rm obs}}\mathrm{d}V \,,
\end{equation}
where we have set the total energy from the decay to be equal to the volume integral of the axion density, i.e., $N_{\rm a}E_{\rm obs} = \int \rho_{\rm a}c^2\mathrm{d}V$. Since we are interested in the flux due to the resonant axion-photon decay at a distance $z_{\rm c}$ over a thin shell of width $\Delta z_{\rm c}$, we have that
\begin{equation} 
 \int \rho_{\rm a} \mathrm{d}V = \int \rho_{\rm c} z_{\rm c}^2\Delta z_{\rm c} \mathrm{d}\Omega \,,
\end{equation} 
where $\rho_{\rm c}$ is the density of the axions in the resonant conversion region. If $v_{\rm c}$ is the velocity of the axions at $z_{\rm c}$, there is then a characteristic time scale over which the axions traverse the width of the shell, $T_{\rm c} = \Delta z_{\rm c}/v_{\rm c}$. Substituting these expressions into \eqref{eqn:FluxNS}, the flux density can be expressed as
\begin{equation}\label{eqn:STc}
 S = \frac{c^2}{4\pi \Delta f_{\rm obs}r(z)^2}\int \rho_{\rm c}z_{\rm c}^2 v_{\rm c}\frac{T_{\rm c}}{\tau_{\rm obs}}\mathrm{d}\Omega \,,
\end{equation}
where we identify $T_{\rm c}/\tau_{\rm obs} \equiv P_{\rm a\rightarrow\gamma}$. Therefore, we have a pleasing interpretation of the probability of conversion as the ratio of the resonant crossing time $T_{\rm c}$ to the decay time $\tau_{\rm obs}$, which means that when the two timescales are equal, the probability of conversion becomes unity. This implies that the integral in eq.~(\ref{eqn:STc}) is equivalent to a specific intensity integrated over the area of the source associated with this decay, $\int I \mathrm{d}A_{\rm source} \approx 4\pi z_{\rm c}^2\,I$, which is consistent with previous work, where this quantity was viewed as the power radiated by the flux of photons sweeping across the resonance shell at a velocity $v_{\rm c}$ \cite{ref:NS-Hook}. The estimated decay time is, therefore,
\begin{equation}\label{eqn:decaytime}
 \tau_{\rm obs} = T_{\rm c}/p_{\rm a\rightarrow \gamma} = \frac{\Delta z_{\rm c}/v_{\rm c}}{p_{\rm a\rightarrow\gamma}} \approx 190~{\rm s} \,,  
\end{equation}
assuming that $p_{\rm a\rightarrow \gamma} \approx 10^{-8}$. For comparison, the axion decay time derived by \cite{ref:Sigl} is given by 
\begin{align}
 \tau_{\rm obs}^{\rm Sigl} \approx & 2\times 10^4~{\rm s} \left(\frac{B_0}{10^{14}~\rm G}\right)^{-2}\left(\frac{g_{\rm a\gamma\gamma}}{10^{-12}~\rm GeV^{-1}}\right)^{-2}\times \nonumber\\
 & \left(\frac{r_0}{10~\rm km}\right)^{-3}\left(\frac{z_{\rm c}}{200~\rm km}\right)^{3}\,.
\end{align}
This expression has been derived directly from the rate of conversion of axions to photons in an astrophysical magnetic field using a non-resonant perturbative calculation and it is two orders of magnitude larger than the decay time in (\ref{eqn:decaytime}).

With current, and realistically possible, telescopes, it is impossible to resolve objects on the scales of $z_{\rm c}$. Therefore, we assume that the neutron star is a point source and hence, in contrast the resolved sources discussed in section~\ref{sect:decay} is better to talk in terms of the flux density rather than the brightness temperature. To determine whether it is possible to detect this conversion, we estimate the flux density 
\begin{equation}
 S \approx \frac{c^2}{\Delta f_{\rm obs}} \frac{\rho_{\rm c} z_{\rm c}^2 v_{\rm c}}{4\pi r(z)^2} \,,
\end{equation}
where the total flux is given by integrating the specific intensity over the solid angle subtended by the source, $S = \int I \mathrm{d}\Omega$, We note that in \cite{ref:NS-Hook}, it was assumed that $\Delta f \approx m_{\rm a}v_0^2/c^2$. We have shown that the broadening of the signal is dominated by the relative motion of the critical surface with respect to the observer (see sect.~\ref{subsec:ResSensVelocity}). Therefore,  using \eqref{doppler:estimate} we can deduce that
\begin{equation}
\begin{split}
 \Delta f_{\rm obs} = 7 \,{\rm MHz}&\left(\frac{\Omega}{1\,\rm Hz}\right)^{4/3}\left(\frac{m_{\rm a}c^2}{6.6\,{\rm \mu eV}}\right)^{1/3} \\
 &\left(\frac{B_0}{10^{14}\,{\rm G}}\right)^{1/3}\epsilon^2 \,.
 \end{split}
\end{equation} 
In the subsequent projections we will use $\epsilon^2=1$ and 0.1 as spanning the likely range of values for this geometrical factor.

 If we now define the dimensionless quantities $\tilde{\rho} = \rho_{\rm c}/\rho_0$ and $\tilde{v} = v_{\rm c}/v_0$, where $\rho_0$ and $v_0$ are the density and velocity of the axions in the neighbourhood of the neutron star, we can write $S = \tilde{S}\tilde{v}\tilde{\rho}$, where $\tilde{S}$ is a characteristic flux density given by 
\begin{align}
  \tilde{S} = &\, \frac{c^2}{\Delta f_{\rm obs}}\frac{\rho_0 v_0 z_{\rm c}^2}{4\pi r(z)^2}\frac{T_{\rm c}}{\tau_{\rm obs}} \,, \\
   = &\, 1.6\,{\rm \mu Jy}\left(\frac{\rho_0}{\rm GeV~cm^{-3}}\right)
   \left(\frac{r(z)}{300~ \rm pc}\right)^{-2}\times 
   \nonumber\\
   &\, \left(\frac{P_{\rm a\rightarrow\gamma}}{10^{-8}}\right) \left(\frac{z_{\rm c}}{224~\rm km}\right)^{-3}\left(\frac{\Delta f_{\rm obs}}{7\,\rm MHz}\right)^{-1}\left(\frac{m_{\rm a}c^2}{6.6\,{\rm \mu eV}}\right)^{-1}\, \nonumber \\
   =&\, 1.6\,{\rm \mu Jy}\left(\frac{\rho_0}{\rm GeV~cm^{-3}}\right) \left(\frac{\Omega}{1\rm Hz}\right)^{-7/3}\times\nonumber\\
   &\, \left(\frac{m_{\rm a}c^2}{6.6\,{\rm \mu eV}}\right)^{2/3}
   \left(\frac{B_0}{10^{14}\,{\rm G}}\right)^{2/3}\left(\frac{r(z)}{300~ \rm pc}\right)^{-2}\times\nonumber\\
   &\, \left(\frac{g_{\rm a\gamma\gamma}}{10^{-12}\,{\rm GeV^{-1}}}\right)^2\,.\nonumber 
\end{align}

The velocity of the axions near the neutron star, $v_{\rm c}$, can be estimated in terms of the dark matter virial velocity and the neutron star mass from energy conservation, i.e., the kinetic energy of the axion far away from the neutron star must be equal to its total energy near the neutron star. According to this argument, one finds that \cite{ref:NS-Hook}
\begin{equation}
 v_{\rm c}^2 = v_0^2 + \frac{2GM}{z_{\rm c}} \approx \frac{2GM}{z_{\rm c}} \,,
\end{equation}
since the escape velocity from the neutron star is much larger than the background virial velocity, $v_0$. It is easy to see that $v_{\rm c}$ is roughly 10\% of the speed of light and therefore $v_{\rm c} \gg v_0$. This implies that the axion velocity in the conversion region is non-relativistic suggesting that the width of the spectral line is likely to be dominated by the Doppler broadening effect. A direct consequence of this is the signal being enhanced by 2 orders of magnitude, since $\tilde{v} \approx150$.

In \cite{ref:NS-Hook}, the authors estimate the dark matter density at $z_{\rm c}$ from Liouville's theorem for the distribution function in the phase-space. Assuming a time-independent Maxwell-Boltzmann distribution function $f(\textbf{v})$ for the dark matter velocity, one may obtain an expression for $\rho_{\rm c}$ integrating the distribution function by expanding in the small parameter $v_0^2/v_{\rm c}^2 = \tilde{v}^{-2} \ll 1$. The result is that the density at the resonant conversion region is enhanced by a factor $\tilde{v}$, which means a further 2 orders of magnitude increase in the flux. Under these assumptions, we obtain a flux of about $0.04\,\mu$Jy. We note that the integration time required to detect this flux using the a GBT-like instrument assuming a bandwidth of about 7~MHz is $\approx 640$ years. If one were to consider the Arecibo telescope instead, the collecting area being approximately a factor of 9 larger, the total integration times decreases by a factor of $\approx 80$. Clearly, this decrease cannot make this signal detectable. 

However, we note that the flux increases with axion mass. If we assume that $z_{\rm c}$ scales as $m_{\rm a}^{-2/3}$ [see (\ref{eq:deltazc})], then the resonance occurs closer to the neutron star and the magnetic field at the resonant conversion region will be stronger. However, the resonant shell has a smaller radius, which means the density at $z_{\rm c}$ is integrated over a smaller volume for larger masses. The increase in the magnetic field dominates over the decrease in volume for a dipole magnetic field that scales as $1/z^3$. An upper mass limit exists due to the condition that $z_{\rm c} \geq R^{\ast}$. This hard limit obviously varies for different neutron stars. We note the subtlety that while low-period neutron stars are preferred since $S\propto \Omega^{-7/3}$, the larger the period of the neutron star, the smaller the range of masses one can probe in a radio observation. Therefore, we conclude that the period of the neutron star is perhaps not the best parameter to optimise an experiment for.  

\subsubsection*{Potential Neutron Star Targets}
Our analysis until now has suggested that the decay due to neutron stars cannot be detected at the level of $g_{\rm a\gamma\gamma} \lesssim 10^{-12}\, {\rm GeV^{-1}}$, as claimed in \cite{ref:NS-Hook}. Using eq.~\eqref{Eqn:Radiometer}, we estimate the integration time for an axion mass of around 82.5 $\mu$eV/$c^2$, assuming the collecting area of the Arecibo telescope, to be

\begin{equation}
\begin{split}
    t_{\rm int} = 50\,{\rm days}& \left(\frac{\tilde{S}}{8.5 \times 10^{-6}\,{\rm \mu Jy}}\right)^{-2}\left(\frac{\tilde{\rho}}{150}\right)^{-2}\left(\frac{\tilde{v}}{150}\right)^{-2}\times \\ 
    &\left(\frac{\Delta f_{\rm obs}}{17 \,{\rm MHz}}\right)^{-1}\left(\frac{A_{\rm eff}}{50000\,{\rm m^2}}\right)^{-2}\left(\frac{T_{\rm sys}}{30\,{\rm K}}\right)^2 \,.
\end{split}
\end{equation}
Note that this value of the axion mass corresponds to about 10 GHz, which is the largest frequency the Arecibo telescope can presently operate at. With the larger frequency coverage of the GBT, one may probe axion masses in the range 1 - 825 $\mu$eV$c^2$. For the fiducial mass we used in sect.~\ref{sect:decay} of 250 $\mu$eV$c^2$, we obtain an integration time of around 3.5 years with a GBT-like instrument. Clearly, to probe larger axion masses with current telescopes, one would have to design an optimisation procedure that might alleviate the difficulties to some extent.

Assuming that $\tilde{v}$ is set by the mass of the neutron star via its gravitational potential and is therefore fixed, one would require a larger $\tilde{\rho}$ to enhance the signal enough for detection. In other words, a simple optimisation procedure would be to look for low-period neutron stars in regions where $\rho_{\rm DM}$ is several orders of magnitude larger than the background value. Another scheme of detection could be to target neutron stars of the largest magnetic fields, like magnetars which are associated to magnetic fields of up to $10^{15}\,{\rm G}$ \citep{Mori2013,Kennea2013,Shannon2013,Eatough2013}. As mentioned in \cite{ref:NS-Hook}, such a candidate magnetar exists near the density spike due to the black hole Sagittarius $\rm A^{\ast}$ at the galactic centre (see sect.~\ref{sect:decay}), for which $\tilde{\rho}$ could be $\approx 10^9$. Of course, with the advent of the SKA2 interferometer with a collecting area of $10^6\,{\rm m^2}$, one could think of probing down to model sensitivities for KSVZ and DFSZ axions. 

\begin{figure*}
 \centering
 \includegraphics[width = 0.49\textwidth]{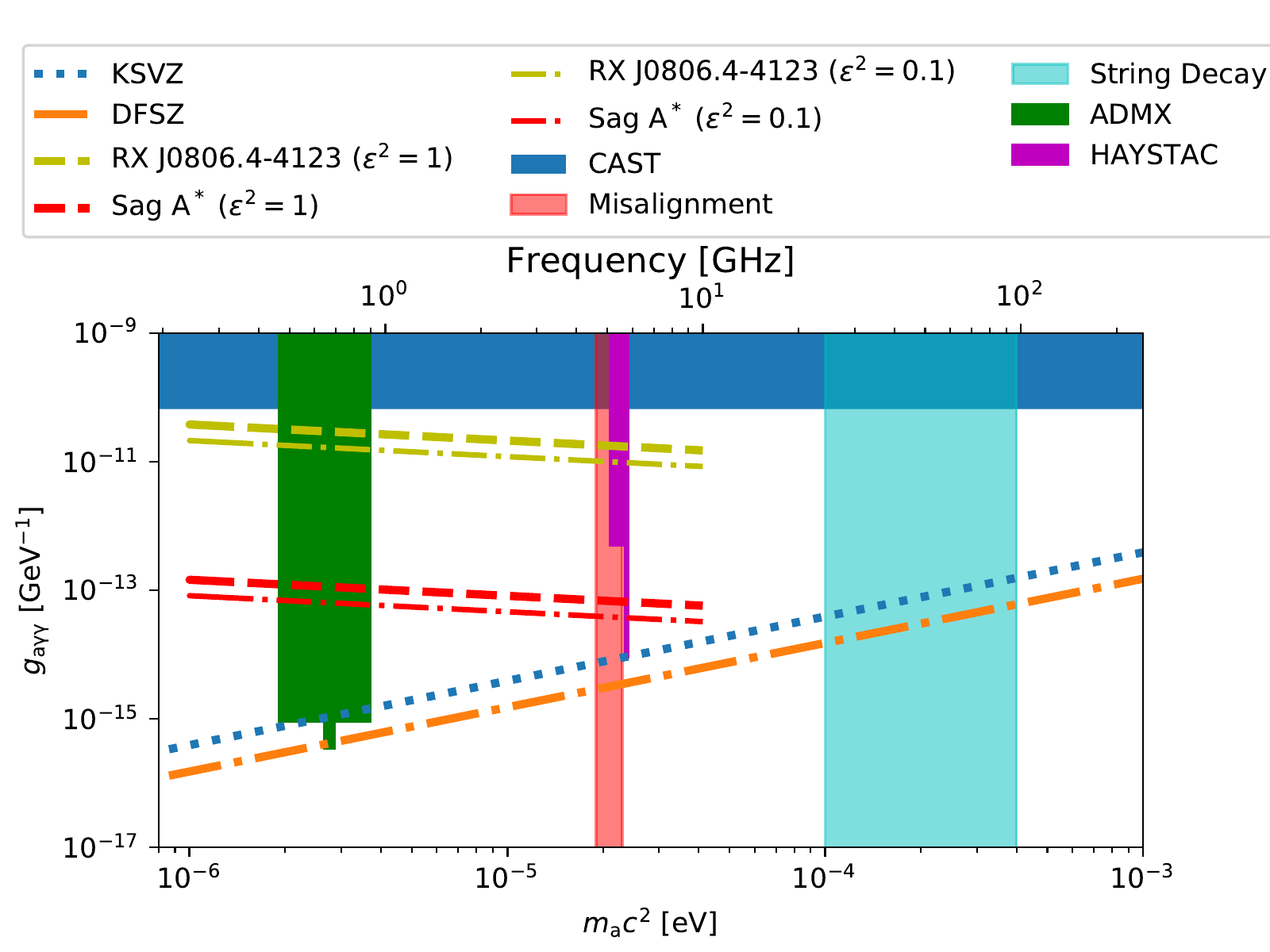}
 \includegraphics[width = 0.49\textwidth]{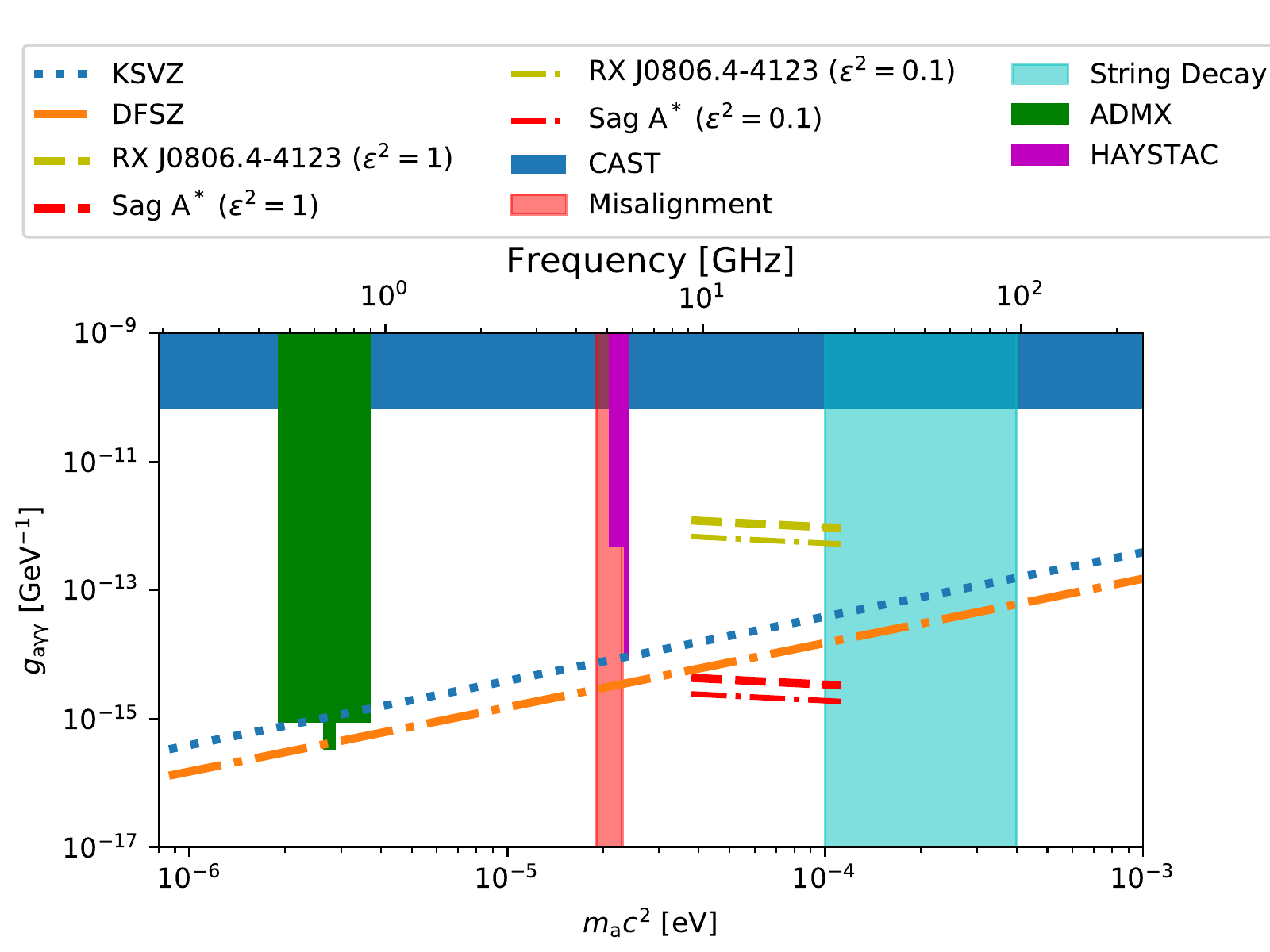}
 \caption{The sensitivity of radio telescopes to the resonant axion-photon decay. In the left panel, we plot the sensitivity of a single dish telescope between 0.3 and 10 GHz, assuming a system temperature of $30\,{\rm K}$, a diameter $D_{\rm tel} = 300\,{\rm m}$ and efficiency $\eta = 0.5$. These numbers are representative of an Arecibo-like single dish system. In the right panel, we plot the sensitivity representative of  SKA2:Band 5, assuming $\eta = 0.7$, $T_{\rm sys} = 30\,{\rm K}$ and $A_{\rm eff} = 10^6\,{\rm m^2}$. The yellow line is for the isolated neutron star RX J0806.4-4123, while the red line is for the neutron star observed to be near Sagittarius $\rm A^{\ast}$ with $\tilde{\rho} \approx 10^9$. We have use dashed lines for the case where the geometrical factor $\epsilon^2=1$ in (\ref{doppler:estimate}), that is, where the Doppler broadening is maximal and dot-dashed lines for when $\epsilon^2 = 0.1$. We remark that these estimates have been chosen to be representative of the typical sensitivity one might expect to achieve. In reality, one would need to take into account the variation in system temperature as a function of frequency.}
 \label{fig:NS_Sensitivity}
\end{figure*}

In fig.~\ref{fig:NS_Sensitivity}, we plot the sensitivity of the Arecibo telescope (left panel) and the SKA2:Band 5 (right panel) assuming a system temperature of 30 K and pulsar mass and radius of $1 \,M_{\odot}$ and $10\,{\rm km}$, respectively. For our standard sensitivity estimate, we consider the pulsar RX J0806.4-4123 \cite{Kaplan2009}. For this pulsar, $\Omega \approx 0.5\, {\rm Hz}$, $B_0 \approx 2.5\times 10^{13}$ G and $r(z) \approx 250$ pc. We also consider the magnetar near the galactic centre SGR J1745-2900, for which $B_0 \approx 1.4 \times 10^{14}\,{\rm G}$ and $\Omega \approx 1.67\,{\rm Hz}$. Note that our sensitivity estimates are more than 2 orders of magnitude weaker than those of reference \cite{ref:NS-Hook}. This is because our estimate of the bandwidth is approximately 2 orders of magnitude larger. Furthermore, no radio telescope is 100\% efficient and therefore the system-equivalent-flux-density (SEFD) for the Arecibo telescope is actually a little larger than 2 Jy. We would like to stress that radio observations of the axion-photon decay are most useful when they are complementary to the haloscope searches, which cannot probe arbitrarily high axion masses\footnote{We note that the MADMAX axion haloscope \citep{Majorovits2017} is sensitive to axions masses predicted by the string decay mechanism ($100~\mu{\rm eV/c^2}\lesssim m_{\rm a}\lesssim 400\mu{\rm eV/c^2}$).}. 
 
\section{Summary and Discussion}\label{sect:conclusions}

In this work we have clarified and extended the analysis of spontaneous decays and resonant conversion of dark matter axions, with an extensive discussion of both theory and observations. Axion masses larger than 100 ${\rm \mu eV}/c^2$ cannot be probed by axion haloscopes and these masses have been motivated by studies of axion string decay \cite{ref:Battye, ref:WS} and the non-linear substructure formed as a result \cite{ref:KhlopovArchioles}. In the case of the detection of the spontaneous decay, previous work \citep{ref:Caputo,ref:Caputo1} has suggested that nearby dwarf spheroidal galaxies are ideal candidates to observe under the claim they maximise the flux density. 
In our analysis, we argue that a procedure to maximise the flux density signal can be non-trivial. This is because the resolution of most single-dish radio telescopes is such that the beam size is smaller than the apparent size of these dwarf galaxies, resolving them. As a result, it becomes confusing to optimise an experiment where one is interested in maximising the flux-density signal since it is difficult to observationally determine the virial radius of dwarf galaxies \cite{ref:DwarfSpheroidalDiff}. 

Our analysis highlights the fact that one need not carry out a matching procedure of sources to the resolution of the telescope. Indeed, the relevant quantity that determines the specific intensity is the ratio of the surface-mass density to the velocity dispersion, $\Sigma_{\rm beam}/\Delta v$. Our results show that, except a weak trend in the halo concentration parameter with respect to the mass, this ratio is independent of the halo mass. We infer that a high resolution is in fact desirable, since the surface-mass density along the line of sight is enhanced for a more concentrated beam. This result motivates a search for structures that are characterised by large values of $\Sigma_{\rm beam}$. 

An important point that was first studied in \cite{ref:Caputo, ref:Caputo1} is the enhancement due to simulated emission, since photons and axions are both described by Bose-Einstein statistics and are therefore indistinguishable. Thus, the presence of an ambient radiation field at the same frequency as that of the axions results in an effective enhancement of the decay. This is quantified by the photon occupation number.

Our analysis of the sensitivity to virialised objects shows that it is virtually impossible to design a conventional interferometer that can constrain the axion-photon coupling below the CAST limit. This is due to the fact that the sensitivity of an interferometer to any brightness temperature signal is weakened by a filling factor that increases the integration time to unachievable values. On the other hand, one may use an interferometer as a ``light-bucket", where all dishes are used in single-dish auto-correlation mode. We show that even in this case, one would require 4 days of on-source integration time with band 5 of SKA2 observing the Virgo cluster to improve on the CAST constraints \cite{ref:CAST} on the axion-photon coupling.

Our previous results indicate that the ideal source for detecting the spontaneous decay is characterised by large values of $\Sigma_{\rm beam}$ and large amounts of ambient radiation at the same frequency - corresponding to ${\cal F}_{\rm eff}\gg 1$. Our order of magnitude estimate of the sensitivity to the galactic centre due to synchrotron emission (using the Planck Point Source catalogue \cite{ref:PSC}) from Sagittarius $\rm A^{\ast}$ shows that this may be an ideal target to improve the CAST limit. This motivates a further detailed study of the radiation field in the galactic centre, which in principle could include components from Anomalous Microwave Emission (AME) (dust), free-free emission as well as synchrotron radiation.

For the resonant conversion in neutron star magnetospheres, we began our analysis with a careful discussion axions in magnetised plasmas. Our results on mixing in 1D formalise many of the aspects of axion-photon conversion in magnetised plasmas, where we derived a controlled WKB expansion, mass-shell conditions in inhomogeneous backgrounds, adiabatic/non-adiabatic limits and their application to the neutron star case. We also hope that our arguments lay bare the precise geometrical assumptions needed to simplify the axion-Maxwell equations to a simple 1D form involving only the parallel electric component and the axion. It seems that these assumptions seem not always transparent in the literature.

Our results on 3D mixing, though preliminary, raise some interesting questions about the role geometry can play in exciting additional plasma modes and highlights the qualitative importance of both the perpendicular photon polarisations $\textbf{E}_\perp$ and the emergence of a longitudinal mode. Of course, one might be tempted to argue that it is a well-known and unsurprising fact that $\textbf{E}_\perp$ also becomes active in a magnetised plasma due to mixing between $\textbf{E}_\perp$ and $\textbf{E}_\parallel$. This is a typical feature of anisotropic media of which a magnetic background is but one example. Formally, this can be seen to happen in the off-diagonal components of the conductivity \eqref{sigma} (see also equation (29) of ref.~\cite{ref:LaiHeyl}). However, our argument is more subtle than this. For the neutron star case, one is typically in the high-magnetisation limit where the cyclotron frequency greatly exceeds that of the photon, $\omega_{\rm B} \gg \omega$, thus conductivity-induced mixing is, in fact, switched off, but nonetheless a mode $\textbf{E}_\perp$ is activated. For example, the Gauss equation shows the complications arising from longitudinal modes in the high-magnetisation limit
\begin{equation}
 \left(1 - \frac{\omega_{\rm pl}^2}{\omega^2}\right)\nabla \cdot \textbf{E}_\parallel + \nabla \cdot \textbf{E}_\perp = - g_{\rm a \gamma \gamma} \nabla a \cdot \textbf{B}\,,
\end{equation}
which clearly couples $\textbf{E}_\perp$, $\textbf{E}_\parallel$ and $a$ in a non-trivial way with the axion fundamentally changing the form of Maxwell's equations. Given the arguments of subsec.~\ref{sec:1DLimitations}, it seems that the only way to prevent the additional modes becoming active and to completely decouple $\textbf{E}_\perp$ and the effects of $\nabla \cdot \textbf{E} \neq 0$ terms, is to choose a very peculiar geometry in which the momentum of the axion and photon are normal to the magnetic field and all background gradients are dominated by a particular preferred direction parallel to the momentum. For a neutron star magnetosphere, this is clearly too simplistic and the plasma will exert gradient forces on the photon causing it to deflect away from the initial axion trajectory. 

It is also worth noting that in going from the planar 1D setup to 3D backgrounds, results are not only modified by the presence of additional polarisations, but also the nature of mixing itself may change. Specifically, any analytic formulae for conversion probabilities must be sensitive to the dimensionality of the underlying differential equations to be solved. For example, an analytic expression such as the Landau-Zener formula \eqref{eq:LZ} is a 1D result and is expected to be modified in higher dimensions regardless of which polarisations are most important.

The ultimate goal is clearly to determine \textit{quantitatively} what are the effects of 3D geometries since this is the most relevant aspect from an observational perspective. This will either entail more work to obtain analytic results in 3D and/or cross-checking these results against numerical simulations. Of course, the latter option presents some numerical challenges, as explained in sec.~\ref{eq:2DExample} owing principally to the fact wavelengths are much smaller than background scales over which the equations must be integrated, in contrast to \cite{Knirck:2019eug} where laboratory halosccopes contain only a few wavelengths. Clearly this limit strongly suggests the problem should be amenable to a WKB expansion, which would require performing a gradient expansion of the axion-Maxwell's equations in 3D as done in 1D in appendix \ref{Density}. The resulting 3D transport equations may allow one to circumvent the issue resolving the wavefront structure and track only field amplitudes/number densities relevant for computing the flux. 

We showed that Doppler broadening of the signal due to the relative motion of the neutron star results in a bandwidth that is at least two orders of magnitude larger than the value estimated in reference \cite{ref:NS-Hook} if the axion velocity is non-relativistic in the resonant conversion region. As a result, our sensitivity estimates are significantly weaker in the mass range $1-10\,{\rm \mu eV}$. Therefore, we emphasise the need to identify candidate pulsars that are located in regions of dark matter density peaks large enough to make the flux density detectable. We find that, in possibly realistic scenarios, it is possible to significantly improve on the CAST constraints from 4 days of observations using the Arecibo telescope. We also find that the SKA2:Band 5 can possibly rule out DFSZ axions of $m_{\rm a}c^2 \geq 20\,{\rm \mu eV}$, assuming an enhancement of $10^9$ in the dark matter density at the location of the magnetar near the galactic centre.

\begin{center}
\textbf{Acknowledgements}
\end{center}
JIM would like to thank Carlos Tamarit and Francesca Chadha-Day for general discussions and Georg Raffelt for useful conversations on wave optics and mixing in higher dimensions as well as Juan Cruz for help with the cluster. This work is supported by an Alexander von Humboldt Fellowship and by the Collaborative Research Centre SFB 1258 of the Deutsche Forschungsgemeinschaft. RAB and FP acknowledge support from Science and Technology Facilities Council (STFC) grant ST/P000649/1. SS is supported by a George Rigg Scholarship from the University of Manchester. SS would like to thank Keith Grainge, Dominic Viatic, Joel Williams and Joshua Hayes for useful advice.

\appendix

\section{Mass inside the beam radius}\label{App:MassinBeam}
Consider a halo density profile $\rho(r)=\rho_{\rm s}F(r/r_{\rm s})$ for $r<R_{\rm vir}$ and zero otherwise. In the function $F(y)$, $r_{\rm s}$ is the scale radius, $R_{\rm vir}$ the virial radius and the ratio of the two ${\hat c}=R_{\rm vir}/r_{\rm s}$ is the concentration parameter. For the specific case of an Navarro-Frenk-White (NFW) profile \cite{ref:NFW}, $F(y)=y^{-1}(1+y)^{-2}$. The mass inside the virial radius is given by
\begin{equation}
\label{mvir}
M_{\rm vir}=4\pi\rho_{\rm s}r_{\rm s}^3\int_0^{\hat c} x^2F(x)dx\,,
\end{equation}
and the surface mass density at some radius, $R$, is 
\begin{equation}\label{surfacemass}
\Sigma(R)=2r_{\rm s}\rho_{\rm s} \int_{{\bar R}}^{\hat c}\frac{yF(y)dy}{\sqrt{y^2-{\bar R}^2}}\,,
\end{equation} 
where ${\bar R}={\hat c}R/R_{\rm vir}$. Both expressions (\ref{mvir}) and (\ref{surfacemass}) converge for $\bar{R}\rightarrow 0$.

In this work, we are particularly interested to the mass inside the radius of a telescope and defined by $R_{\rm beam}$. We can evaluate this from 
\begin{align}
M_{\rm beam} = &\, 2\pi\int_0^{R_{\rm beam}}R\Sigma(R)\mathrm{d}R\,,\\
= &\, M_{\rm vir}\frac{\int_0^{{\bar R}_{\rm beam}}x\mathrm{d}x \int_x^{\hat{c}}{\frac{yF(y)dy}{\sqrt{y^2-x^2}}}}
      {\int_0^{\hat{c}} x^2F(x)\mathrm{d}x}\,,\nonumber
\end{align}
where $\bar{R}_{\rm beam}=\hat{c}R_{\rm beam}/R_{\rm vir}$. By manipulating the double integral, we can deduce that 
\begin{equation}\label{eqn:Exact}
\frac{M_{\rm beam}}{M_{\rm vir}} = 1-\frac{\int_{\bar{R}_{\rm beam}}^{\hat{c}}\sqrt{y^2-\bar{R}_{\rm beam}^2} F(y)\mathrm{d}y}{\int_0^{\hat{c}}x^2F(x)\mathrm{d}x}\,.
\end{equation}

For an NFW profile \cite{ref:Coe} 
\begin{equation}\label{eqn:Massinbeam}
 \begin{split}
  \frac{M_{\rm beam}}{M_{\rm vir}} \approx \frac{1}{f({\hat c})} & \left[ \log\left(\frac{{\bar R}_{\rm beam}}{2}\right) + \right.\\
  & \left. \,\, \frac{1}{\sqrt{1-{\bar R}_{\rm beam}^2}}\cosh^{-1}\frac{1}{{\bar R}_{\rm beam}}\right]\,,
 \end{split}
\end{equation}
where $f(x) = \log(1+x) - \frac{x}{1+x}$. For small ${\bar R}$ this is given by
\begin{equation}\label{eqn:SmallR}
 \frac{M_{\rm beam}}{M_{\rm vir}} = -{{\bar R}_{\rm beam}^2\frac{\log\left({\frac{{\bar R}_{\rm beam}}{2}}\right)}{2f({\hat c})}}\,.
\end{equation}
The analytic approximation  \eqref{eqn:SmallR} and the exact results \eqref{eqn:Exact} are shown respectively as solid/dotted lines in fig.~\ref{fig:VirgoMass}.

\section{Density matrix and Landau-Zener}\label{Density}
One can define a density matrix for the system by writing:
\begin{equation}
 \boldsymbol{ \rho}(z_1,z_2) = \left(a(z_1) , \mathcal{E}(z_1) \right) \otimes \left( a(z_2) , \mathcal{E}(z_2) \right)^\dagger\,,
\end{equation}
which satisfies
\begin{equation}
\left(
\begin{array}{cc}
 \partial_{z_1}^2 - m_{\rm a}^2 + \omega^2 & \quad 
 \omega g_{\rm a \gamma \gamma} B(z)\\ 
 \omega g_{\rm a \gamma \gamma} B(z) & 
 \partial_{z_1}^2 - \omega_{\rm pl}^2(z) + \omega^2
\end{array}
\right)
\boldsymbol{\rho}(z_1,z_2) = 0\,.
\end{equation}
One can introduce a local phase-space by performing a 1D Wigner transformation defined by
\begin{equation}
\boldsymbol{\rho}(k,z) = \int \mathrm{d}y \, \boldsymbol{\rho}\left( z + \frac{y}{2}, z -\frac{y}{2} \right) e^{-\imath k y}\,,
\end{equation}
with $y = z_1 - z_2$ and $z = (z_1+z_2)/2$, and using temporal translation invariance, one arrives at \cite{Prokopec:2003pj,Garbrecht:2002pd}
\begin{equation}\label{WigEq}
 \left[\omega^2 - k^2 + \frac{1}{4}\partial^2_z - \imath k \partial_z - \textbf{M}^2(z) \,e^{\frac{\imath}{2} \overleftarrow{\partial_z} \partial_k } \right] \rho (k,z) =0\,,
\end{equation}
where the Hermitian mass-mixing matrix is given by
\begin{equation}\label{eq:flavourMass}
 \textbf{M}^2  = \left(
\begin{array}{cc}
m_{\rm a}^2 &  \omega g_{\rm a\gamma \gamma} B(z)\\
\omega g_{\rm a\gamma \gamma}B(z) & \omega_{\rm pl}^2(z)
\end{array}
\right)\,,
\end{equation}
whose mass eigenvalues are
\begin{equation}
 \begin{split}
  M_{1,2}^2 = \frac{1}{2}\Big\{& m_{\rm a}^2 + \omega_{\rm pl}^2 \pm\\
  &\, \left.\left[ (m_{\rm a}^2 - m_{\rm p}^2(z))^2 + 4 B^2g_{\rm a\gamma\gamma}^2 \omega^2 \right]^{1/2} \right\}\,. \label{m12}
 \end{split}
\end{equation}
Since local physical states are mass-diagonal states, in order to extract useful dispersion information, we first convert to the local mass basis:
\begin{equation}
 \textbf{M}_d^2 = U M^2 U^{\dagger}\,, \qquad 
 \boldsymbol{\rho}_d = U \boldsymbol{\rho} U^{\dagger}\,, 
\end{equation}
where
\begin{equation}\label{eq:angle}
U = 
\left(
\begin{array}{cc}
\cos \theta & \qquad -\imath \sin{\theta} \\
- \imath \sin{\theta} & \cos{\theta}
\end{array}
\right)\,,
\end{equation}
with $\tan{2\theta} = \tfrac{\omega B(\textbf{x})g_{\rm a \gamma \gamma}}{m_{\rm a}^2 - m_\gamma^2}$, diagonalises the mass matrix, which amounts to the replacement
\begin{equation}
 \partial \rightarrow D_z = \partial_z - \imath \left[\Xi , \cdot \right]\,, \qquad \Xi = \imath U \partial_z U^{\dagger}\,,
\end{equation}
in eqs.~\eqref{WigEq}, leading to
\begin{equation}
\left[\omega^2 - k^2 + \frac{1}{4}D^2_z - \imath k \partial_z - \textbf{M}^2_d(z) \,e^{\frac{\imath}{2} \overleftarrow{D}_z \partial_k } \right] \boldsymbol{\rho}_d (k,z) =0\,, \label{WigEq2}
\end{equation}
where $\textbf{M}_d^2 = \text{diag}(M_1^2,M_2^2)$. Taking the hermitian and antihermitian parts of \eqref{WigEq} gives
\begin{align}
& \left(\omega^2 - k^2  + \frac{1}{4} D^2_z \right)\boldsymbol{\rho} - \frac{1}{2}\left\{\textbf{M}_c^2, \boldsymbol{\rho} \right\} + \frac{\imath}{2}\left[\textbf{M}_s^2, \boldsymbol{\rho} \right] = 0\,,\label{Constraint}\\
& k \,D_z \boldsymbol{\rho} + \frac{1}{2} \left\{\textbf{M}^2_s, \boldsymbol{\rho}\right\} - \frac{\imath}{2} \left[\textbf{M}_c^2, \boldsymbol{\rho} \right] = 0\,,\label{Kinetic}
\end{align}
where
\begin{subequations}
\begin{align}
 \textbf{M}_c^2 = &\, \textbf{M}^2 \cos{\left( \frac{1}{2} \overleftarrow{D}_z \partial_k\right)}\,, \\ 
 \textbf{M}_s^2 = &\, \textbf{M}^2\sin{\left( \frac{1}{2} \overleftarrow{D}_z \partial_k\right)}\,.
\end{align}
\end{subequations}
These are known as the \textit{constraint} and \textit{kinetic} equations respectively. The first contains information about dispersion relations and imposes appropriate mass-shell constraints, whilst the second controls the evolution of number densities. To leading order in gradients, the constraint equation~(\ref{Constraint}) implies the following Ansatz for the mass-basis density matrix $\boldsymbol{\rho}_d$
\begin{subequations}\label{Ansatz}
 \begin{align}
 \rho_{d, ii}(z,k) = n_i(z,k) \,\delta(\omega^2 -k^2 - M_i )\,, \\
 \rho_{d, ij} (z,k)= n_{ij}(z,k) \delta(\omega^2 - k^2 - \bar{M}^2)\,,
 \end{align}
\end{subequations}
where $\overline{M} = (M_1^2 + M_2^2)/2$ is the average mass and any total derivatives in $k$ drop out upon integration. In the present setup
\begin{equation}
\Xi = \left( 
\begin{array}{cc}
 0 & -\theta^{\prime} \\
 - \theta^{\prime} & 0
\end{array}
\right)\,,
\end{equation}
so that inserting (\ref{Ansatz}) into eq.~(\ref{Kinetic}) and integrating over $k$ to put all states on shell, leads to the following equations
\begin{subequations}
 \begin{align}
  \imath \frac{\mathrm{d} \textbf{N}(z)}{\mathrm{d}z} = &\, \left[ \textbf{H} , \textbf{N}(z) \right]\,,\\
  \textbf{H} = &\, 
  \left(
  \begin{array}{cc}
   M_1^2/2\bar{k} & \theta^{\prime} \\
   \theta^{\prime} & M^2_2/2\bar{k}
  \end{array}
  \right)\,, \\
  \textbf{N} = &\,
  \left( 
  \begin{array}{cc}
   n_{\rm 1} & n_{\rm 12}\\
   n_{\rm 21} & n_2
  \end{array}
  \right)\,,
 \end{align}
\end{subequations}
where $\bar{k}^2 = \omega^2 - \bar{M}^2$ is the ``average momentum" arising from the off-diagonal coherence terms. Reverting to the flavour basis, one finds
\begin{equation}
 i \frac{\mathrm{d}\textbf{N}_{\rm f}}{\mathrm{d}z} = \frac{1}{2\bar{k}}\left[\textbf{M}^2 ,\textbf{N}_{\rm f}\right]\,,
\end{equation}
where $\textbf{M}^2$ is the flavour mass matrix \eqref{eq:flavourMass}. For ``pure state" solutions, the system can be realised via a wavefunction $\textbf{N}_{\rm f} = \boldsymbol{\Psi} \otimes \boldsymbol{\Psi}^\dagger$, where $\boldsymbol{\Psi}=(\psi_{\rm a}, \psi_\gamma)$, corresponding to an auxilliary Sch{\"o}dinger-like equation
\begin{equation}\label{eq:FirstOrderMixing}
\imath \frac{\mathrm{d}}{\mathrm{d}z}
\left( 
\begin{array}{c}
\psi_{\rm a}\\
\psi_\gamma
\end{array}
\right) = \frac{1}{2 \bar{k}(z)} \left(
\begin{array}{cc}
m_{\rm a}^2 & \omega g_{\rm a\gamma \gamma} B(z)\\
\omega g_{\rm a\gamma \gamma}B(z) & \omega_{\rm pl}^2(z)
\end{array}
\right)
\left( 
\begin{array}{c}
\psi_{\rm a}\\
\psi_\gamma
\end{array}
\right)\,.
\end{equation}
For a problem in which the mass-splitting varies linearly with the integration parameter such that the mass-mixing takes the form:
\begin{equation}
\textbf{M}^2(z) = \left(
\begin{array}{cc}
\epsilon_1 + \lambda_1 z & v^{\ast}\\
v & \epsilon_2 + \lambda_2 z
\end{array}
\right)\,,
\end{equation}
where $\epsilon_1, \lambda_1 \in \mathbb{R}$ and $v \in \mathbb{C}$ are constants, the $S$-matrix for conversion probabilities is given by the well-known Landau-Zener formula \cite{Brundobler}
\begin{equation}
S_{\rm LZ} = 
\left(
 \begin{array}{cc}
  p & \quad q \\
  q & \quad p \\
 \end{array}
\right)\,,
\end{equation}
where $p = e^{-\pi \gamma}$, $q = \sqrt{1 - p^2}$ and $\gamma = |v|^2/|\lambda_2 - \lambda_1|$. 
Thus, by linearising the plasma frequency in \eqref{eq:FirstOrderMixing} about $z=z_{\rm c}$ with $\omega_{\rm pl}^2 \simeq m_{\rm a}^2 + (z-z_{\rm c})(\omega_{\rm pl}^2)^{\prime}(z_{\rm c}))$, we can immediately read off the form of $\gamma$, leading to 
\begin{equation}
\gamma = \frac{g^2_{\rm a \gamma \gamma} B^2(z_{\rm c}) \omega^2/2\bar{k}}{\left|(\omega_{\rm pl}^2)^{\prime}(z_{\rm c})\right|}.
\end{equation}
The conversion probability is then given by the squared S-matrix elements:
\begin{equation}
P_{\rm a\rightarrow \gamma} = 1 - e^{-2 \pi \gamma}\,, \qquad 
\gamma \simeq \frac{\Delta M^2(z_{\rm c})/2 \bar{k}}{4\left|\theta^{\prime}(z_{\rm c})\right|}\,,
\end{equation}
where we used the definitions \eqref{eq:angle} and \eqref{m12} to parameterise the probability in terms of the mass-splitting $\Delta M^2 = M_1^2 - M_2^2$ and mixing angle gradients, evaluated at the resonance, where we neglected gradients in $B(z)$ and $\bar{k}(z)$.

\bibliographystyle{apsrev4-1}
\bibliography{Ref.bib}

\label{lastpage}
\end{document}